\documentclass[iop,apj,twocolappendix,numberedappendix,appendixfloats]{emulateapj}

\usepackage[T1]{fontenc}
\usepackage{ae,aecompl}

\usepackage{natbib}
\usepackage{graphicx}	
\usepackage{amsmath}	
\usepackage{amssymb}	
\usepackage{multirow}
\usepackage{txfonts}

\usepackage[breaklinks,colorlinks,citecolor=blue,linkcolor=magenta]{hyperref} 
\usepackage[all]{hypcap}

\begin{document}

\newcommand{\jhst}{$J_{\rm 125}$}	
\newcommand{\jhhst}{$JH_{\rm 140}$}
\newcommand{\hhst}{$H_{\rm 160}$}
	

\shorttitle{Environment in the 3D-HST fields}
\shortauthors{M.Fossati et al.}

\title[Environment in the 3D-HST fields]{Galaxy environment in the 3D-HST fields. \\ Witnessing the onset of satellite quenching at $z\sim 1-2$.}

\author{M.~Fossati\altaffilmark{1}$^{,}$\altaffilmark{2}, D.J.~Wilman\altaffilmark{1}$^{,}$\altaffilmark{2},
J.T.~Mendel\altaffilmark{2}$^{,}$\altaffilmark{1}, 
R.P.~Saglia\altaffilmark{2}$^{,}$\altaffilmark{1}, 
A.~Galametz\altaffilmark{2}$^{,}$\altaffilmark{1},
A.~Beifiori\altaffilmark{1}$^{,}$\altaffilmark{2}, 
R.Bender\altaffilmark{1}$^{,}$\altaffilmark{2}, 
J.C.C.~Chan\altaffilmark{1}$^{,}$\altaffilmark{2}, 
M.~Fabricius\altaffilmark{2},
K.~Bandara\altaffilmark{2}, 
G.B.~Brammer\altaffilmark{3}, 
R.~Davies\altaffilmark{2}, 
N.M.~F{\"o}rster Schreiber\altaffilmark{2}, 
R.~Genzel\altaffilmark{2}$^{,}$\altaffilmark{4}$^{,}$\altaffilmark{5}, 
W.~Hartley\altaffilmark{6}, 
S.K.~Kulkarni\altaffilmark{2}, 
P.~Lang\altaffilmark{2}, 
I.G.~Momcheva\altaffilmark{6}, 
E.J.~Nelson\altaffilmark{7}$^{,}$\altaffilmark{2}, 
R.~Skelton\altaffilmark{8}, 
L.J.~Tacconi\altaffilmark{2}, 
K.~Tadaki\altaffilmark{2}, 
H.{\"U}bler\altaffilmark{2}, 
P.G. van Dokkum\altaffilmark{7},
E.~Wisnioski\altaffilmark{2}, 
K.E.~Whitaker\altaffilmark{9}$^{,}$\altaffilmark{$*$}, 
E.~Wuyts\altaffilmark{2}, 
S.~Wuyts\altaffilmark{10}}

\altaffiltext{1}{Universit{\"a}ts-Sternwarte M{\"u}nchen, Scheinerstrasse 1, D-81679 M{\"u}nchen, Germany}
\altaffiltext{2}{Max-Planck-Institut f{\"u}r Extraterrestriche Physik, Giessenbachstrasse, D-85748 Garching, Germany}
\altaffiltext{3}{Space Telescope Science Institute, 3700 San Martin Drive, Baltimore, MD 21218, USA}
\altaffiltext{4}{Department of Physics, Le Conte Hall, University of California, Berkeley, CA 94720, USA}
\altaffiltext{5}{Department of Astronomy, Hearst Field Annex, University of California, Berkeley, CA 94720, USA}
\altaffiltext{6}{Department of Physics \& Astronomy, University College London, Gower Street, London, WC1E 6BT, UK}
\altaffiltext{7}{Astronomy Department, Yale University, New Haven, CT 06511, USA}
\altaffiltext{8}{South African Astronomical Observatory, P.O. Box 9, Observatory, Cape Town, 7935, South Africa}
\altaffiltext{9}{Department of Astronomy, University of Massachusetts, Amherst, MA 01003, USA}
\altaffiltext{10}{Department of Physics, University of Bath, Claverton Down, Bath, BA2 7AY, UK}
\altaffiltext{$*$}{Hubble Fellow}

\email{mfossati@mpe.mpg.de}

\begin{abstract}
We make publicly available a catalog of calibrated environmental measures for galaxies in the five 3D-HST/CANDELS 
deep fields. Leveraging the spectroscopic and grism redshifts from the 3D-HST survey, multi wavelength 
photometry from CANDELS, and wider field public data for edge corrections, we derive densities in fixed apertures 
to characterize the environment of galaxies brighter than $JH_{140} < 24$ mag in the redshift range $0.5<z<3.0$. 
By linking observed galaxies to a mock sample, selected to reproduce the 3D-HST sample selection and redshift accuracy, 
each 3D-HST galaxy is assigned a probability density function of the host halo mass, and a probability that is a central or a 
satellite galaxy. The same procedure is applied to a $z=0$ sample selected from SDSS. We compute the fraction of passive 
central and satellite galaxies as a function of stellar and halo mass, and redshift, and then derive the fraction of galaxies 
that were quenched by environment specific processes. Using the mock sample, we estimate that the 
timescale for satellite quenching is $t_{\rm quench} \sim 2-5$ Gyr; longer at lower stellar mass or lower redshift, but remarkably 
independent of halo mass. This indicates that, in the range of environments commonly found within the 3D-HST 
sample ($M_h \lesssim 10^{14} M_\odot$), satellites are quenched by exhaustion of their gas reservoir in absence of cosmological 
accretion. We find that the quenching times can be separated into a delay 
phase during which satellite galaxies behave similarly to centrals at fixed stellar mass, and a phase where the star formation rate 
drops rapidly ($\tau_f \sim 0.4-0.6$ Gyr), as shown previously at $z=0$. We conclude that this scenario requires satellite 
galaxies to retain a large reservoir of multi-phase gas upon accretion, even at high redshift, and that this gas sustains 
star formation for the long quenching times observed.
\end{abstract}


\maketitle


\section{Introduction}
It has long been known that galaxies are shaped by the environment in which they reside.  Works 
by e.g., \citet{Oemler74}, \citet{Dressler80}, and \citet{Balogh97} showed that galaxies in high-density environments are 
preferentially red and early-type compared to those in lower density regions. The more recent advent 
of large scale photometric and spectroscopic surveys confirmed with large statistics those early findings
\citep{Balogh04, Kauffmann04, Baldry06}.
Meanwhile, space and ground based missions have probed the geometry of our Universe. Those observations 
coupled to cosmological models have built the solid Lambda cold dark matter ($\Lambda$CDM) framework \citep{White78, Perlmutter99}, 
in which lower mass haloes are the building blocks of more massive structures. One of the major tasks for modern studies of 
galaxy formation is therefore to understand how and when galaxy evolution is driven by internal 
processes or the evolving environment that each galaxy experiences during its lifetime. 
While internal mechanisms, including ejective feedback from supernovae or active galactic nuclei, are deemed 
responsible for suppressing star formation in all galaxies \citep{Silk98, Hopkins08}, 
a galaxy can also directly interact with its environment when falling into a massive, gas- and galaxy-rich 
structure such as a galaxy cluster.

At low redshift detailed studies of poster child objects \citep{Yagi10, Fossati12, Fossati16, Merluzzi13, Fumagalli14, Boselli16} coupled with 
state-of-the-art models and simulations \citep{Mastropietro05, Kapferer09, Tonnesen10} have started to explore the rich 
physics governing those processes \citep[e.g.,][for reviews]{Boselli06, Boselli14c, Blanton09}. 
Broadly speaking, they can be grouped into two classes. The first of them includes gravitational interactions between
cluster or group members \citep{Merritt83} or with the potential well of the halo as a whole \citep{Byrd90}, 
or their combined effect known as ``galaxy harrassment'' \citep{Moore98}. The second class includes hydrodynamical
interactions between galaxies and the hot and dense gas that permeates massive haloes. 
This class includes the rapid stripping of the cold gas via ram pressure as the galaxy passes through the hot gas 
medium \citep{Gunn72}.  Ram-pressure stripping is known to effectively and rapidly suppress star formation in cluster 
galaxies in the local Universe \citep{Solanes01, Vollmer01, Gavazzi10, Gavazzi13a, Boselli08, Boselli14b}. 

Less directly influencing the galaxy's current star formation, the multi-phase medium 
(e.g. warm, hot gas) associated to the galaxy (known as the ``reservoir'') should be easier to strip than the cold 
gas. Even easier, the filamentary accretion onto the galaxy from the surrounding cosmic web will be truncated 
as the galaxy is enveloped within the hot gas of a more massive halo \citep{White91}. Both of these processes 
will suppress ongoing accretion onto the cold gas disk of the galaxy and lead to a more gradual 
suppression of star formation, variously labelled ``strangulation'' or ``starvation'' \citep[e.g.][]{Larson80, Balogh97} 
These processes are complicated in nature and the exact details of their efficiency and dynamics are still poorly understood. 
The situation is even more complicated when several of those processes are found to act together \citep{Gavazzi01,Vollmer05}.

A different approach to disentangle the role of environment from the secular evolution is to study large samples
of galaxies and correlate their properties (e.g. star formation activity) to internal properties (e.g. stellar mass) and environment.
In the local Universe, the advent of the Sloan Digital Sky Survey (SDSS) has revolutionized the field of large statistical studies
and allowed for the effects of the environment on the galaxy population as a whole to be studied
\citep{Kauffmann04,Baldry06, Peng10, Peng12, Wetzel12, Wetzel13, Hirschmann14}. One of the main results is that 
environmental quenching is a separable process that acts on top of the internal processes that regulate the star 
formation activity of galaxies. 
A crucial parameter to understand the collective effect of the several environmental processes is the timescale over which the
star formation activity is quenched. Several authors took advantage of excellent statistics to estimate the average timescale 
for environmental quenching, accounting for internal quenching processes, and found that in the low redshift 
Universe this is generally long \citep[$\sim 5-7$ Gyr;][]{McGee09, DeLucia12, Wetzel13, Hirschmann14}, while possibly 
shorter in clusters of galaxies \citep[$\sim 2-5$ Gyr;][]{Haines15, Paccagnella16} .

At higher redshift, the situation is made more complex due to the more limited availability of spectroscopic redshifts which are
paramount to depict an accurate picture of the environment. In the last decade, several 
ground based redshift surveys started to address this issue \citep{Wilman05, Cooper06}. By exploiting the multiplexing 
of spectroscopic instruments at 8-10 meter class telescopes \citep[e.g. VIMOS and GMOS,][]{Lilly07, Kurk13, Balogh14}, 
these works showed that the environment plays a role in quenching the star formation activity of galaxies accreted onto
massive haloes (satellite galaxies) up to $z\sim 1$ \citep{Muzzin12, Quadri12, Knobel13, Kovac14, Balogh16}, although the samples 
are limited to massive galaxies or a small number of objects.

Low-resolution space-based slitless spectroscopy is revolutionising this field providing deep and highly complete 
spectroscopic samples. The largest of those efforts is the 3D-HST survey \citep{Brammer12} which, by combining a 
large area, deep grism observations and a wealth of ancillary photometric data, provides accurate redshifts to $\Delta z/(1+z) \sim 0.003$ 
\citep{Bezanson16} for a large sample of objects down to low stellar masses ($\sim 10^{9} M_\odot$, and $\sim 10^{10} M_\odot$ 
at $z\sim1$ and $z\sim2$ respectively). The public release of their spectroscopic observations \citep{Momcheva16}, 
in synergy with deep photometric observations \citep{Skelton14} has opened the way to an accurate quantification 
and calibration of the environment over the redshift range $z\sim 0.5-3$. 

Another source of uncertainty in the interpretation of correlations of galaxy properties with environment is the inhomogeneity 
of methods used for different surveys \citep[e.g.,][]{Muldrew12, Haas12, Etherington15} and the lack of calibration of
important parameters such as halo mass. In \citet{Fossati15}, we studied how to link a purely observational 
parameter space to physical quantities (e.g., halo mass, central/satellite status) by analysing a stellar mass limited sample 
extracted from semi-analytic models of galaxy formation. To do so, we computed a projected density field in the 
simulation box and we tested different definitions of density at different redshift accuracy. Our method is Bayesian in 
nature (galaxies have well-defined observational parameters, while the calibration into physical parameters is probabilistic). 
This approach is best suited to statistical studies where the application of selection functions and observational 
uncertainties can be fully taken into account. 

In this paper, we extend this method to the 3D-HST survey by building up an environment catalogue which we make available to the 
community with this work\footnote[1]{\url{http://dx.doi.org/10.5281/zenodo.168056}}. 
We then explore the efficiency and timescales for quenching of satellite galaxies over cosmic time ($z\sim 0-2$)
by combining the 3D-HST data at high redshift with SDSS data in the local Universe in a homogeneous way. We also address the long
standing issue of impurity and contamination of the calibrated parameters (the fact that the observations do not perfectly constrain the 
halo mass of the parent halo for each galaxy or its central/satellite status) by recovering the ``pure'' trends 
using the mock sample as a benchmark.

The paper is structured as follows. In Section \ref{obssample}, we introduce the 3D-HST dataset. In Section \ref{sec_environment},
we derive the local density for 3D-HST galaxies including accurate edge corrrections. Section \ref{sec_overdensities} presents the
range of environments in the 3D-HST area and how they compare to known galaxy structures from the literature. In Section
\ref{sec_models}, we introduce the mock galaxy sample and how we calibrate it to match the 3D-HST sample.  We then  
link models and observations in Section \ref{sec_halomass}, and assign physical quantities to observed galaxies. 
In Section \ref{sec_quenching} we study the quenching of satellite galaxies at $0 < z < 2.5$, and derive quenching efficiency 
and timescales. Lastly, we discuss the physical implications of our findings in Section \ref{sec_discussion} and summarize our work 
in Section \ref{sec_conclusions}.

All magnitudes are given in the AB system \citep{Oke74} and we assume a flat $\Lambda$CDM Universe with
$\Omega_M = 0.3$, $\Omega_\Lambda = 0.7$, and $H_0 = 70~\rm{km~s^{-1}~Mpc^{-1}}$ unless otherwise specified.
Throughout the paper, we use the notation $\log(x)$ for the base 10 logarithm of $x$. 

\section{The observational sample} \label{obssample}
In this work we aim at a quantification and calibration of the local environment for 
galaxies in the five CANDELS/3D-HST fields \citep{Grogin11, Koekemoer11, Brammer12} namely 
COSMOS, GOODS-S, GOODS-N, AEGIS and UDS.
The synergy of these two surveys represents the largest effort to
obtain deep space-based near-infrared photometry and spectroscopy in those fields.
For a description of the observations and reduction techniques, we refer the reader to 
\citet{Skelton14} and \citet{Momcheva16} for the photometry and spectroscopy respectively. 
The CANDELS observations provide HST/WFC3 near infrared imaging in the F125W and F160W filters (\jhst \ and \hhst \ hereafter) 
for all the fields, while 3D-HST followed-up a large fraction of this area with the F140W filter (\jhhst \ hereafter) and 
the WFC3/G141 grism for slitless spectroscopy. The novelty of this approach is to obtain low resolution 
($R\sim100$) spectroscopy for all the objects in the field. Taking advantage of the low background of the 
{\it HST} telescope, it is possible to reach a depth similar to traditional slit spectroscopy from 10m class telescopes 
on Earth. Hereafter, we use the term ``3D-HST'' sample to refer to the combination of CANDELS and all the 
other space- and ground-based imaging datasets presented in \citet{Skelton14}, plus the grism 
spectroscopy of the 3D-HST program.

The 3D-HST photometric catalog \citep{Skelton14} used \hhst\ or \jhhst\ as detection bands and its depth 
varies from field to field and across the same field due to the observing strategy of CANDELS. 
However, even in the shallowest portions of each field, the 90$\%$ depth confidence level is 
$H_{\rm 160} \sim 25$ mag. Beyond this magnitude limit, the star/galaxy classification (which is
a key parameter for the environment quantification) becomes uncertain. 

The 3D-HST spectroscopic release \citep{Momcheva16} provides reduced and extracted spectra down to\footnote[2]{The 3D-HST
spectroscopic catalog is based on $JH_{\rm IR}=J_{125}+JH_{140}+H_{160}$ magnitudes. We therefore quote limits in this band when
referring to their catalog. However in this paper the sample selection is performed in \jhhst \ to ensure that the footprint of our
sample is entirely covered by G141 grism observations and direct imaging in F140W.  }
$JH_{\rm IR} = 26$ mag. The spectra are passed through the EAZY template fitting code \citep{Brammer08} along with
the extensive ground- and space-based multiwavelength photometry. This results in ``grism'' redshifts, which
are more accurate than photometric redshifts thanks to the wealth of stellar continuum and emission line 
features present in the spectra.
However, only objects brighter than $JH_{\rm  IR} = 24$ mag have been visually inspected, and have a \texttt{use\_grism}
flag that describes if the grism spectrum is used to compute the redshift. Incomplete masking of contaminating 
flux from nearby sources in the direction of the grism dispersion, residuals from spectra of bright stars, and
corrupted photometric measurements can lead to this flag being set to 0 (``bad'').

We include in the present analysis all galaxies brighter than $JH_{\rm  140} = 24$ mag, therefore limiting our 
footprint to the regions covered by grism and \jhhst \ observations. We limit the redshift range to $0.5 < z < 3.0$.
The lower limit roughly corresponds to the redshift where the $\rm H\alpha$ line enters the grism coverage and the upper
limit is chosen such that the number density of objects above the magnitude cut allows a reliable estimate of the 
environment. It also allows follow-up studies targeting the rest-frame optical features from ground
based facilities in the $J$, $H$, and $K$ bands (e.g. $\rm{KMOS}$, \citealt{Sharples13} and MOSFIRE, \citealt{McLean12})

We exclude stars by requiring \texttt{star\_flag} to be 0 or 2 (galaxies or uncertain classification). We do not 
use the \texttt{use\_phot} 
flag because it is too conservative for our goals. Indeed this flag requires a minimum of 2 exposures in the 
F125W and F160W filters, and the object not being close to bright stars. The quantification of environment 
requires a catalog which is as complete as possible 
even at the expenses of more uncertain photometry (and photo-z) for the objects that do not meet those cuts.
Nonetheless a $JH_{\rm 140} = 24$ mag cut allows a reliable star/galaxy separation for 99\% of the objects and 
is at least 1 mag brighter than the minimum depth of the mosaics, thus alleviating the negative effects of nearby stars on 
faint sources. The final sample is made of 18745 galaxies.

As a result of the analysis in \citet{Momcheva16}, each galaxy is assigned a ``best'' redshift. This is:
\begin{enumerate}
\item a spectroscopic redshift from a ladder of sources as described below. 
\item a grism redshift if there is no spectroscopic redshift and \texttt{use\_grism = 1}
\item a pure photometric redshift if there is no spectroscopic redshift and \texttt{use\_grism = 0}.
\end{enumerate}
A \texttt{zbest\_type} flag is assigned to each galaxy based on the conditions above. 
The best redshift is the quantity used to compute the environment for each galaxy in the 3D-HST fields.

Spectroscopic redshifts are taken from the compilation of \citet{Skelton14} which we complement with newer data.
For the COSMOS field we include the final data release of the zCOSMOS bright survey \citep{Lilly07}. We find
253 new sources with reliable redshifts in the 3D-HST/COSMOS footprint mainly at $z<1$. 
In COSMOS and GOODS-S ,we include 95 objects from the DR1 \citep{Tasca16} of the VIMOS Ultra Deep 
Survey \citep[VUDS, []{LeFevre15}.  This survey mainly targets galaxies at $z>2$ therefore complementing 
zCOSMOS. We include 105 redshifts from the MOSFIRE Deep Evolution Field Survey \citep[MOSDEF, ][]{Kriek15} 
which provides deep rest frame optical spectra of galaxies selected from 3D-HST. 
For the UDS field, we also include 164 redshifts from VIMOS spectroscopy in a narrow slice of redshift 
($0.6 < z < 0.7$, Galametz et al.  in prep.) 
Lastly, we include 376 and 33 secure spectroscopic redshifts from $\rm{KMOS^{3D} }$ \citep {Wisnioski15} and VIRIAL
\citep{Mendel15} respectively. Those large surveys use the multiplexing capability of the 
integral field spectrometer KMOS on the ESO Very Large Telescope to follow-up 3D-HST selected objects. 
The former is a mass selected survey of emission line galaxies at $0.7<z<2.7$, while the latter observed passive 
massive galaxies at $1.5<z<2.0$. 

In the selected sample, $20\%$ of the galaxies have a spectroscopic redshift, 
$64\%$ have a grism redshift, and only $16\%$ have a pure photometric redshift.
In the next Section, we explore the accuracy of the grism and photometric redshifts as a function of the 
galaxy brightness and the $S/N$ of emission lines in the spectra.

\begin{figure*}
\centering
\includegraphics[scale=.85]{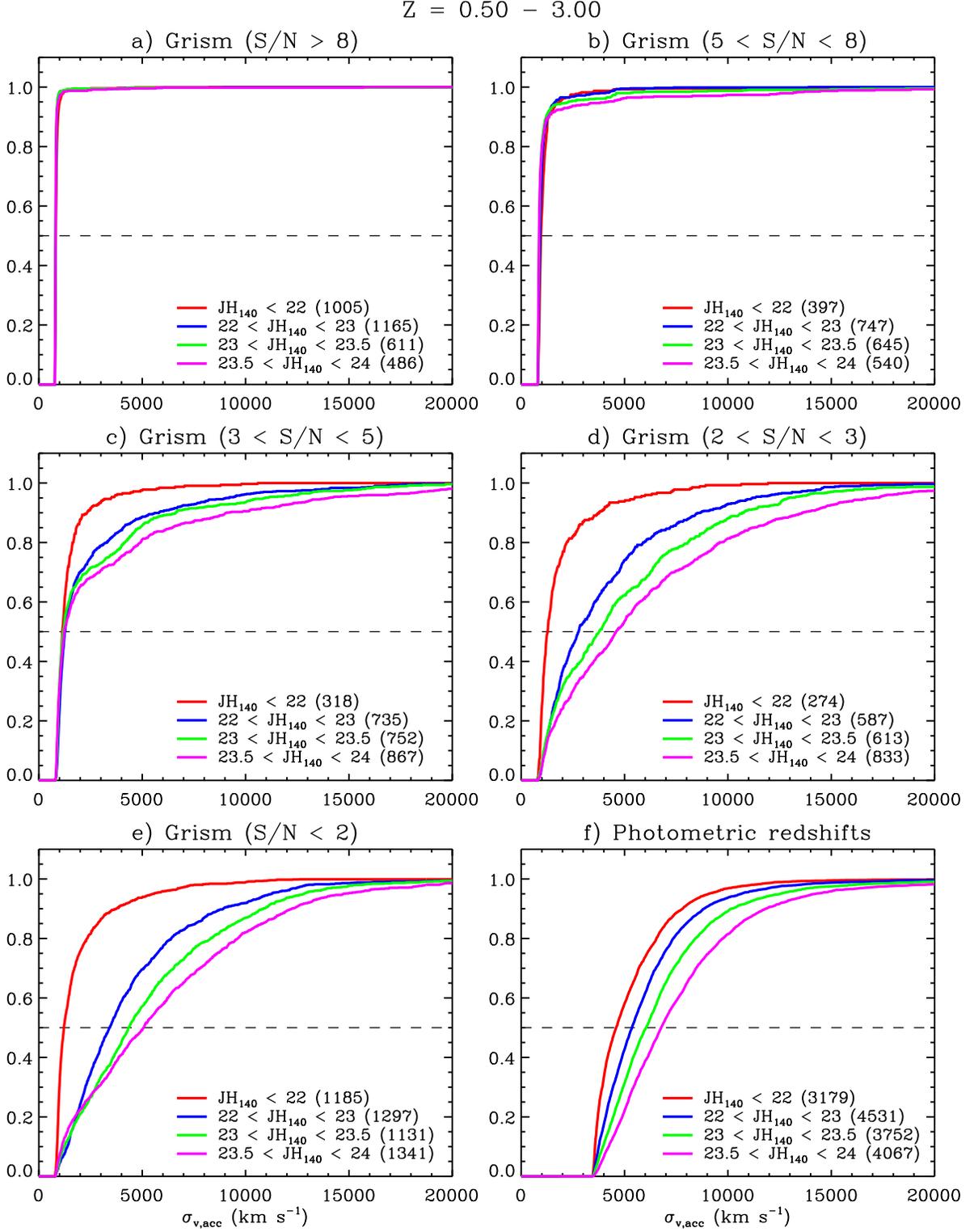}
\caption{Cumulative distributions of the redshift accuracy ($\sigma_{v,{\rm acc}}$ in $\rm{km~s^{-1}}$) for galaxies at $0.5 < z < 3.0$ and 
$JH_{\rm 140} < 24$ mag. Panels from a) to e) are for grism redshifts in bins of $S/N$ of the strongest emission line in the spectra.
In each panel the different lines are for different bins of \jhhst\ total magnitude. The panel f) is for pure photometric redshifts (including 
galaxies which have grism redshifts).}
\label{fig1}
\end{figure*}

Stellar masses and stellar population parameters are estimated using the FAST code \citep{Kriek09}, coupled with
\citet{Bruzual03} stellar population synthesis models. Those models use a \citet{Chabrier03} initial mass 
function (IMF) and solar metallicity. The best redshift is used for each galaxy together with the available space- and
ground-based photometry. The star formation history is parametrized by an exponentially declining function
and the \citet{Calzetti00} dust attenuation law is adopted.

\subsection{Redshift accuracy} \label{3dhstzaccuracy}
A careful quantification of the grism and photometric redshift accuracy is paramount for a good calibration of
the environmental statistics into physically motivated halo masses. In Section \ref{sec_halomass}, we will show
how these masses are obtained from mock catalogues selected to match the number density and redshift uncertainty 
of 3D-HST galaxies. 

The low-resolution spectra cover different spectral features as a function of galaxy properties and redshift. 
The most prominent features are emission lines, which are however limited to star forming objects. On the other 
hand, stellar continuum features (Balmer break, absorption lines) are present in the spectra of all galaxies with a 
$S/N$ that depends on the galaxy magnitude. Because all those features contribute to the redshift fitting procedure, 
we explore their impact on the redshift accuracy in bins of $S/N$ of the strongest emission line in the spectrum 
and \jhhst\ total magnitude.
Given the limited spectral coverage of the G141 grism, it is common to find only one prominent emission line 
feature in the spectrum \citep{Momcheva16}; this justifies our approach of using the $S/N$ of the strongest line. 
{We define the redshift accuracy ($\sigma_{v,{\rm acc}}$) as half the separation of the 16$^{\rm th}$ 
and 84$^{\rm th}$ confidence levels obtained from the probability density function (PDF) of grism redshifts 
as derived from the EAZY template fitting procedure. In the case of fits obtained without including the 
spectral information, it becomes a pure photometric redshift uncertainty.}
A comparison of grism redshifts to spectroscopic redshifts shows that  $\sim 800~\rm{km~s^{-1}}$
should be added to the formal uncertainty on the grism redshifts to obtain a scatter in $\Delta z/\sigma(z)$ with a 
$1\sigma$ width of unity. This ``intrinsic grism'' uncertainty can arise from morphological effects, i.e. the 
light-weighted centroid of the gas emission can be offset from that of the stars 
\citep[see][]{Nelson16a,Momcheva16}. In this analysis, we added the intrinsic uncertainty of the grism 
data in quadrature to the formal uncertainty from the fitting process. 

Figure \ref{fig1} shows $\sigma_{v,{\rm acc}}$, for bins of emission line $S/N$ and \jhhst \ magnitude. The bottom
right panel (f) shows the accuracy of photometric redshifts for the same sources 
highlighting the significant improvement on the redshift quality when the spectra are included.
From the top panels of Figure \ref{fig1}, it is clear how an emission line detection narrows the redshift PDF to the 
intrinsic uncertainty, irrespective of the stellar continuum features. At $S/N$ where the emission line becomes less dominant, we start 
to witness a magnitude dependence of the redshift accuracy. Brighter galaxies have better continuum detections and 
therefore a more accurate redshift. Even when there is no line detection (Panel e), the typical redshift 
uncertainties are a factor 2-3 lower than pure photometric redshifts. The inclusion of the spectra helps the determination
of the redshifts even when the spectra are apparently featureless. {The grism redshift accuracy is comparable 
to the pure photometric redshift accuracy only for the faintest objects (\jhhst $>$ 23 mag) with no emission line detection ($S/N<2$),
a population which accounts for $\sim 10\%$ of our grism sample.}

As a final note of caution, we highlight that whenever the information in the spectra is limited, the final grism redshift accuracy
depends largely on the photometric data, whose availability depends on the field. Indeed, COSMOS and GOODS-S have been
extensively observed with narrow or medium band filters \citep{Taniguchi07,Cardamone10, Whitaker11} resulting in better 
photometric redshifts compared to the other fields. However, as shown in Section \ref{sec_halomass} these field-to-field 
variations have negligible effects on our calibration of halo mass.

\section{Quantification of the environment} \label{sec_environment}
There are many ways to describe the environment in which a galaxy lives \citep[e.g.,][]{Haas12, Muldrew12, Etherington15}. 
In this work, we apply to observational data the method we explored and calibrated in \citet{Fossati15} and based on the 
work of \citet{Wilman10}.
We use the number density of neighbouring galaxies within fixed cylindrical apertures because it is more sensitive to high 
overdensities, less biased by the viewing angle, more robust across cosmic times, and easier to physically interpret and calibrate 
than the N$^{\rm th}$ nearest neighbour methods \citep{Shattow13}.

\subsection{Density} \label{subsec_density}
We consider all 3D-HST galaxies selected in Section \ref{obssample} to be part both of the primary (galaxies for which the 
density is computed) and neighbour samples. We calculate the projected density $\Sigma_{r_{\rm ap}}$ in a combination of 
circular apertures centered on the primary galaxies with radii $r_{\rm ap}$. The apertures range from 0.25 to 1.00 Mpc
in order to cover from intra-halo to super-halo scales. 

For a given annulus defined by $r_{\rm ap}$, the projected density is given by
\begin{equation}
\Sigma_{r_{\rm ap}} = \frac{w_{r_{\rm ap}}}{\pi \times r_{\rm ap}^2}
\label{eqSigma}
\end{equation}

where $w_{r_{\rm ap}}$ is the sum of the weights of galaxies in the neighbour sample falling at a projected distance on the sky 
$r < r_{\rm ap} $ from the primary galaxy and within a relative rest-frame velocity $\pm dv$. For the 3D-HST galaxies with a 
grism or spectroscopic redshift, the weights are set to unity (non weighted sum), while for galaxies with pure photometric
redshifts, we apply a statistical correction for the less accurate redshifts as described in Section \ref{edgecor}. The primary galaxy is
not included in the sum therefore isolated galaxies have $\Sigma = 0$.

We set the velocity cut at $dv = 1500 \rm{km~s^{-1}}$. This value is deemed appropriate for surveys with complete spectroscopic 
redshift coverage \citep{Muldrew12} and for 3D-HST given the quality of grism redshifts shown in Figure \ref{fig1}. A small
value of $dv$ avoids the peaks in the environmental density to be smoothed by interlopers in projection along the redshift axis. 
On the other hand, if only less accurate redshifts are available, a larger cut must be used to collect all the signal from overdense
regions which is artificially dispersed along the redshift axis \citep[see Figure 4 in][]{Fossati15, Etherington15}. 

\begin{figure*}
\centering
\includegraphics[scale=.86]{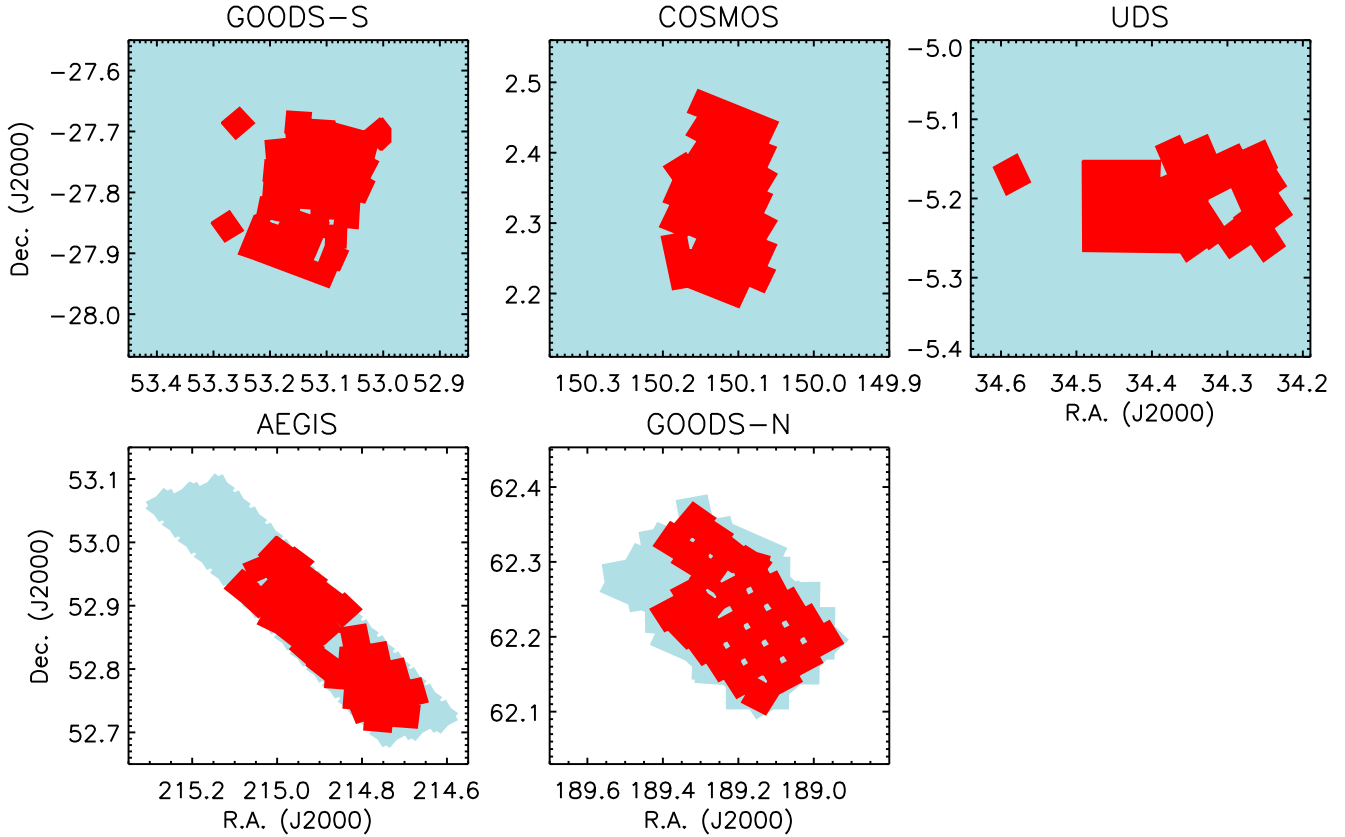}
\caption{Footprints of the 3D-HST grism observations (red areas) in the five fields studied in this work and of the extended area 
catalogues (blue areas) used for the edge corrections. White areas have no photometric coverage.}
\label{figfootprints}
\end{figure*}

Because the mean number density changes continuously with redshift, it is not possible to compare the local density ($\Sigma$)
across time. Instead we define a relative overdensity $\delta$, which is given by:
\begin{equation}
\delta{r_{\rm ap}} = \frac{\Sigma_{r_{\rm ap}} - \Sigma_{\rm mean}(z)}{\Sigma_{\rm mean}(z)}
\label{eqoverdens}
\end{equation}
where $\Sigma_{\rm mean}(z)$ is the average surface density of galaxies at a given redshift.
This is obtained by computing the volume density of galaxies (per Mpc$^3$) in the whole survey and parametrising 
the redshift dependence with a third degree polynomial. This value is multiplied by the depth of the cylindrical aperture 
at redshift $z$ to obtain the surface density $\Sigma_{\rm mean}(z)$. 
Throughout the paper, we will mainly use the overdensity in terms of the logarithmic density contrast 
defined as $\log(1+\delta{r_{\rm ap}})$.

\subsection{Edge corrections} \label{edgecor}

The calculation of the environment of primary galaxies at the edges of the 3D-HST footprint (see Figure \ref{figfootprints}, red areas) 
suffers from incomplete coverage of neighbours that results into an underestimated density in the considered aperture. 
In large scale surveys \citep[e.g. SDSS,][]{Wilman10}, it is common
practice to remove galaxies too close to the edges of the observed field. In the case of deep fields, however, the observed area 
is relatively small and the removal of such galaxies would reduce total number of objects significantly.
One possible solution is to normalize the densities by the area of the circular aperture which is within the survey footprint 
in equation \ref{eqSigma}.
Although this is a simple choice, it assumes a constant density field and neglects possible overdense structures just beyond 
the observed field. A more accurate solution consists of building up galaxy catalogues for a more extended area than 3D-HST 
and then use galaxies within these areas as ``pure neighbours'' for the environment of the primary 3D-HST
galaxies. Given the amount of publicly available data, this is possible in GOODS-S, COSMOS and UDS (see Figure 
\ref{figfootprints}, blue areas). 
In appendix \ref{edgecor_cats} we describe the data, depth and redshift quality of the catalogues we built in those fields. 
Here we present the edge correction method we developed and how it was tuned to perform the edge corrections 
in the other two fields GOODS-N and AEGIS. 

\begin{figure*}
\centering
\includegraphics[scale=.82]{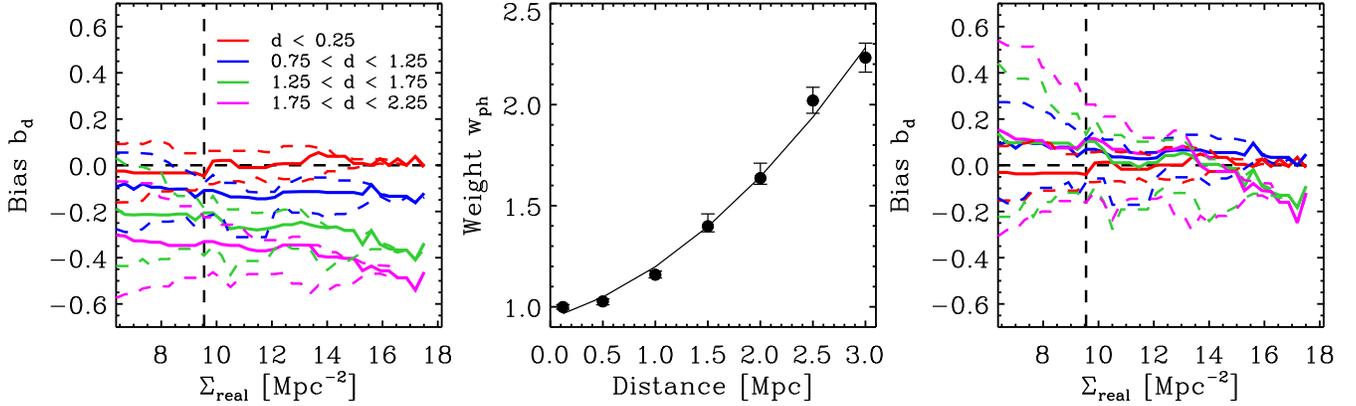}
\caption{Left Panel: median bias $b_d$ in the density field introduced when photo-z are replaced with spectroscopic redshifts of sources
at $d < 0.25$ Mpc (Red), $d \sim 1.0$ Mpc (Blue), $d \sim 1.5$ Mpc (Green), and $d \sim 2.0$ Mpc (Magenta). Dashed lines mark the 
$10^{\rm th} - 90^{\rm th}$ percentiles of the distributions. The vertical dashed line marks the density above which the bias is computed
and converted into a weight. Middle Panel: median weight as a function of the distance of the neighbour with a 
spectroscopic (or grism) redshift. The solid line is the best fit quadratic polynomial we use for the statistical correction (see equation \ref{edgecorweights}). 
Right Panel: median bias $b_d$ after the statistical correction, color coded as in the left panel. The median
values are consistent with no bias for all the distance bins.}
\label{figedgecor}
\end{figure*}

\subsubsection{Edge correction method for GOODS-S, COSMOS, and UDS} \label{Edgecormethod}
The availability of spectroscopic redshifts in the extended area catalogs is limited (from $\sim5\%$ in COSMOS and UDS to $\sim 15\%$ 
in GOODS-S). We thus need to deal with the limited accuracy of photometric redshifts for the galaxies in those fields. 
The photo-z accuracy, which varies from field 
to field and depends on the redshift, brightness and color of the objects \citep{Bezanson16}, is such that most of the sources 
which are part of the same halo in real space
would not be counted as neighbours of a primary galaxy, simply due to the redshift uncertainty. \citet{Fossati15} show that 
increasing the depth of the velocity window would recover most of the real neighbours but at the expense of a larger fraction
of interlopers (galaxies which are not physically associated to the primary). Here, we thus exploit a different method. We assume
that galaxies which are at small angular separation and whose redshifts are consistent within the uncertainties are, with a high probability, 
physically associated \citep[e.g.,][]{Kovac10, Cucciati14}. If one of them has a secure spectroscopic redshift, we  
assign this to the others.

Our method works as follows:
\begin{itemize}
\item For each galaxy with a photometric redshift, we select all neighbours with a redshift within $dv_{\rm phot} = \pm10.000~\rm{km~s^{-1}}$.
This value is chosen to recover most of the real neighbours given the average photo-z uncertainties. 
\item Among those neighbours, we select the closest (in spatial coordinates) which has a secure spectroscopic (or grism) redshift. 
Here we assume grism redshifts to have a negligible uncertainty compared to photo-z. 
\item We replace the photo-z of the galaxy of interest with this spec-z (or grism-z). Since the statistical validity of the assumption
of physical association depends on the distance of the neighbour, for increasing distances we underestimate the true clustering. 
We correct for the bias by assigning a weight $w_{ph}$ to each galaxy.
\end{itemize}

The weight is evaluated on a training sample made of galaxies in 3D-HST with $JH_{\rm 140} < 23$ mag. 
For each galaxy in the three fields, we compute the ``real'' density
($\Sigma_{\rm real}$ in a 0.75Mpc radius and $dv = \pm1500~\rm{km~s^{-1}}$) using spec-z or grism-z from 3D-HST.  
We then take for each galaxy its 
photometric redshift, and follow the procedure described above, but, instead of choosing the closest neighbour with a 
secure redshift, we select a random neighbour in different bins of projected sky distance (from 0 to 3 Mpc in bins of 0.5 Mpc width). 
Then we compute densities with each of those distance replacements separately and the fractional bias ($b_d$) as:
\begin{equation}
b_d = \frac{\Sigma_{d}- \Sigma_{\rm real} }{\Sigma_{\rm real}}
\label{eqedgecor}
\end{equation}
where the $d$ subscript denotes the replacement with a spec-z of a galaxy found at distance $d$. By using the 3D-HST data, we 
make sure that there are always a large number of neighbours with a secure redshift, and we repeat this procedure 1000 times in 
order to uniformly sample the neighbours. Figure \ref{figedgecor} left panel shows $b_d$ as a function of the real density in four bins
of $d$. Clearly, the larger $d$ is, the more underestimated the real density will be, due to a decreasing fraction of correct redshift
assignments. 

We then derive the median weight $w_{ph,d} = {\rm med}((b_d+1)^{-1})$ where the median
is computed among all galaxies that have $\Sigma_{\rm real} > 9.5 {\rm Mpc^{-2}}$ (see the vertical dashed line 
in Figure \ref{figedgecor} left panel). The density dependence of $b_d$ is negligible at these densities, therefore by 
avoiding underdense regions (where the uncertainty on $b_d$ is large) we obtain a robust determination of $w_{ph,d}$.
Figure \ref{figedgecor} middle panel shows $w_{ph,d}$ versus $d$, which we fit with a quadratic relation obtaining: 
\begin{equation}
w_{ph,d} = 9.66\times10^{-2} \times d^2 + 0.155 \times d + 0.946
\label{edgecorweights}
\end{equation}
with the additional constraint that $w_{ph,d} \ge 1$ which corresponds to $w_{ph,d} = 1$ for $d<0.29$ Mpc. We tested that this relation, 
although obtained combining all fields, holds within the uncertainties when each field is considered separately. Lastly we show in 
Figure \ref{figedgecor} right panel how the systematic bias is removed when the weight is applied to all neighbours when computing the 
density. This is consistent with no bias within the uncertainties for all the distance bins.

\subsubsection{Edge correction method for GOODS-N and AEGIS}

The GOODS-N and AEGIS fields do not have deep and extended near-infrared public catalogues that can be used to derive the edge 
corrections as presented above.
As shown in Figure \ref{figfootprints} (light blue shaded areas) the 3D-HST/CANDELS
footprint slightly extends beyond the area covered by G141 grism observations (the main requirement for our primary sample). 
Therefore the 3D-HST/CANDELS catalogue itself can be used to perform edge corrections.
We derive \jhhst\ magnitudes from the \jhst \ magnitudes using a linear function derived from the five 3D-HST fields
($JH_{140} = 1.000 \times J_{125} - 0.295$).
We then use 3D-HST photometric redshifts (or spec-z where available) and apply the method described in Section \ref{Edgecormethod}.

\begin{figure}
\includegraphics[width=8cm]{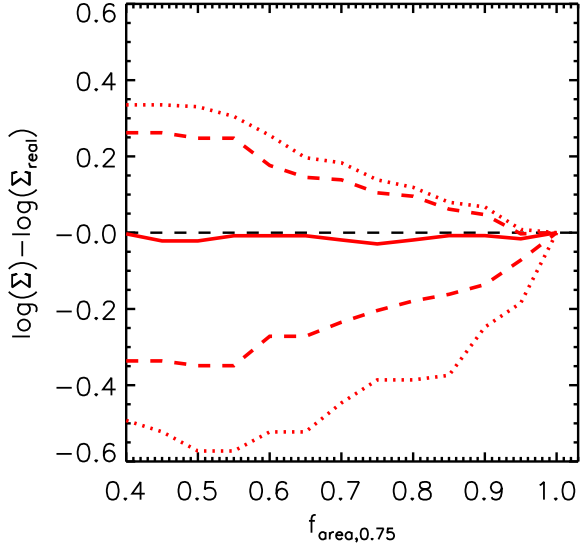}
\caption{{Logarithmic offset between the density ($\Sigma_{\rm real}$) in a 0.75Mpc aperture measured 
using the extended catalogues for COSMOS, GOODS-S, UDS and the density ($\Sigma$) measured using the fraction of the aperture 
in the 3D-HST footprint ($f_{\rm area,0.75}$) as a function of the latter quantity}. The solid line is the median while dashed and dotted lines mark the 1$\sigma$
and 2$\sigma$ confidence intervals respectively.  The offset between the two methods is zero, with a scatter which
increases with decreasing fraction of the aperture in the 3D-HST footprint. }
\label{edgecor_frac}
\end{figure}

However, the 3D-HST/CANDELS photometric catalogues do not extend enough beyond the primary sample area to ensure the apertures used to
compute the density are entirely covered by the photometric catalog footprint. For this reason, we compute the densities using the area of the circular 
aperture within the photometric catalogue. We test this method by comparing the density ($\Sigma_{\rm real}$) in a 0.75Mpc aperture measured 
using the extended catalogues for COSMOS, GOODS-S, UDS and the density ($\Sigma$) measured correcting for the fraction of the aperture 
($f_{\rm area,0.75}$) in the 3D-HST/CANDELS footprint. The result is shown in Figure \ref{edgecor_frac}. We note that although the median 
(red solid line) is consistent with no bias, the area correction introduces a scatter (dotted and dashed lines) which increases by decreasing 
the fraction of the aperture in the footprint. 

In conclusion, the environment catalogue released with this work includes all the primary galaxies in the five 3D-HST fields. The structure of the
catalogue is described in Appendix \ref{envcatalogue}. However, in the rest of this work we only include galaxies for which $f_{\rm area,0.75} > 0.9$ 
for the GOODS-N and AEGIS fields. The total number of objects in the primary 3D-HST sample with a robust determination of the environmental 
density is therefore reduced to 17397 ($93\%$ of the original sample). 

\section{Overdensities in the 3D-HST deep fields} \label{sec_overdensities}

\begin{figure*}
\includegraphics[width=18cm]{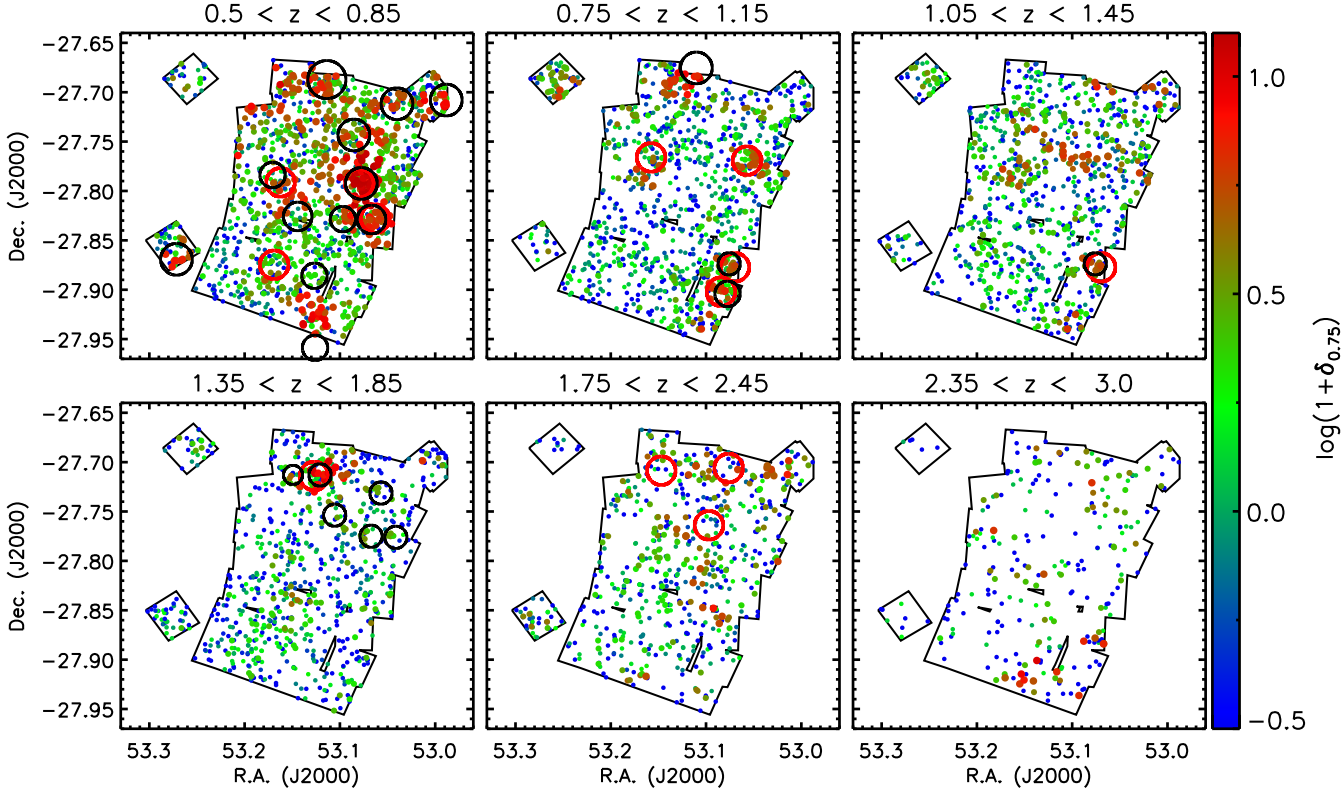}
\caption{Gallery of the 3D-HST galaxies in the GOODS-S field in different redshift slices. Points are color coded by their overdensity in a 0.75 Mpc
aperture. The size of the points also scales with overdensity. This figure demonstrates the large dynamic range in environments found in the
CANDELS deep fields.  Black circles mark the position of X-Ray extended emission from \citet{Finoguenov15}, the size of the circle representing the extension of the 
emission (R$_{200}$). Red circles mark the position of galaxy overdensities from \citet{Salimbeni09} who used a smoothed 3D density technique
from the GOODS-MUSIC catalogue to search for overdensities (the size of the circle is arbitrary and fixed).} 
\label{densityGOODSS}
\end{figure*}

\begin{figure*}
\centering
\includegraphics[width=17cm]{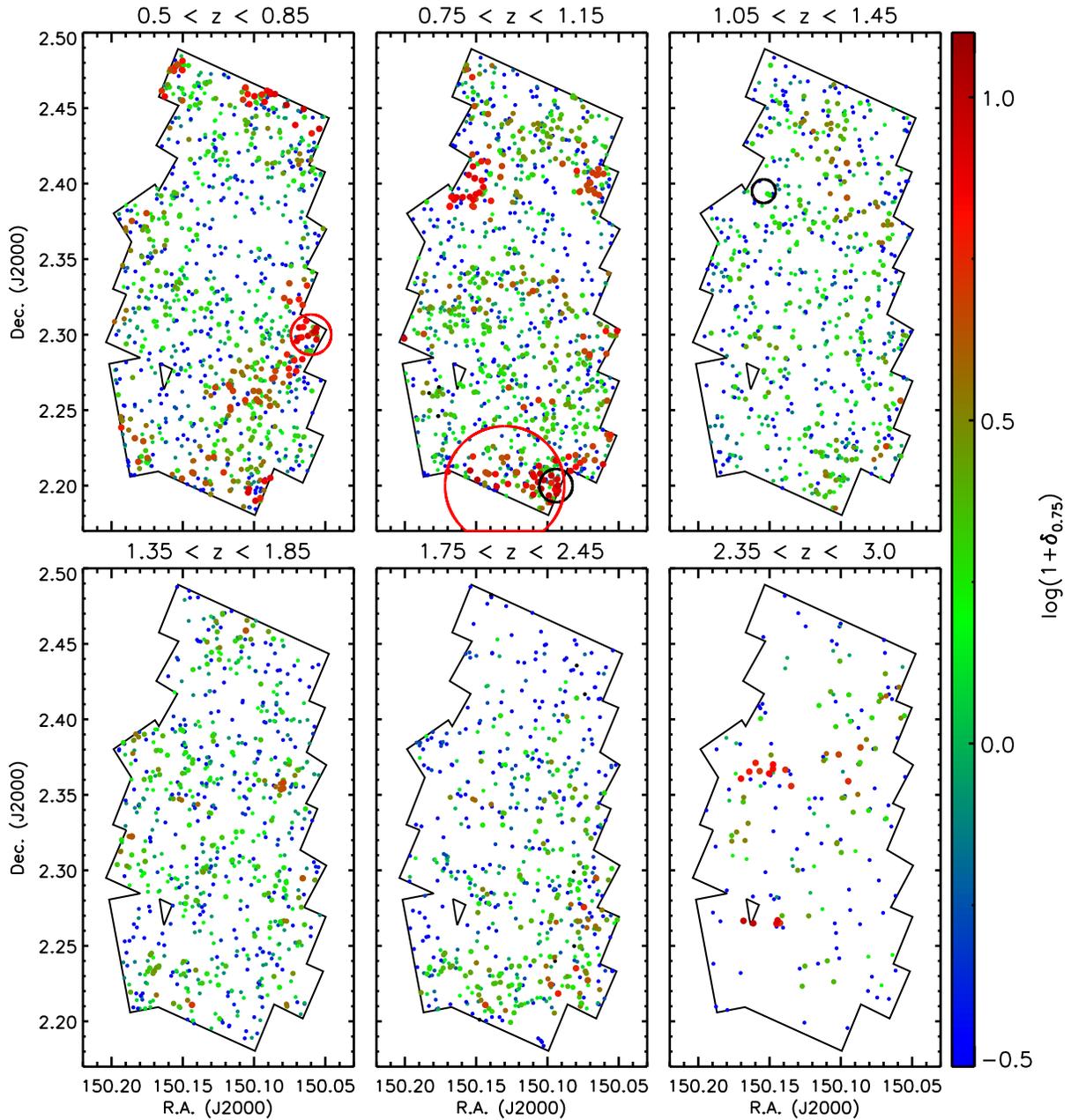}
\caption{Same as Figure \ref{densityGOODSS} but for the 3D-HST COSMOS field. The X-Ray extended emission circles (black) are from 
\citet{Finoguenov07}, and galaxy density based large scale structures (red) from \citet{Scoville07} using pure photometric redshifts up to $z\sim1$.}
\label{densityCOSMOS}
\end{figure*}

\begin{figure*}
\centering
\includegraphics[width=18cm]{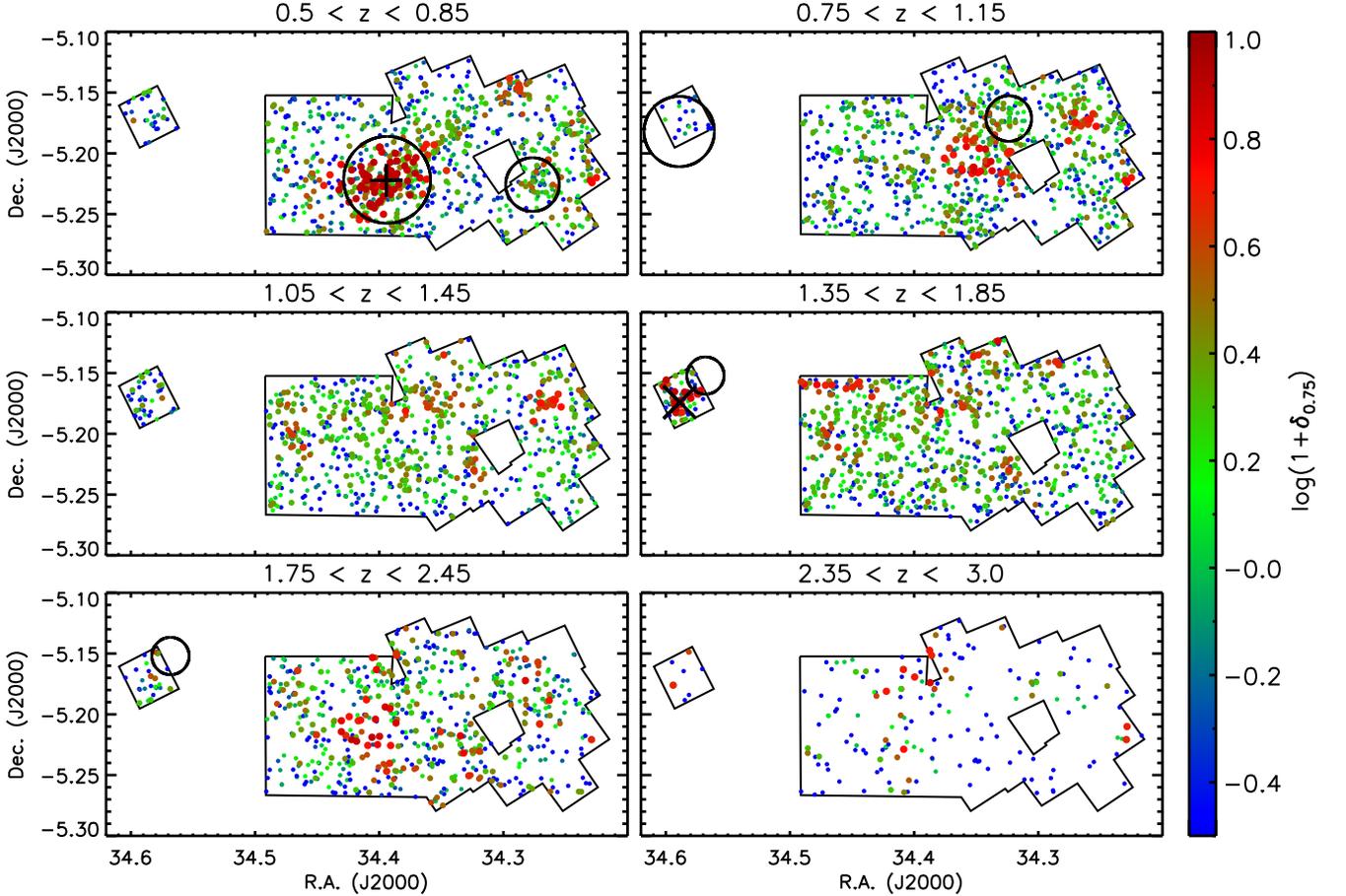}
\caption{Same as Figure \ref{densityGOODSS} but for the 3D-HST UDS field. The X-Ray extended emission circles (black) are from 
\citet{Finoguenov10}. The black $+$ and $\times$ symbols mark the center of known clusters at $z=0.65$ (\citealt{Geach07}, Galametz et al. in prep.) 
and $z=1.62$ \citep{Papovich10,Tanaka10} respectively. }
\label{densityUDS}
\end{figure*}

\begin{figure*}
\centering
\includegraphics[width=18cm]{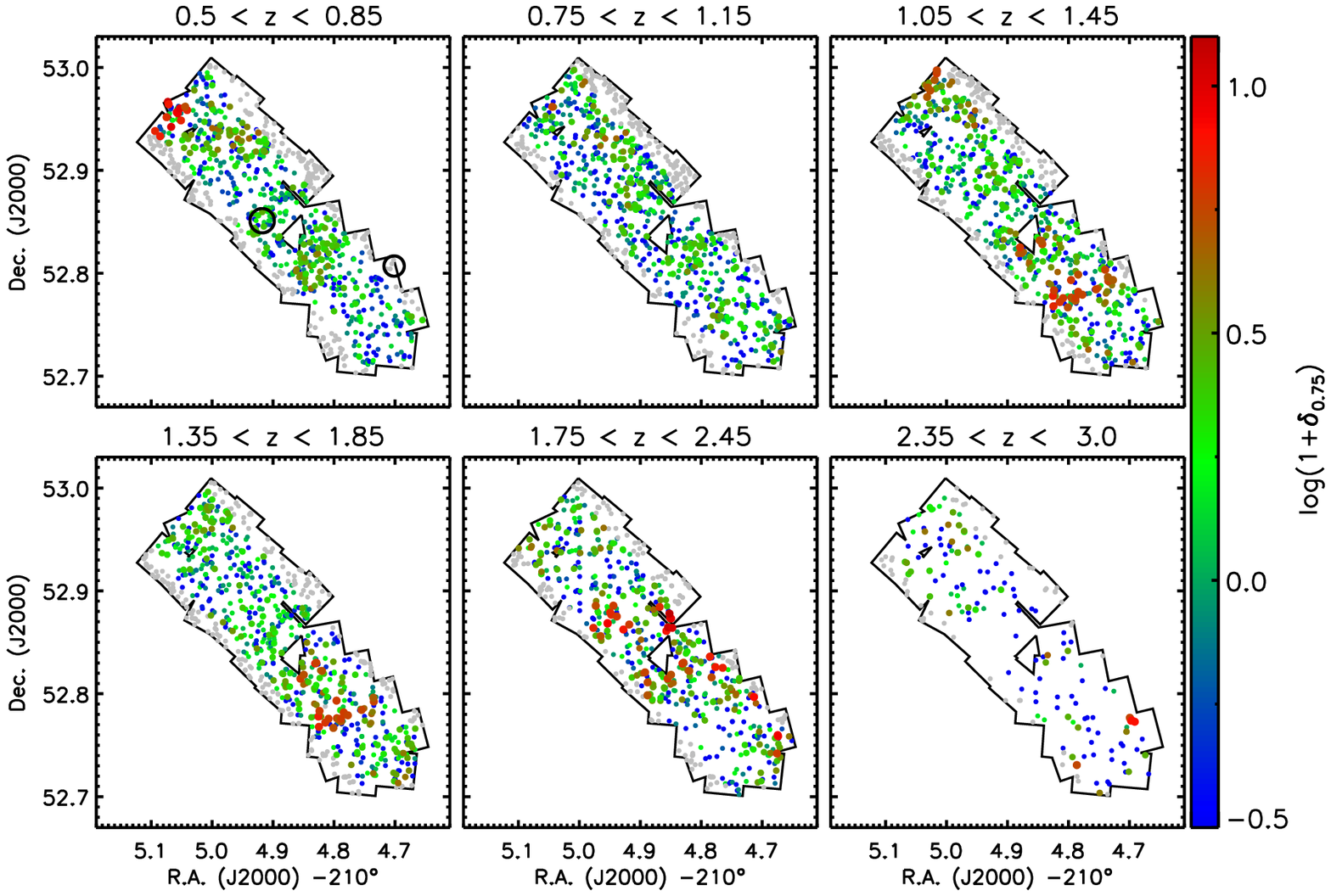}
\caption{Same as Figure \ref{densityGOODSS} but for the 3D-HST AEGIS field. The gray dots are for galaxies for which the aperture where the density is computed is 
within the photometric footprint by less than 90\%. The X-Ray extended emission circles (black) are from 
\citet{Erfanianfar13}.}
\label{densityAEGIS}
\end{figure*}

\begin{figure*}
\centering
\includegraphics[width=18cm]{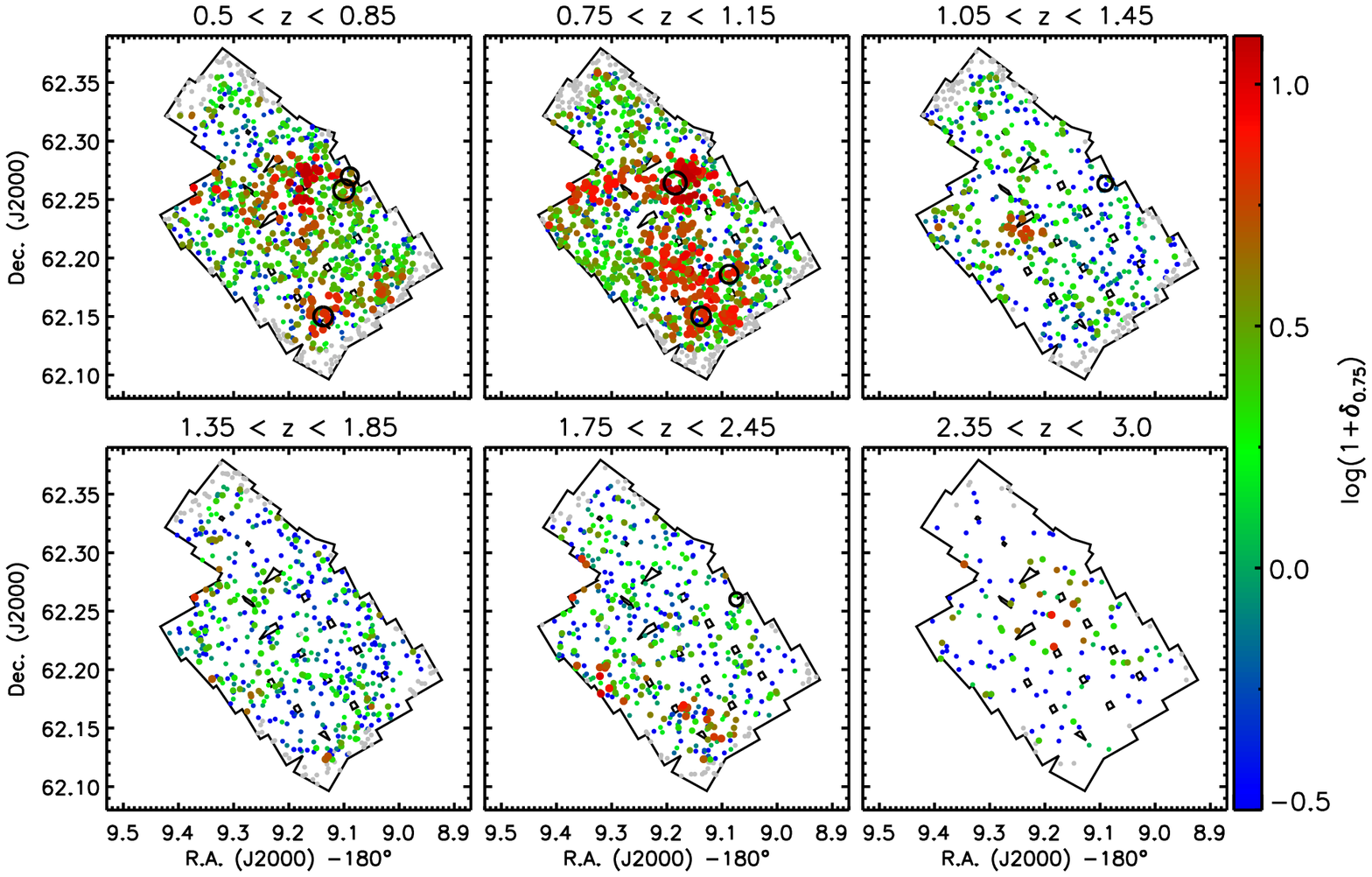}
\caption{Same as Figure \ref{densityGOODSS} but for the 3D-HST GOODS-N field. he gray dots are for galaxies for which the aperture where the density is computed is 
within the photometric footprint by less than 90\%. The X-Ray extended emission circles (black) are from 
A. Finoguenov (private comm.).}
\label{densityGN}
\end{figure*}

In order to explore correlations of galaxy evolution with environment, we need to make sure the 3D-HST fields span a wide range of 
galaxy (over-)densities, and use known structures as a sanity check of our density estimates. 
Figures \ref{densityGOODSS},  \ref{densityCOSMOS}, \ref{densityUDS}, \ref{densityAEGIS}, and \ref{densityGN} 
present the primary sample of 3D-HST galaxies in the five fields color coded by their overdensity in the 0.75 Mpc aperture in 
different redshift slices. This aperture corresponds to the typical virial radius of massive haloes ($M_h>10^{13.5} M_\odot$) in the redshift
range under study. The range of density probed is wide and spans from isolated galaxies to objects for which the 
local number of neighbours is up to ten times larger than the mean at that redshift, reaching the regime of 
clusters or massive groups. 

In each Figure, we overplot the position and extent of X-Ray extended emission from the hot intragroup (and intracluster) medium
that fills massive haloes. The exquisite depth of X-Ray data in the deep fields \citep{Finoguenov07, Finoguenov10, Finoguenov15, Erfanianfar13} 
allows the detection of the hot gas from haloes down to $M_{\rm h} \sim 10^{13} M_\odot$. We find a very satisfactory agreement between 
our overdensities and the X-Ray emission position. Indeed, most of the X-Ray groups are coincident with large overdensities in 
our maps. On the other hand, not all the overdense structures identified in our work are detected in X-Ray. We speculate this is 
mainly due to the presence of more than one massive structure along the line of sight or that low mass groups may not yet be virialized. 
Lastly, we note that the redshift of the X-Ray 
emission is assigned based on the photometric or spectroscopic information available at the epoch of the publication of the catalogue;
these data might not have been as accurate as the density field reconstruction performed in this work. Our analysis has therefore 
the potential to spectroscopically confirm more X-Ray groups and improve the quality of previous redshift assignments.

Several other works have also analysed, with different techniques, the presence of overdense structures in the deep fields. We 
overplot on Figure \ref{densityGOODSS} the position of overdensities found in the GOODS-S field by \citet{Salimbeni09}. These have been derived from 
the GOODS-MUSIC catalogue using a smoothed 3D density searching algorithm. These data have 15\% spectroscopic redshifts and 
photometric redshifts for the remaining fraction. Because the smoothing technique is less able to constrain the size of the structure,
we plot circles with an arbitrary radius. The structures within the 3D-HST footprint (except those at $z>2$) are confirmed with 
our data to be at least a factor of $2-3$ denser than the mean. The differences in samples and techniques hamper a 
more quantitative comparison. Our data confirm with a high degree of significance the detection of two well known super-structures, 
one at redshift $z=0.73$ \citep{Gilli03, Adami05, Trevese07} and one at redshift $z=1.61$ first detected by \citet{Kurk09}. The latter is
made of 5 peaks in the photo-z map (which correspond to putative positions for the X-Ray emission, see Table 1 in \citealt{Finoguenov15}). 
The main structure is robustly recovered by our analysis while the other sub-structures are only mild ($\log(1+\delta_{0.75}) \sim 0.5$) overdensities.

In the COSMOS field (see Figure \ref{densityCOSMOS}), \citet{Scoville07} applied an adaptive smoothing technique 
\citep[similar to][]{Salimbeni09} to find large scale 
structures at $z<1$. While their results do not constrain the size of the structure and are less sensitive to very compact overdensities,
we do find that their detections in the 3D-HST footprint correspond to high overdensities in our work.

Similarly in the UDS field, we do detect a very massive cluster surrounded by filaments and less massive groups (upper
left panel of Figure \ref{densityUDS}) at $z=0.65$ (Galametz et al. in prep.). Another well known structure in this field is located at $z=1.62$ 
\citep{Papovich10,Tanaka10}. Despite being only partially covered by the 3D-HST grism observations (isolated pointing on the left of the contiguous field),
we do find it corresponds to a large overdensity of galaxies thanks to our accurate edge corrections using UKIDSS-UDS photometric data.

In summary, our reconstruction of the density field in the 3D-HST deep fields recovers the previously known 
massive structures across the full redshift range analysed in this work.

\section{The model galaxy sample} \label{sec_models}
The goal of this work is to understand the environment of galaxies in the context of a hierarchical Universe. To reach this goal, we need
to calibrate physically motivated quantities using observed metrics of environment by means of semi-analytic models (SAM) of galaxy 
formation. 
We make use of light cones from the latest release of the Munich model presented by \citet{Henriques15}.
This model is based on the Millennium N-body simulation \citep{Springel05} which has a size of $500 h^{-1}$ Mpc.
The simulation outputs are scaled to cosmological initial conditions from the \textit{Planck} mission (Planck Collaboration XVI, \citeyear{Planck14}): 
$\sigma_8 = 0.829, H_0 = 67.3 \rm{km~s^{-1}~Mpc^{-1}}, \Omega_\Lambda = 0.685,  \Omega_M =0.315$. Although those values are 
slightly different from those used in our observational sample, the differences in cosmological parameters have a much smaller effect 
on mock galaxy properties than the uncertainties in galaxy formation physics \citep{Wang08, Fontanot12, Guo13a}.

This model includes prescriptions for gas cooling, size evolution, star formation, stellar and active galactic nuclei feedback and 
metal enrichment as described by e.g. \citet{Croton06, DeLucia07, Guo11}. The most significant updates concern the reincorporation 
timescales of galactic wind ejecta that, together with other tweaks in the free parameters, reproduce observational data on
the abundance and color distributions of galaxies from $z=0$ to $z=3$ \citep{Henriques15}. Our choice of this model is therefore 
driven by those new features which are critical for an accurate quantification of the environment.

We make use of the model in the form of 24 light cones, which are constructed by replicating the simulation box evaluated at multiple
redshift snapshots. Before deriving the density for the light cones, as described in Section \ref{subsec_density},
we first match the magnitude selection and redshift accuracy of the 3D-HST survey.

\subsection{Sample selection} 
SAMs are based on N-body dark matter only simulations. Therefore (and opposite to observations), the galaxy stellar masses are 
accurate quantities, while observed magnitudes are uncertain and rely on radiative transfer and dust absorption recipes implemented in the
models. On the other hand, magnitudes are direct observables in a survey (like 3D-HST) are therefore known with a high 
degree of accuracy. 

To overcome these limitations and the fact that \jhhst\ magnitudes are not given in  \citet{Henriques15} cones,
we employ a method that generates observed magnitudes for SAM galaxies by using observational constraints from 3D-HST.
Each model galaxy is defined by its stellar mass ($M_{*,\rm{mod}}$), $U-V$ rest frame color ($(U-V)_{\rm{mod}}$) and 
redshift ($z_{\rm{mod}}$).
Similarly, 3D-HST galaxies are defined by stellar mass ($M_{*,\rm{obs}}$),  $U-V$ rest frame color ($(U-V)_{\rm{obs}}$), 
redshift ($z_{\rm{obs}}$), and magnitude ($JH_{\rm{obs}}$).
The method works as follows:
\begin{itemize}
\item For each bin of stellar mass (0.25 dex wide) and redshift (0.1 wide) we select all the model and 3D-HST galaxies.
\item For each model galaxy in this bin we rank the $(U-V)_{\rm{mod}}$ and we find the $(U-V)_{\rm{obs}}$ that corresponds
to the same ranking. 
\item We assign to the model galaxy a randomly selected stellar mass-to-light ratio $(M_*/L_{JH})_{\rm obs}$ at $(U-V)_{\rm{obs}}\pm0.05$
drawn from the distribution of 3D-HST galaxies in the stellar mass and redshift bin of the mock galaxy of interest.
\item From $(M_*/L_{JH})_{\rm obs}$, $M_{*,\rm{mod}}$, and $z_{\rm{mod}}$, we compute $JH_{\rm{mod}}$ for the model galaxy.
\end{itemize}

This method generates \jhhst\ magnitudes for all the model galaxies down to $10^{8} M_\odot$. This is much deeper than the
3D-HST magnitude limit even at the lower end of our redshift range. We then select model galaxies down to a $JH_{\rm{mod,lim}}$
magnitude that matches the total number density of the primary targets (\jhhst$<24$ mag) in the five 3D-HST fields to that in the 24 
lightcones. This protects us from stellar mass function mismatches between the models and the observations (although 
those differences are very small in \citealt{Henriques15}). We employ a  $JH_{\rm{mod,lim}} = 23.85$ mag, which is very close to 24 mag further 
supporting the quality of the stellar mass functions in the models.

\subsection{Matching the redshift accuracy} \label{subsec_redshiftaccuracy}
After the model sample is selected, the next goal is to assign to each galaxy a redshift accuracy that matches as closely
as possible to the one in 3D-HST. To do so, we should not only assign the correct fraction of spec-z, grism-z and photo-z as a function
of observed magnitude but also assign an accuracy for the grism-z and photo-z as a function of physical properties such that the
final distributions resemble those in Figure \ref{fig1}. 
We showed in Section \ref{3dhstzaccuracy} that the grism redshift accuracy depends on the signal-to-noise of the strongest 
emission line and the galaxy magnitude. For the latter, we use \jhhst\ as derived above, while the former 
quantity needs to be parametrized in terms of other quantities available in the models.

\begin{figure}
\includegraphics[width=9cm]{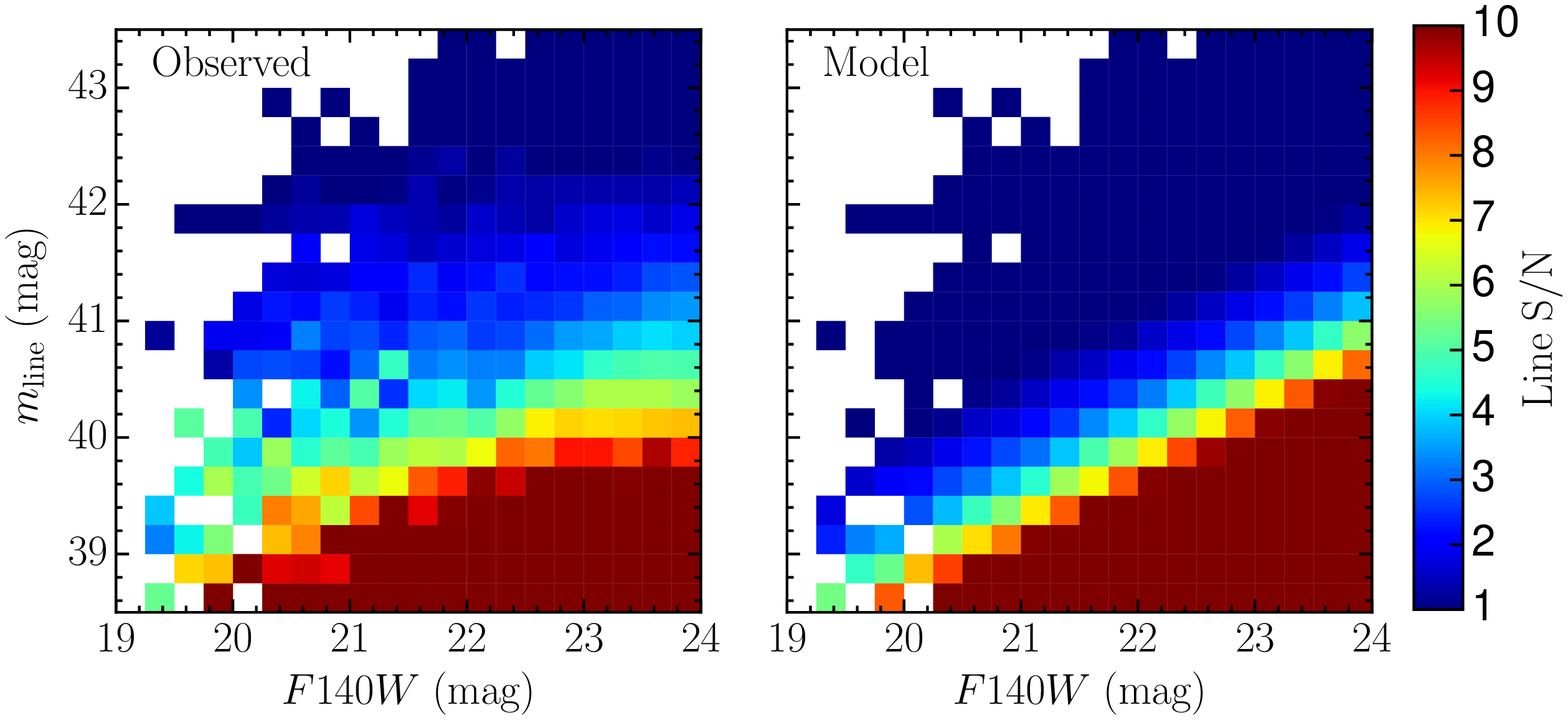}
\caption{Left panel: emission line $S/N$ for 3D-HST galaxies as a function of the \jhhst\ magnitude and line magnitude. Right panel:
emission line $S/N$ obtained using the parametrization from equation \ref{eqlineSN}.}
\label{SNgrism}
\end{figure}

Figure \ref{SNgrism} left panel shows how the emission line $S/N$ depends both on the line flux and the \jhhst\ magnitude for
galaxies with a measured line flux. For each galaxy, we take the flux (in units of erg cm$^{-2}$ s$^{-1}$) of the strongest line and 
we define the line magnitude as $m_{\rm{line}} = -2.5 \times \log(f_{\rm{line}})$. At fixed line flux, brighter galaxies have more 
continuum, thus decreasing the line $S/N$.  
This relation is well reproduced by the following parametrization:
\begin{equation}
\log(S/N) = -0.33 \times(2\times m_{\rm{line}} - JH_{\rm 140}) + 19.85
\label{eqlineSN}
\end{equation} 
Figure \ref{SNgrism} right panel shows the line $S/N$ obtained with this equation. The small differences 
between the two panels can be due to additional variables not taken into account (e.g. dust extinction or grism throughput). 
We tested (by perturbing the $S/N$ assigned to each model galaxy) that a more accurate parametrization of this relation is not required
for the purpose of this paper.

In order to obtain a synthetic line $(S/N)_{\rm mod}$ for the model galaxies, we first convert the star formation rate (SFR) of the 
model galaxies into an $\rm{H\alpha}$ flux (or $\rm{H\beta}$ flux where $\rm{H\alpha}$ is redshifted outside the grism 
wavelength range) by inverting the relation given in \citet{Kennicutt98}. We then obtain $S/N_{\rm mod}$ from 
$m_{\rm{line}}$, and $JH_{\rm{mod}}$ using equation \ref{eqlineSN}.
The rank in $(S/N)_{\rm mod}$ (and not the absolute value) is then matched to that in $S/N$ for the 3D-HST galaxies.

Lastly, we assign to each mock galaxy a random grism redshift accuracy such that the observed distributions shown in 
Figure \ref{fig1} are reproduced for the mock sample. A photometric redshift accuracy is also generated using the same distributions 
(as a sole function of $JH_{\rm{mod}}$). 

Each model galaxy is then defined by three redshifts: a spectroscopic redshift which is derived from the geometric redshift ($z_{\rm{GEO}}$) 
of the cones plus the peculiar velocity of the halo, a grism like redshift which is derived from the spec-z plus a random value drawn from a 
gaussian distribution with sigma equal to the grism redshift accuracy derived above, and a photometric redshift derived as the previous but 
using the photometric redshift accuracy.

The last step in this procedure requires that for each galaxy only one of these three redshifts is selected to generate a ``best'' redshift. 
To do so, we work in bins of \jhhst\ magnitude. For each bin of magnitude, the fraction of 3D-HST
galaxies with spec-z, grism-z and photo-z is computed. Then in order of descending $(S/N)_{\rm mod}$, the spec-z is taken for a number of 
galaxies matching the fraction of galaxies with spec-z in the observational catalog, a grism-z is taken for an appropriate number of galaxies and lastly
a photo-z is taken for the galaxies with the lowest $(S/N)_{\rm mod}$ which mimic line non-detections in the grism data. We stress that since
the grism redshift accuracy is a function of $(S/N)_{\rm mod}$, the quality of grism redshifts for objects with marginal line detections is preserved
by this method.

Once a catalog of model galaxies is selected and their redshift accuracy matches the 3D-HST catalog, we compute the environment parameters as 
described in Section \ref{subsec_density}. The only minor difference is that, as the number of model galaxies is
very large, we can remove objects closer than 1.0 Mpc from the edges of the cone, to avoid edge biases.
 
\section{Calibration of physical parameters} \label{sec_halomass}

The local density of galaxies is not the only parameter that describes the environment in which a galaxy lives. Another important
parameter is whether a galaxy is the dominant one within its dark matter halo (Central), or if it orbits within a deeper potential well (Satellite). 
{The definition of centrals and satellites in the mock sample is obtained from the hierarchy of subhaloes (the main units hosting a single galaxy). 
First, haloes are detected using a friends-of-friends (FOF) algorithm with a linking length $b=0.2$ \citep{Springel05}. Then each 
halo is decomposed into subhaloes running the algorithm {\sc SUBFIND} \citep{Springel01}, which determines the self-bound structures within 
the halo. As time goes by, the model follows subhaloes after they are accreted on to larger structures. When two haloes merge, the 
galaxy hosted in the more massive halo is considered the central, and the other becomes a satellite.}

In this Section, we describe how we use the mock catalog to assign a halo mass probability density function (PDF) and a probability
of being central or satellite to 3D-HST galaxies. 
The method builds on the idea of finding all the galaxies in the mock lightcones that match each 3D-HST galaxy in redshift, density,
mass-rank (described below), and stellar mass (within the observational uncertainties). 
The main advantage of using multiple parameters is to break degeneracies which are otherwise dominant if only one parameter is used
\citep[e.g. to account for the role of stellar mass at low density, where halo mass depends more significantly on stellar mass than 
density;][]{Fossati15}.


\subsection{The stellar mass rank in fixed apertures} \label{sec_censat}

\citet{Fossati15} explored how the rank in stellar mass of a galaxy in an appropriate aperture can be a good 
discriminator of the central/satellite status for a galaxy. This method, which complements the one usually used in local large 
scale surveys of galaxies based on halo finder algorithms, is more effective with the sparser sampling of high redshift surveys. 

We refer the reader to \citet{Fossati15} for the details of how this method is calibrated. Here, we recall that we define a galaxy to be
central if it is the most massive (mass-rank = 1) within an adaptive aperture that only depends on the stellar mass. Otherwise, if it is not
the most massive (mass-rank $> 1$), it is classified as a satellite. 

The adaptive aperture is motivated by the fact that ideally, the aperture in which the mass-rank is computed should be as similar as possible
to the halo virial radius to maximize the completeness of the central/satellite separation and reduce the fraction of spurious classifications. 
\citet{Fossati15}, defined this aperture as a cylinder with radius:
\begin{equation}
r_0= 3\times 10^{(\alpha \log M_* + \beta)} ~\rm{[Mpc]}
\end{equation}
where $M_*$ is the stellar mass, $\alpha = 0.25$, and $\beta = -3.40$ are the parameters which describe the dependence of
the virial radius with stellar mass. These values are calibrated using the models \citep[see][]{Fossati15}. We also limit the aperture 
between 0.35 and 1.00 Mpc. The lower limit is set to avoid small apertures which would result in low mass galaxies being assigned mass-rank = 1
even if they are satellites of a large halo. The upper limit is approximately the radius of the largest haloes in the redshift range under study.
The adaptive aperture radius (in Mpc) is therefore defined as:
\begin{equation}
r = \begin{cases} 0.35 & \mbox{if } r_0 < 0.35 \\ r_0 & \mbox{if } 0.35 \le r_0 \le 1.00 \\ 1.00 & \mbox{if } r_0 > 1.00  \end{cases}
\label{eqadaptiveap}
\end{equation}

In this work, we have to consider the variable redshift accuracy of 3D-HST galaxies. Therefore fixing the depth of the cylinder to $\pm1500
\rm{km~s^{-1}}$ does not optimize the central versus satellite discrimination. We set the depth of the adaptive aperture cylinder (in $\rm{km~s^{-1}}$) to:
\begin{equation}
dv= \begin{cases} 1500 & \mbox{if } \sigma_{v,{\rm acc}} < 1500 \\ \sigma_{v,{\rm acc}} & \mbox{if } 1500 \le \sigma_{v,{\rm acc}} \le 7500 \\ 7500 & \mbox{if } \sigma_{v,{\rm acc}} > 7500 \end{cases}
\label{eqadaptivedepth}
\end{equation}
where $\sigma_{v,{\rm acc}}$ is the redshift accuracy of the primary galaxy. By using the mock sample, we tested 
that this combination of upper and lower limits gives a pure yet sufficiently complete sample of central galaxies. 

The simple classification of centrals and satellites based on mass-rank only is subject to a variety of contaminating factors.
For instance in galaxy pairs or small groups (where the mass of the real central and satellites are very close), it is difficult 
to use the stellar mass to robustly define which galaxy is the central. On the other hand, in the infalling regions beyond
the virial radius of massive clusters, many central galaxies would be classified as satellites as analysed in detail in \citet{Fossati15}.
In this work, we go beyond the simple dichotomic definition that each galaxy is either central or satellite using 
the mass-rank only. We combine multiple observables to derive a probability that each 3D-HST galaxy 
is central or satellite by matching observed galaxies to mock galaxies. This probabilistic approach naturally 
takes into account all sources of impurity and is of fundamental importance to separate the effects 
of mass and environment on the quenching of galaxies.

\begin{figure}
\begin{center}
\includegraphics[width = 0.95\columnwidth]{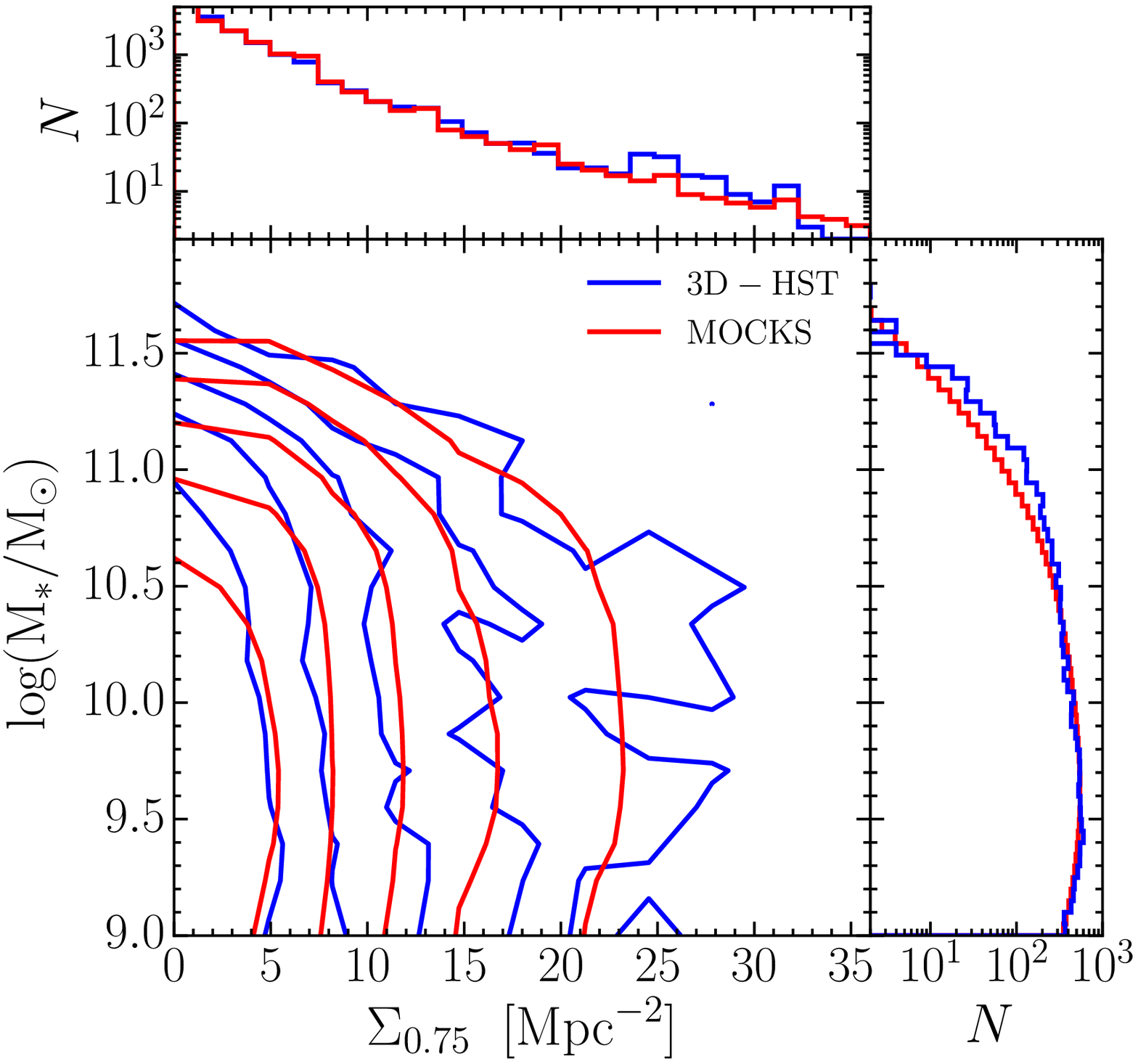}
\end{center}
\caption{Main panel: bivariate distribution of density on the 0.75 Mpc scale and stellar mass for the 3D-HST sample (blue) and
the mock sample (red). The mock contours have been scaled to account for the ratio of volumes between the lightcones and the data.
The contours are logaritmically spaced with the outermost contour at 4 objects per bin and the innermost at 300 objects per bin. 
Upper panels: marginalized distributions of density on the 0.75 Mpc scale for the 3D-HST and the mock samples. The counts refer
to the 3D-HST sample while the mock histogram has been normalized by the ratio of the volumes. Right-hand panel:
same as above but marginalized over the stellar mass. } 
\label{Mstar_dens_compare}
\end{figure}

\subsection{Matching mock to real galaxies}
In this Section, we describe how we match individual 3D-HST galaxies to the mock sample to access physical quantities
unaccessible from observations only.
Our method heavily relies on the fact that the distributions of stellar mass and density (and their bivariate distribution) are 
well matched between the mocks and the observations across the full redshift range. 

In the upper and right panels of Figure \ref{Mstar_dens_compare}, we show the distributions of density and stellar mass 
respectively, while the main panel shows the 2D histogram of both quantities. The overall agreement is very satisfactory and 
relates to the agreement of the observed stellar mass functions to that from \citet{Henriques15}, and to our careful selection of objects. 
The match of the density distributions also confirms that the redshift assignment for mock galaxies is 
accurate enough to reproduce the observed density distributions.
A good match between models and observations is found for other apertures as well. 
In the future, it should be possible to improve our method by combining density information on several scales by 
means of machine learning algorithms. 

\begin{figure}
\begin{center}
\includegraphics[width = 1.00\columnwidth]{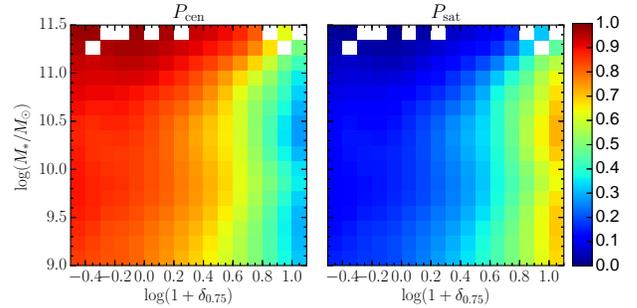}
\end{center}
\caption{Average probability for a 3D-HST galaxy of being central ($P_{\rm{cen}}$, left panel) or satellite ($P_{\rm{sat}}$, right panel) 
in bins of density contrast in the 0.75 Mpc aperture and stellar mass. The trends are consistent with the analysis of mock galaxies which
shows a lower purity for the selection of centrals at high halo masses (large overdensities) and the opposite trend for satellites.} 
\label{PcenPsat}
\end{figure}

To match observed galaxies to mock galaxies, we also require an estimate of the uncertainty on both the density and the 
stellar mass. For the stellar mass, we use $\sigma({\log(M_*)}) = 0.15$ dex \citep{Conroy09, Gallazzi09, Mendel14}. 
For the density, the error budget is dominated by the redshift uncertainty of each galaxy and the fact that for a sample of galaxies 
with given \jhhst\ and emission line $S/N$, the redshift accuracy has a distribution with non zero width. 
This means that the redshift uncertainty of 
mock galaxies can only match the observational sample in a statistical sense. To test how the densities of individual galaxies
are affected by the redshift uncertainty, we repeat 50 times the process of assigning a redshift to mock galaxies described in 
Section \ref{subsec_redshiftaccuracy}. We then compute the density for each of those samples independently and analyse the distribution
of densities for each galaxy. We find that the distribution roughly follows a Poissonian distribution: 
$\sigma(\Sigma_{r_{\rm ap}}) = \sqrt{w_{r_{\rm ap}}}/(\pi \times r_{\rm ap}^2)$
\begin{figure}
\begin{center}
\includegraphics[width = 0.95\columnwidth]{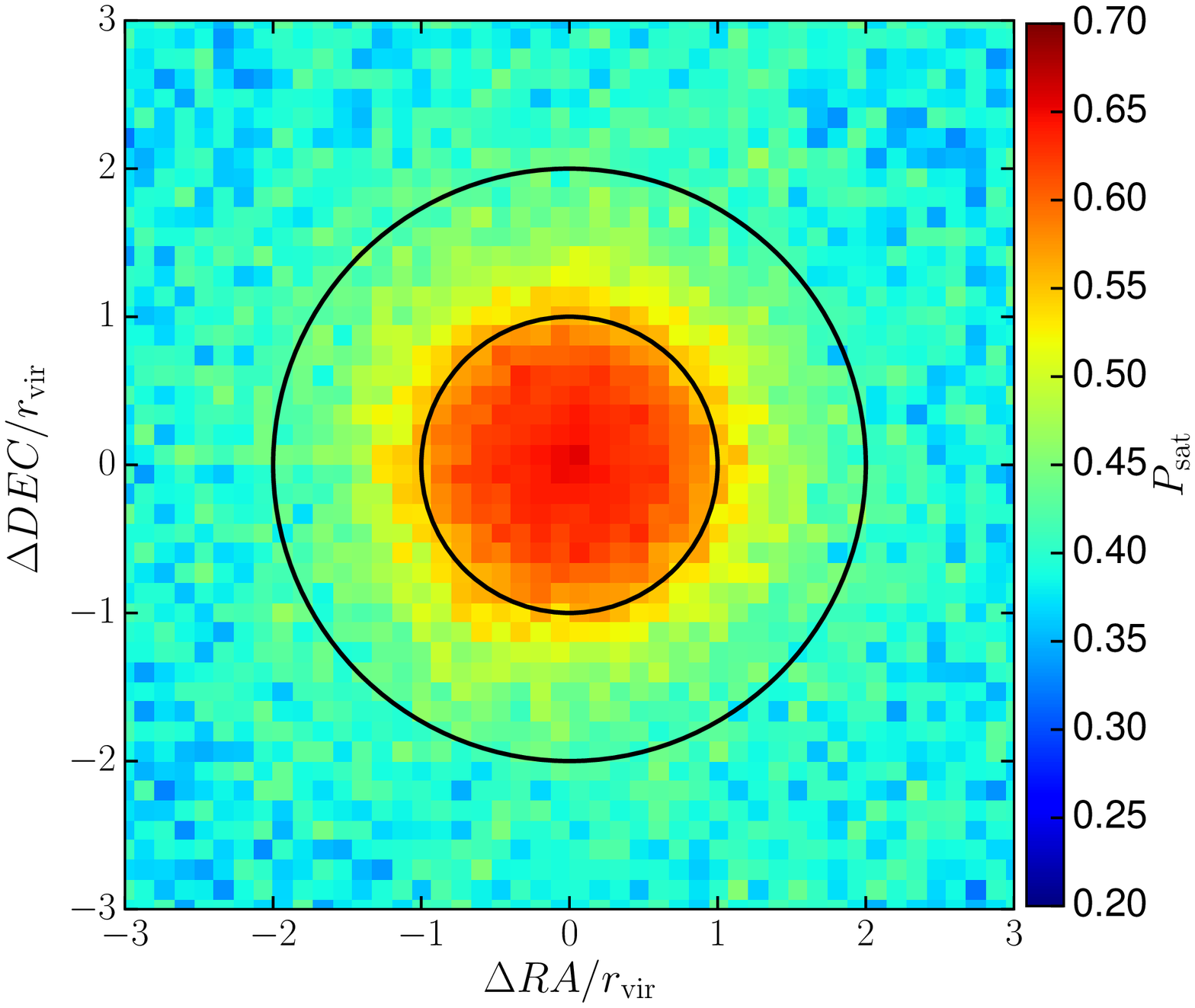}
\end{center}
\includegraphics[width = 0.90\columnwidth]{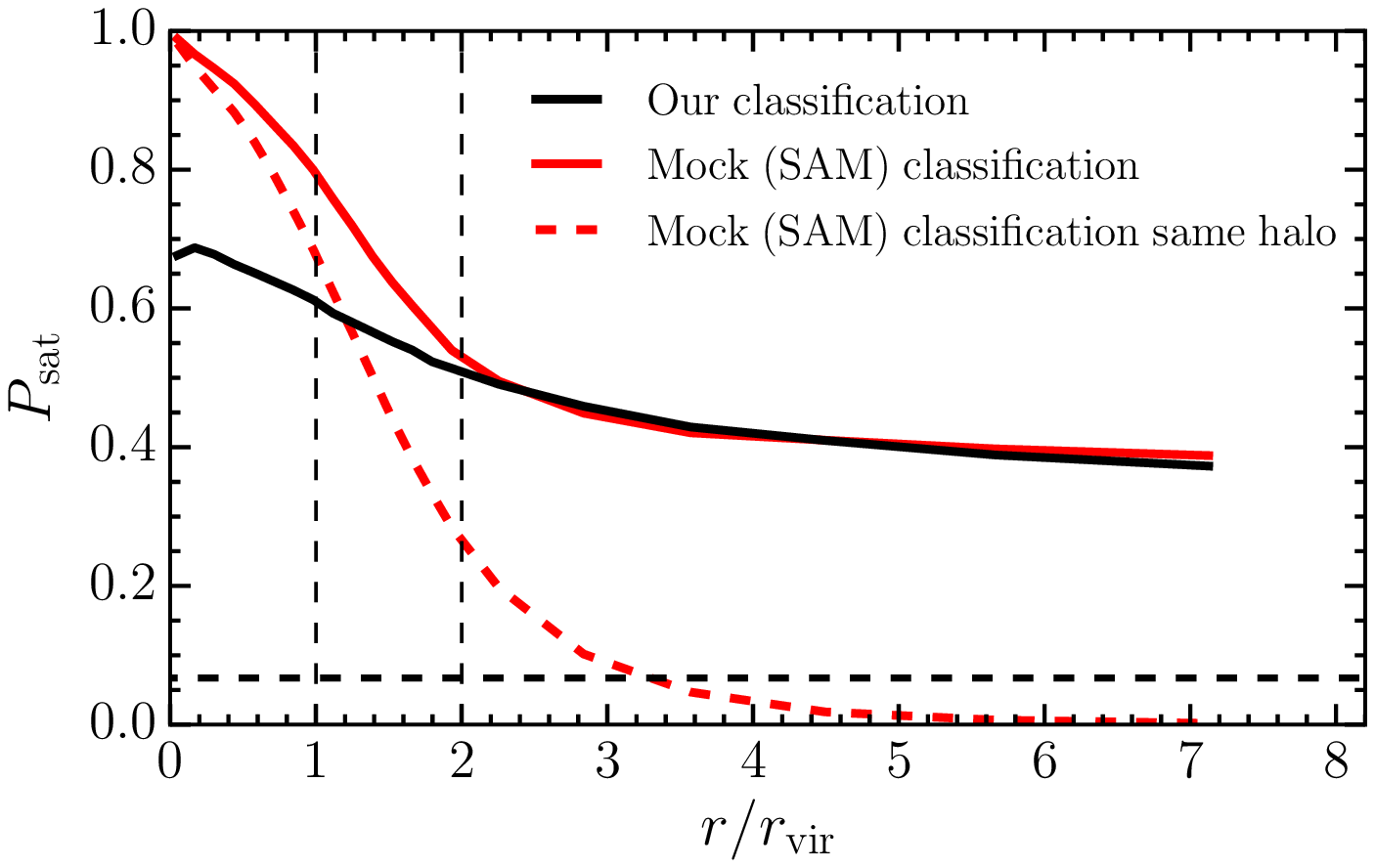}
\caption{Top panel: average $P_{\rm sat}$ for mock galaxies 
as a function of normalized R.A. and Dec. offset from the center of haloes more massive than $10^{13.5} M_\odot$. The black solid 
circles mark $r_{\rm vir}$ and $2\times r_{\rm vir}$. Bottom panel: average $P_{\rm sat}$ for mock galaxies 
using our Bayesian definition (black solid line) or the SAM definition of satellites (red solid line) 
as a function of radial distance from the center of haloes more massive than $10^{13.5} M_\odot$. {The red dashed line
shows the value of $P_{\rm sat}$ obtained from SAM satellites living in the same halo of the central galaxy. The plateau of $P_{\rm sat}$
at large radii is caused by the contribution of satellites from nearby haloes. }The vertical dashed lines
mark $r_{\rm vir}$ and $2\times r_{\rm vir}$. The horizontal dashed line is the value of $P_{\rm{sat}}$ for {a stellar mass and redshift matched sample of} 
galaxies living in average density environments.} 
\label{Psat_2Dstack}
\end{figure}
{Based on this evidence, we match each 3D-HST galaxy to the mock galaxies within $\pm 0.1$ in redshift space and within $\pm \sigma({\log(M_*))}$ and
 $\pm \sigma(\Sigma_{0.75})$ for the stellar mass and density on the 0.75 Mpc scale respectively. 

The local density is a quantity that depends on the redshift accuracy both of the primary galaxy and of the neighbours, which in turn
depends on the emission line strength in the grism data and the galaxy brightness (see Section \ref{3dhstzaccuracy}). As a result the density peaks are 
subject to different degrees of smoothing if the neighbouring galaxies have a systematically poorer redshift accuracy in a given environment. 
Our mock catalogue is a good representation of the observational sample only if the SFR (from which the syntetic line $S/N$ is derived) 
and the stellar mass distributions as a function of environment are well reproduced by the SAM. \citet{Henriques16} have shown that the H15 model 
is qualitatively able to recover the observed trends of passive fraction as a function of environment. By matching  
model galaxies with a redshift accuracy within $\pm 2000 ~\rm{km~s^{-1}}$ to that of the observed galaxy we introduce no
bias in the halo mass distributions: for galaxies with less accurate redshifts, we simply obtain broader PDFs of halo mass. }

Lastly, we restrict the match for the most massive galaxies (mass-rank $=1$) to the most massive mock galaxies.
The rest of the population (mass-rank $>1$) was matched to the same population in the mocks.

\begin{figure*}
\begin{center}
\includegraphics[width = 17.5 cm]{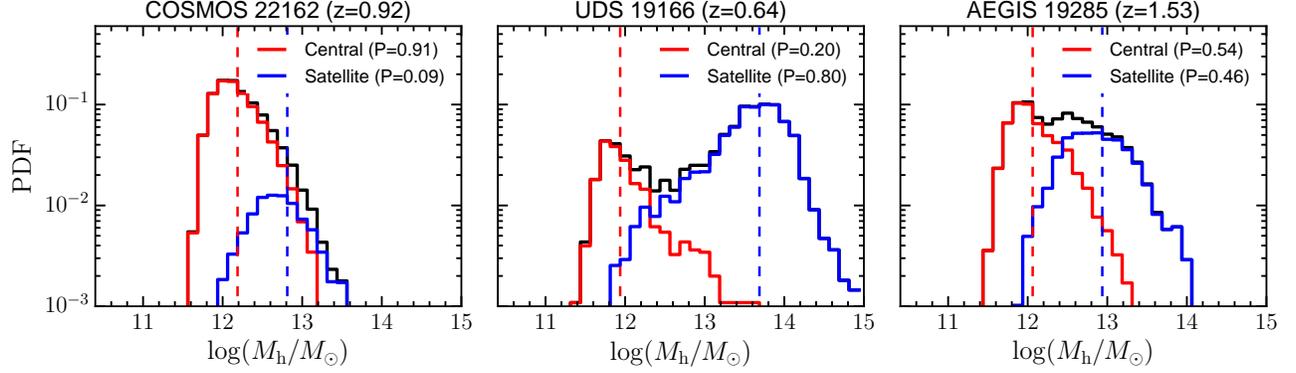}
\end{center}
\caption{Example halo mass PDFs for three 3D-HST galaxies. The left panel shows a galaxy with a high probability of being a central, the middle panel one
with a high probability of being a satellite, and the right panel an object with an almost equal probability of being a central or a satellite. The red 
and blue histograms show the halo mass probability given that the galaxy is a central ($P_{M_h | {\rm{cen}}}$) or a satellite 
($P_{M_h | {\rm{sat}}}$), while the black histogram is the total halo mass PDF. The histograms are normalized such that the
 area under them gives $P_{\rm{cen}}$ and $P_{\rm{sat}}$ respectively. The vertical 
dashed lines mark the median halo mass for a given type.} 
\label{halopdfexamples}
\end{figure*}

\subsubsection{A probabilistic determination of central versus satellite status} \label{pcensatcalib}
{The central and satellite fractions of those matched mock galaxies are used to define a probability that the 3D-HST galaxy under consideration
is central ($P_{\rm{cen}}$) or satellite ($P_{\rm{sat}}$):
\begin{equation}
P_{\rm{cen}} = \frac{N_{\rm matched~cen}}{N_{\rm matched}},~~P_{\rm{sat}} = \frac{N_{\rm matched~sat}}{N_{\rm matched}} = 1-P_{\rm cen}
\end{equation}}

\begin{figure*}
\begin{center}
\includegraphics[width = 17.5cm]{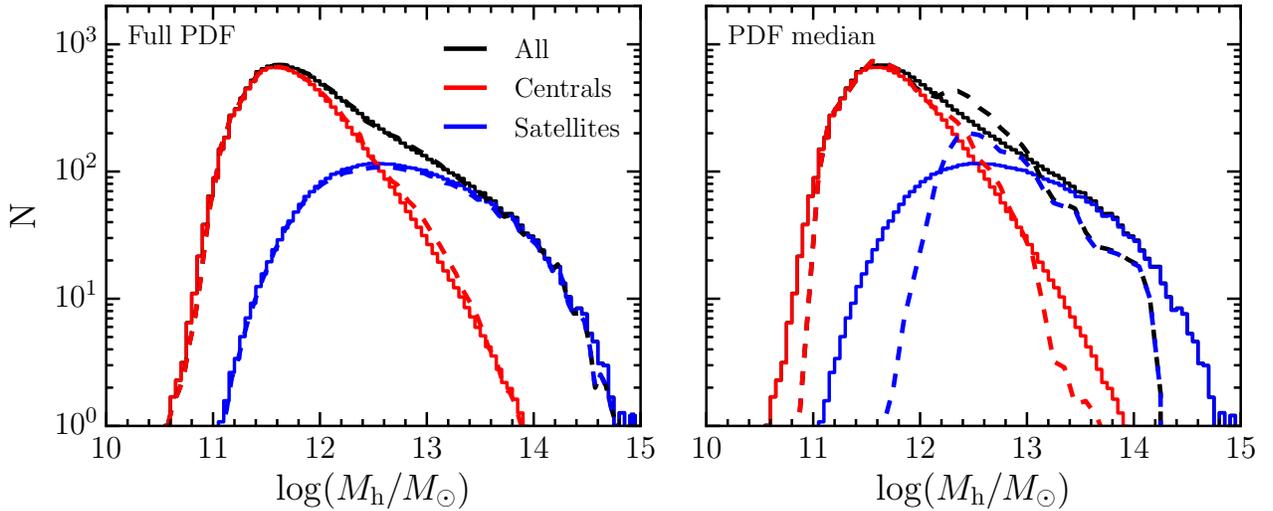}
\end{center}
\caption{Comparison of the halo mass distributions for the mock galaxies (solid histograms) and 
3D-HST galaxies (dashed lines). In the left panel the dashed lines are obtained by summing the full halo mass PDFs for centrals 
($P_{M_h | {\rm{cen}}}$, red) and satellites ($P_{M_h | {\rm{sat}}}$, blue) weighted by $P_{\rm{cen}}$ and $P_{\rm{sat}}$ for each galaxy. 
In the right hand panel the dashed lines are obtained from the single value estimator (median value of the PDF given the type) 
weighted by the probability that a galaxy is of a given type. The black histograms and dashed lines are the sum of the colored. } 
\label{Mhalodistr}
\end{figure*}

Figure \ref{PcenPsat} shows the average values of those quantities in bins of logarithmic density 
contrast (see Section \ref{subsec_density}) in the 0.75 Mpc aperture and stellar mass for all the 3D-HST galaxies included in our 
sample. The average value of $P_{\rm{cen}}$  decreases with increasing density and decreasing stellar mass, and the 
opposite trend occurs for $P_{\rm{sat}}$. High mass haloes (high density regions) are indeed dominated by the satellite 
population, but objects with high stellar masses are more likely to be centrals.
Galaxies in low density environments ($\log(1+\delta_{0.75})<0.2$) are almost entirely centrals. 
However, in the analysis performed in the next sections we use the values of $P_{\rm{cen}}$  and $P_{\rm{sat}}$ computed for each galaxy instead of 
the average values (\citealt{Kovac14} performs instead an average correction as a function of galaxy density). This takes into full account 
possible second order dependencies on mass-rank, redshift, or redshift accuracy.

We also examine how $P_{\rm{sat}}$ varies as a function of the distance from the center of over-dense structures, such as massive groups or 
clusters of galaxies. To do so, we take the haloes more massive than $10^{13.5} M_\odot$ in the mock lightcones.  We then 
select all galaxies in a redshift slice centered on the redshift of the central galaxy and within $\Delta z \leq \pm 0.01$ and compute
their projected sky positions with respect to the central galaxy. We normalize their positions to the virial radius of the halo and remove 
the central galaxy. 

Figure \ref{Psat_2Dstack}, top panel, shows the average value of  $P_{\rm{sat}}$ as a function of normalized R.A. and Dec. offset from the center of the haloes. 
The black solid circles mark $r_{\rm vir}$ and $2\times r_{\rm vir}$. Figure \ref{Psat_2Dstack}, bottom panel, shows the average value of 
$P_{\rm{sat}}$ (black solid line) as a function of radial distance from the center of the haloes. The red solid line shows the fraction of satellites
in the same radial bins but using the mock definition of satellites. {Lastly, the red dashed line shows the value of $P_{\rm sat}$ including 
only SAM satellites living in the same halo of the central galaxy. }

Our Bayesian definition tracks well the SAM definition of satellites as a function of halo mass. However the real trend is smoothed due to both the 
transformation from real to redshift space, and the intrinsic uncertainty of our method to extract $P_{\rm{sat}}$ based on observational parameters. 
{Moreover, $P_{\rm{sat}}$ only drops to $40\%$ at $\sim5\times r_{\rm vir}$. This is caused by satellites from nearby haloes, while the
contribution from satellites belonging to the same halo becomes negligible at $\sim3\times r_{\rm vir}$. Indeed, massive structures are embedded in
filaments and surrounded by groups which will eventually merge with the cluster. Therefore, even at large distances from the center, the density is higher
than the mean density (at $\sim5\times r_{\rm vir}$ the density is $\sim 4$ times higher than the average density)}. As a reference we show in Figure \ref{Psat_2Dstack}, bottom panel, the value of $P_{\rm{sat}}$ for {a stellar mass and redshift matched sample of} galaxies living in average density 
environments ($0.8 < (1+\delta_{0.75}) < 1.2$, horizontal dashed line).

\subsubsection{The halo mass calibration} \label{halomasscalib}
Similarly, we use the halo masses of matched central and satellite model galaxies to generate the halo mass PDFs given their type 
($P_{M_h | {\rm{cen}}}$ and $P_{M_h | {\rm{sat}}}$ respectively). Figure \ref{halopdfexamples} shows three examples of such PDFs for 
one object with high $P_{\rm{cen}}$, one with high $P_{\rm{sat}}$, and one object with an almost equal probability of being a central or a satellite.
The vertical dashed lines mark the median halo mass for a given type. Although the total halo mass PDF can be double peaked (middle panel), the 
degeneracy between the two peaks is broken once the galaxy types are separated, making the median values well determined for each type 
independently.

\subsection{Testing calibrations}
We test the halo mass calibration by comparing the halo mass distributions of the mock sample to the 3D-HST sample. In both panels of Figure 
\ref{Mhalodistr}, we plot the halo mass histograms for centrals and satellites of the entire mock sample. 
The number counts are scaled by the ratio of the volume between the 24 lightcones and the five 3D-HST fields. 

In the left panel of Figure \ref{Mhalodistr} the dashed lines are the halo mass distributions of 3D-HST galaxies obtained by summing the full 
halo mass PDFs for centrals  ($P_{M_h | {\rm{cen}}}$, red dashed) and satellites  ($P_{M_h | {\rm{sat}}}$, blue dashed) weighted by 
$P_{\rm{cen}}$ and $P_{\rm{sat}}$ for each galaxy. The agreement with the mock sample distributions is remarkable. Although 
this is in principle expected because the halo mass PDFs for observed
galaxies are generated from the mock sample, it should be noted that we perform the match in bins of redshift, redshift accuracy, stellar mass, density
and mass-rank. The good agreement for the whole sample between the derived PDFs and the mock distributions (for centrals and satellites separately)
should therefore be taken as an evidence that our method has not introduced any bias in the final PDFs. 

We take 
the median value of the halo mass PDFs given that each galaxy is a central ($M_{\rm h,50|cen}$) or a satellite ($M_{\rm h,50|sat}$) as an 
estimate of the ``best'' halo mass, weighted by $P_{\rm{cen}}$ and $P_{\rm{sat}}$. These values are shown in Figure \ref{Mhalodistr}, right panel.
The agreement with the mock distributions is good. For central galaxies, the shape and extent of the distribution is well preserved. 
For satellite galaxies, the halo mass range is less extended than the one in the mocks; values above $10^{14.2} M_\odot$ 
and below $10^{12} M_\odot$ indeed only contribute through the tails of the PDFs, and therefore do not appear when the median of the PDFs are used.

In the next Section, we make use of the full PDFs to derive constraints on the environmental quenching of satellite galaxies. However,
the satisfactory agreement of single value estimates of halo mass with the mock distributions makes them a valuable and reliable estimate 
in science applications when the use of the full PDFs is not possible or feasible. 

\section{Constraining environmental quenching processes at $z=0.5-2$} \label{sec_quenching}
In this Section, we explore the role of environment in quenching the star formation activity of galaxies over $0.5 < z < 2$ using 3D-HST data. 
It was first proposed by \citet{Baldry06} that the fraction of passive galaxies depends both on stellar mass and environment in a 
separable manner. \citet{Peng10}, using the SDSS and zCOSMOS surveys, extended the independence of those 
processes to $z \sim1$. More recently, \citet{Peng12} interpreted these trends in the local Universe by suggesting that central galaxies 
are only subject to ``mass quenching'' while satellites suffer from the former plus an ``environmental quenching''. 
\citet{Kovac14} similarly found that satellite galaxies are the main drivers of environmental quenching up to $z\sim0.7$ using zCOSMOS data.

Here, we extend these analysis to higher redshift by exploring the dependence of the fraction of passive
galaxies on stellar mass, halo mass and central/satellite status in order to derive the efficiency and timescale of environmental 
quenching. In Appendix \ref{app_passfrac_dens}, we show that we obtain consistent results using the observed galaxy density
as opposed to calibrated halo mass.

\begin{figure}
\includegraphics[width = 8.7cm]{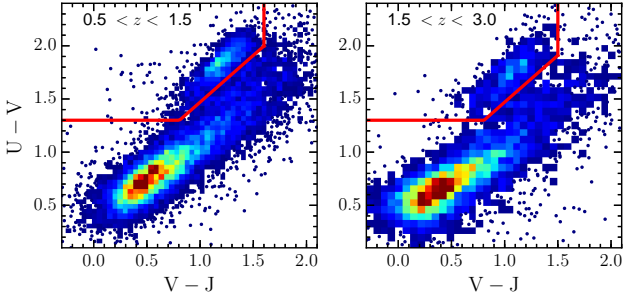}
\caption{Rest-frame UVJ diagram for 3D-HST galaxies in two redshift bins. The color scale represents the density of points. 
Where the density is low we plot individual galaxies. The solid red line indicates the adopted separation between passive galaxies and star-forming 
galaxies.} 
\label{UVJdiagram}
\end{figure}

\begin{figure*}
\begin{center}
\includegraphics[width = 15.5cm]{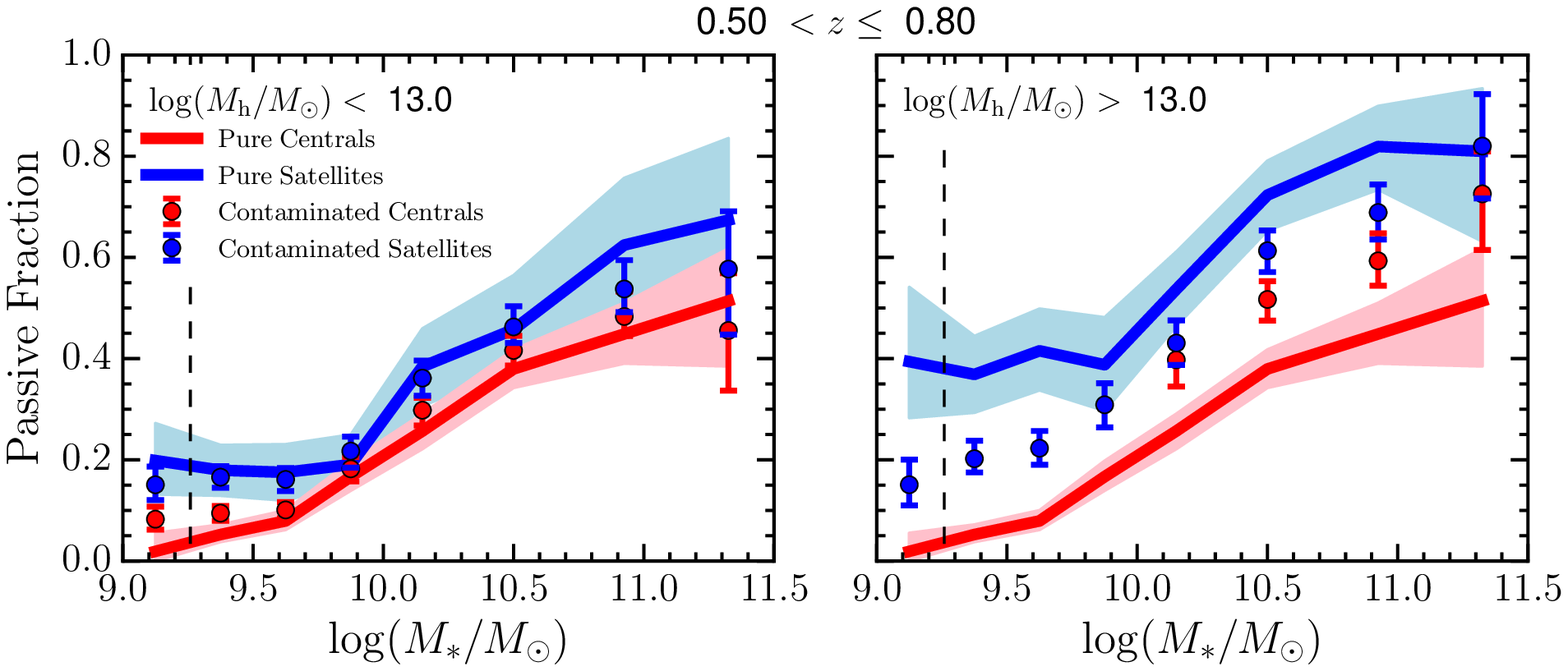} 
\includegraphics[width = 15.5cm]{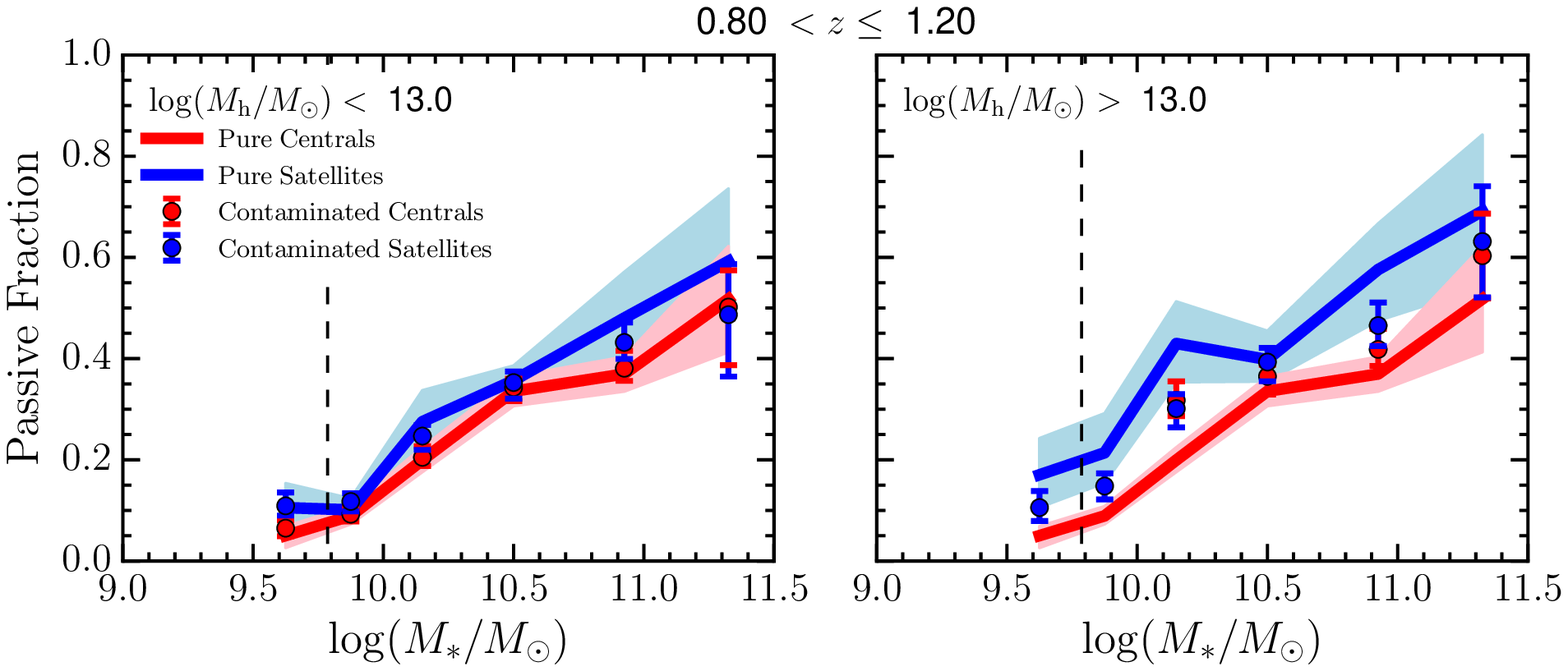} 
\includegraphics[width = 15.5cm]{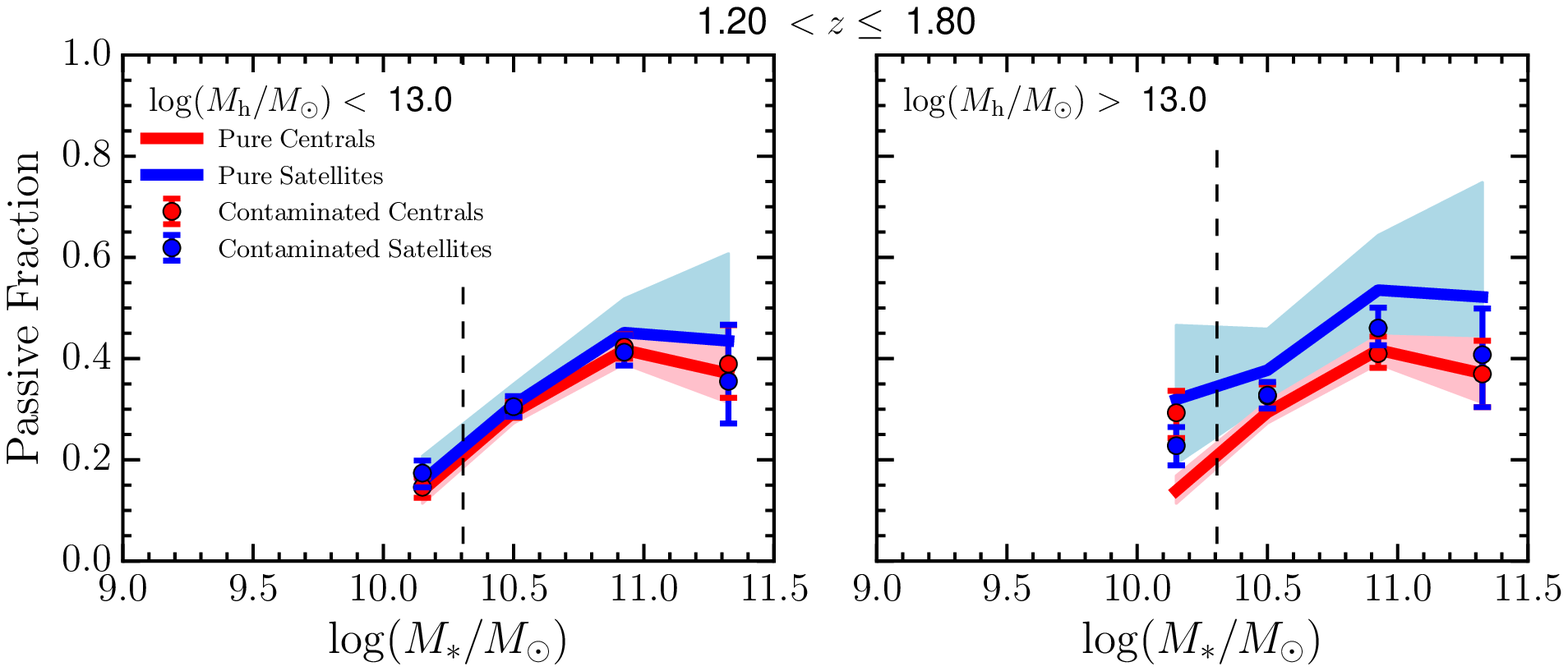} 
\end{center}
\caption{Passive fraction for central and satellite galaxies in bins of  $M_{\rm{*}}$ and $M_{\rm{h}}$ in three independent redshift bins. 
The median (log) halo masses for satellites are 12.36, 13.53 at $z=0.5-0.8$ for the lower and higher halo mass bin respectively, 
12.41, 13.44 at $z=0.8-1.2$, and 12.43, 13.34 at $z=1.2-1.8$. Datapoints show the observed
passive fractions with uncertainties derived from Monte Carlo resampling of the mock sample. The thick red line is the passive 
fraction of a pure sample of central galaxies from the 3D-HST dataset. The thick blue line represents our modelled ``pure'' 
passive fraction of satellites (see Section \ref{sec_decontamination} for the details of the modelling process). In both cases
the shaded regions show the $1\sigma$ confidence intervals. The vertical dashed line marks the stellar mass limit of 
the volume limited sample. } 
\label{Passfrac_Mhalo}
\end{figure*}

\subsection{Passive fractions} \label{sec_passfrac_mhalo}
The populations of passive and star-forming galaxies are typically separated either by a specific star formation rate cut \citep[e.g.][]{Franx08, 
Hirschmann14, Fossati15} or by a single color or color-color selection \citep[e.g.][]{Bell04, Weiner05, Whitaker11, Muzzin13, Mok13}. 
In this work, we use the latter method and select passive and star forming galaxies based on their position in the rest-frame UVJ 
color-color diagram \citep{Williams09}. 
Following \citet{Whitaker11}, passive galaxies are selected to have:
\begin{equation}
(U -V) > 0.88\times(V -J)+0.59
\end{equation}
\begin{equation}
(U-V)>1.3,(V -J)<1.6~[0.5<z<1.5]
\end{equation}
\begin{equation}
(U-V)>1.3,(V -J)<1.5~[1.5<z<2.0]
\end{equation}
where the colors are rest-frame and are taken from \citet{Momcheva16}.
Figure \ref{UVJdiagram} shows the distribution of 3D-HST galaxies in the rest frame UVJ color-color plane. The red solid line
shows the adopted division between passive and star forming galaxies.

The fractions of passive centrals and satellites in bins of $M_{\rm{*}}$ and $M_{\rm{h}}$ are computed as the 
fraction of passive objects in a given stellar mass bin where each galaxy is weighted by its probability of being central
or satellite and the probability of being in a given halo mass bin for its type. Algebraically:
\begin{equation}
f_{\rm{pass|ty}} = \frac{\sum_i \left( \delta_{\rm pass,i} \times \delta_{M_* \rm{,i}} \times P_{\rm{ty,i}} \times \int_{M_{\rm h}} P_{M_{\rm h},i | ty} dM_{\rm h}\right )}{\sum_i \left(\delta_{M_* \rm{,i}} \times P_{\rm{ty,i}} \times \int_{M_{\rm h}} P_{M_{\rm h},i | ty} dM_{\rm h}\right)}
\label{eqpassive_mhalo}
\end{equation}
where ty refers to a given type (centrals or satellites), $\delta_{\rm pass,i}$ is 1 if a galaxy is UVJ passive and 0 otherwise, 
$\delta_{M_* \rm{,i}}$ is 1 if a galaxy is in the stellar mass bin and 0 otherwise, $P_{\rm{ty,i}}$ is the probability that a 
galaxy is of a given type {(see Section \ref{pcensatcalib})} and $\int_{M_{\rm h}} P_{M_{\rm h},i | ty} dM_{\rm h}$ is the halo mass PDF given the type
integrated over the halo mass bin limits {(see Section \ref{halomasscalib})}. 

The data points in Figure \ref{Passfrac_Mhalo} show the passive fractions in two bins 
of halo mass (above and below $10^{13} M_\odot$) and in three independent redshift bins. 
The median (log) halo masses for satellites are 12.36, 13.53 at $z=0.5-0.8$ for the lower and higher halo mass bin respectively; 
12.41, 13.44 at $z=0.8-1.2$; and 12.43, 13.34 at $z=1.2-1.8$.

The uncertainties on the data points cannot be easily
evaluated assuming Binomial statistics because the number of galaxies contributing to each point is not a priori known. Indeed,
$P_{\rm{ty,i}}$ and $\int_{M_{\rm h}} P_{M_{\rm h},i | ty} dM_{\rm h}$ act as weights and all galaxies with a stellar mass 
within the mass bin do contribute to the passive fraction. To assess the uncertainties we use the mock lightcones (where each mock galaxy has 
been assigned a $P_{\rm{cen}}$ and $P_{\rm{sat}}$ and halo mass PDFs as if they were observed galaxies). 
In a given stellar mass bin we assign each model galaxy to be either passive or active such that the fraction of passive
galaxies matches the observed one. Then we randomly select a number of model galaxies equal to the number of observed galaxies 
in that bin and we compute the passive fraction of this subsample using equation \ref{eqpassive_mhalo}. We repeat this procedure
50000 times to derive the $1\sigma$ errorbars shown in Figure \ref{Passfrac_Mhalo}. This method
accounts for uncertainties in the estimate of $P_{\rm{ty,i}}$ and $\int_{M_{\rm h}} P_{M_{\rm h},i | ty} dM_{\rm h}$ as well as cosmic
variance. 

{The vertical dashed lines, in Figure \ref{Passfrac_Mhalo}, mark the stellar mass completeness 
limit derived following \citet{Marchesini09}. In brief, we use the 3D-HST photometric catalog (down to \jhhst\ = 25 mag) and 
we scale the stellar masses of the galaxies as if they were at the spectroscopic sample limit of \jhhst = 24 mag (which defines the 
sample used in this work). The scatter of the points is indicative of the $M/L$ variations in the population at a given redshift. 
We then take the upper $95^{th}$ percentile of the distributions as a function of redshift as the stellar mass limit, which
is approximately $\sim10^{9.5}$ and $\sim 10^{10.5}$ for old and red galaxies at $z=1$ and $z=2$ respectively.  }
Below this mass we limit the upper edge of the redshift slice such that all galaxies in the stellar mass bin are included in a mass 
complete sample. A stellar mass bin is included only if the covered volume is greater than $1/3$ of the total volume of the redshift slice. 
This typically results in only one stellar mass bin below the completeness limit being included in the analysis.

In the highest halo mass bin of Figure \ref{Passfrac_Mhalo} at $z=0.5-0.8$, the satellite passive fraction (integrated over all galaxies) is
higher than the central passive fraction, with a marginal significance. The same trend can be
observed in the other halo mass and redshift bins, although the separation of the observed satellite and central passive 
fractions becomes more marginal.

In each redshift bin we also identify a sample of ``pure'' central galaxies ($P_{\rm cen} > 0.8$, irrespective of overdensity or halo mass),
which provides a reference for the passive fraction of galaxies subject only to mass-quenching. 
The passive fraction of this sample $f_{\rm{pass|cen,pure}}(M_*)$ of centrals (which has an average $P_{\rm cen} = 0.95$) is 
shown as the thick red line in both halo mass bins. 

The separation of the observed satellite passive fraction from that of the pure sample of
centrals is more significant (especially at $z<1.2$).  Indeed, the passive fractions derived using equation \ref{eqpassive_mhalo} 
can be strongly affected by impurities in the central/satellite classification and by cross-talk between the two halo mass bins, given that each galaxy 
can contribute to both bins and types (see equation \ref{eqpassive_mhalo}). 
Any contribution of central galaxies to the satellite passive fraction, and vice versa, will reduce the observed difference between the
two populations with respect to the ``pure'', intrinsic difference. 

\subsection{Recovering the ``pure'' passive fractions for satellite galaxies}\label{sec_decontamination}
In order to recover the ``pure'' passive fraction for satellite galaxies as a function of halo mass, we 
perform a parametric model fitting to our dataset.

We start by parametrizing the probability of a satellite galaxy being passive
independently in each stellar mass bin as a function of log halo mass, using a broken function 
characterized by a constant value ($P_{\rm{pass,lo}}$) below the lower break ($M_{\rm br,lo}$) and 
another constant value ($P_{\rm{pass,hi}}$) above the upper break ($M_{\rm br,hi}$). In between the
breaks, the passive fraction increases linearly.
Algebraically, this 4-parameter function is defined as:
\begin{equation}
P_{\rm{pass|sat}}(M_{\rm h}) =  \begin{cases} P_{\rm{pass,lo}} & \mbox{if } M_{\rm h} \leq M_{\rm br,lo} \\ m \times (\log \frac{M_{\rm h}}{M_{\rm br,lo}})+P_{\rm{pass,lo}} & \mbox{if } M_{\rm br,lo} < M_{\rm h} \leq M_{\rm br,hi}  \\ P_{\rm{pass,hi}} & \mbox{if } M_{\rm h}>M_{\rm br,hi} \end{cases}
\label{eqpassive_satmodel}
\end{equation}
where $m = (P_{\rm{pass,hi}}-P_{\rm{pass,lo}})/(\log(M_{\rm br,hi})-\log(M_{\rm br,lo}))$.

This function is chosen to allow for a great degree of flexibility.
We make the assumption that satellite galaxies are not subject to environmental quenching below $M_{\rm br,lo}$,
 and therefore treat $P_{\rm{pass,lo}}$ as a nuisance parameter of the model with a 
Gaussian prior centered on the observed passive fraction of pure centrals $f_{\rm{pass|cen,pure}}(M_*)$ and a sigma 
equal to its uncertainty.  For $P_{\rm{pass,hi}}$, instead we assume a semi-Gaussian prior with the same center and sigma 
as above, but only extending below the observed passive fraction of central galaxies (this implies that satellites are affected by 
the same mass-quenching as centrals). Above this value we assume a uniform prior. For the break masses we assume uniform priors.
Table \ref{modelpars} summarizes the model parameters, their allowed range, and the number of bins in which the range is 
divided to compute the posterior.

\begin{table}
\begin{center}
  \begin{tabular}{c c c c}
    \hline
        Parameter & Range & Nbins &  Prior  \\
    \hline
        $\log M_{\rm br,lo}$        & 11,15  & 80 & Uniform \\
        $\log M_{\rm br,hi}$        & 11,15  & 80 & Uniform \\
        \multirow{2}{*}{$P_{\rm{pass,hi}}$} & \multirow{2}{*}{0.0,1.0} & \multirow{2}{*}{100}  & Gaussian (if $P_{\rm{pass,hi}} \leq f_{\rm{pass|cen,pure}}$)\\
        &&&   Uniform (if $P_{\rm{pass,hi}} > f_{\rm{pass|cen,pure}}$) \\
    \hline
  \end{tabular}
  \caption{Table of the model parameters. }
  \label{modelpars}
\end{center}
\end{table}

The probability that each 3D-HST galaxy, $i$ is passive is:
\begin{equation}
P_{\rm{pass,i}} = P_{\rm{cen,i}} \times P_{\rm{pass|cen}}+P_{\rm{sat,i}} \times \int_{M_{\rm h}} P_{M_{\rm h}i | {\rm sat}} \times P_{\rm{pass|sat}} dM_{\rm h}
\end{equation}
where $P_{\rm{pass|sat}}$ is from equation \ref{eqpassive_satmodel} and $P_{\rm{pass|cen}}=f_{\rm{pass|cen,pure}}$.

The likelihood space that the star forming or passive activity of 3D-HST galaxies in a stellar mass 
bin is reproduced by the model is computed as follows:
\begin{equation}
\mathcal{L} =  \prod_{i} \begin{cases} P_{\rm{pass,i}} & \mbox{if } i {\rm ~is~UVJ~passive}  \\ 1-P_{\rm{pass,i}} & \mbox{if } i {\rm ~is~not~UVJ~passive} \end{cases}
\end{equation}

We compute the posterior on a regular 
grid covering the parameter space. We then sample the posterior distribution and we apply the model 
described in equation \ref{eqpassive_satmodel} to obtain the median value of $P_{\rm{pass|sat}}$ and
its $1\sigma$ uncertainty as a function of halo mass. 
Lastly, we assign the probability of being passive to mock satellites in each stellar mass bin 
according to their model halo mass, and we compute the average passive fraction
in the two halo mass bins (above and below $10^{13}M_\odot$). This results in the thick blue 
($f_{\rm{pass|sat,pure}}(M_*)$) lines with $1\sigma$ confidence intervals 
plotted as shaded regions in Figure \ref{Passfrac_Mhalo}.  
We illustrate in Appendix \ref{app_decontamination} an example of this procedure applied to a single
redshift bin. 

We verify that the separation seen in the pure passive fractions in Figure \ref{Passfrac_Mhalo} is 
real. To do so we randomly shuffle the position in the UVJ diagram for galaxies in each stellar mass bin 
(irrespective of environmental properties) to break any correlation between passive fraction and environment. 
Then we compute the observed passive fractions of centrals and satellites, and for the pure
sample of centrals and we perform again the model fitting procedure. 

At $0.5 < z < 0.8$ we find that 
the pure satellite passive fraction is inconsistent with the null hypothesis (no satellite quenching) at a $\gtrsim 2 \sigma$ level 
in each stellar bin at high halo mass and 7 out of 8 stellar mass bins at low halo mass. The combined probability of the null hypothesis is 
$P <10^{-10}$ in either halo mass bin. The difference is smaller, but still very significant ($P \lesssim 10^{-5}$) at $0.8<z<1.2$.
At $1.2<z<1.8$ the hypothesis of no satellite quenching is acceptable ($P \sim 0.4$)  in the low halo mass bin, while it can be 
ruled out ($P \lesssim 10^{-5}$) at higher halo mass.

\defcitealias{van-der-Burg13}{Van der Burg et al.}

\citetalias{van-der-Burg13} \citeyearpar{van-der-Burg13}, \citet{Kovac14}, and \citet{Balogh16} have found that the environment 
plays an important role in determining the star formation activity of satellites, at least up to $z \sim 1$. However these works have 
only probed relatively massive haloes ($M_{\rm h} \gtrsim 10^{13} M_\odot$). 
The depth of the 3D-HST sample allows us, for the first time, to extend these results to higher redshift, to lower mass galaxies 
and to lower mass haloes.

\begin{figure*}
\begin{center}
\includegraphics[width = 15.5cm]{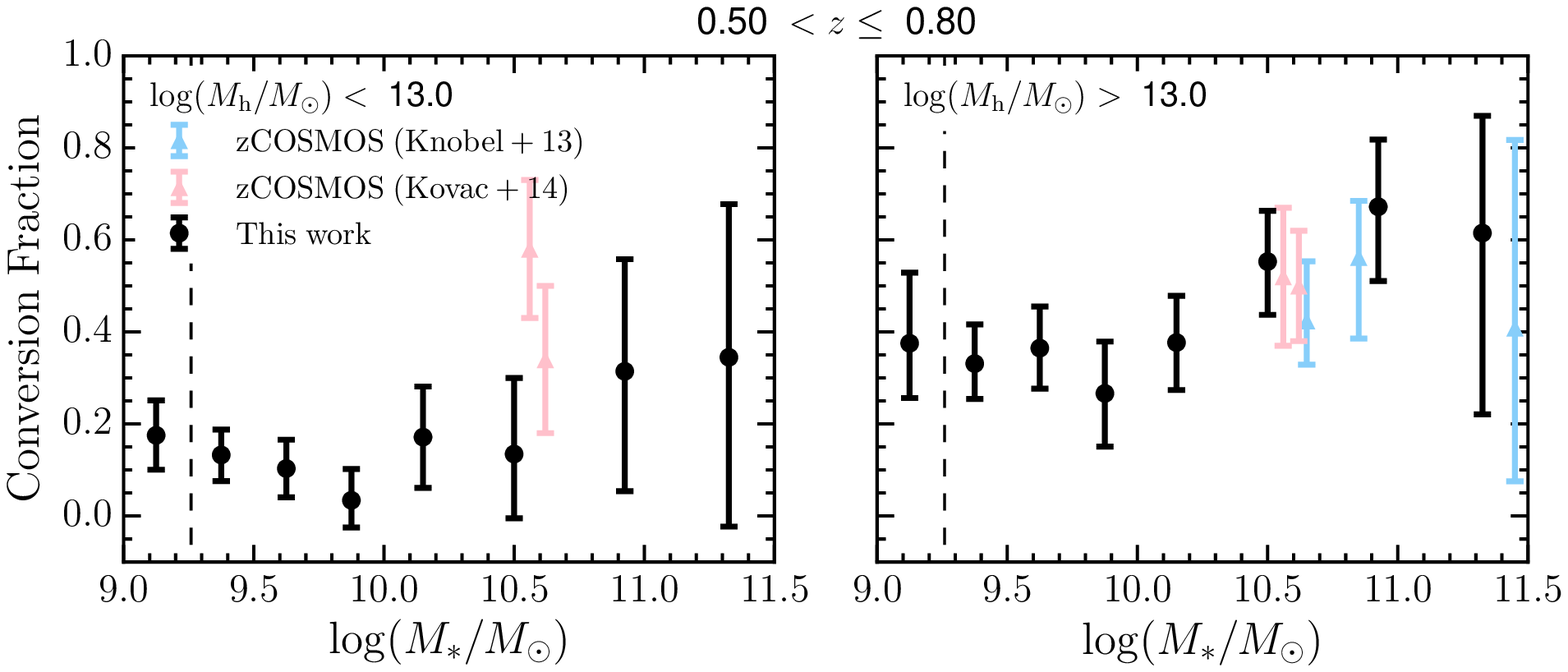}
\includegraphics[width = 15.5cm]{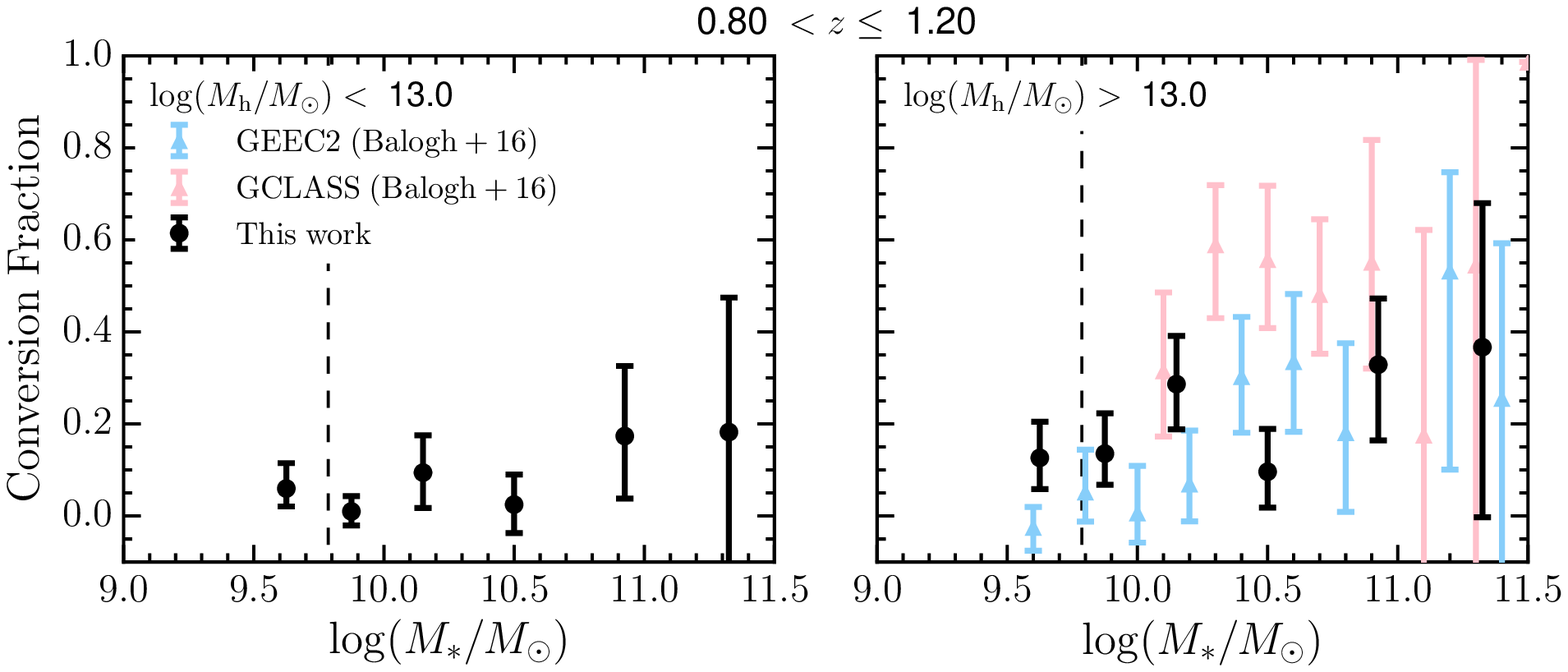}
\includegraphics[width = 15.5cm]{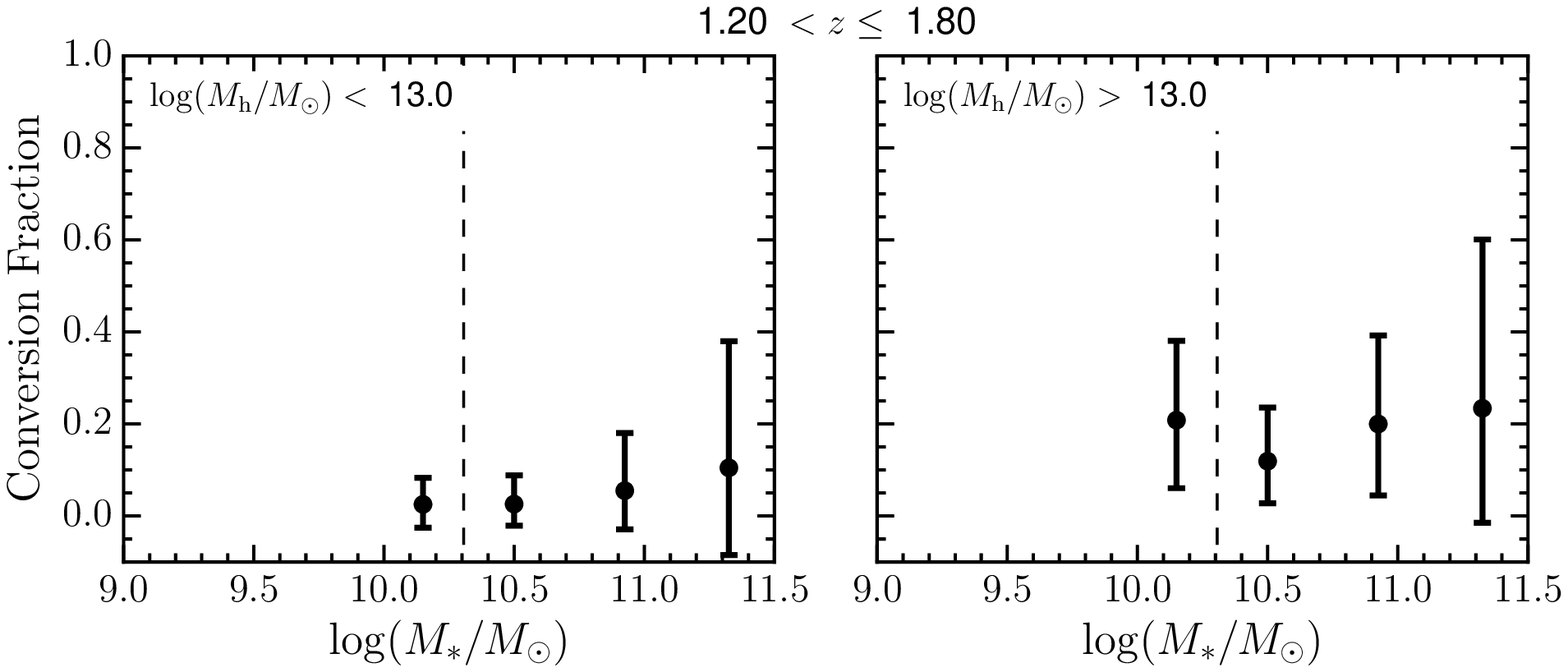}
\end{center}
\caption{Conversion fractions for satellite galaxies in bins of  $M_{\rm{*}}$ and $M_{\rm{h}}$ obtained from equation \ref{eqConvfrac}
in three independent redshift bins. Black points are from this work and the $1\sigma$ errorbars are propagated from the uncertaintines on the 
passive fractions using a Monte Carlo technique. Colored points are from previous studies in the same redshift range. We note a
good agreement with our measurements despite different passiveness criteria and environment estimates. The vertical dashed 
line marks the stellar mass limit of the volume limited sample. } 
\label{Convfrac_Mhalo}
\end{figure*}

\begin{figure*}
\begin{center}
\includegraphics[width = 15.5cm]{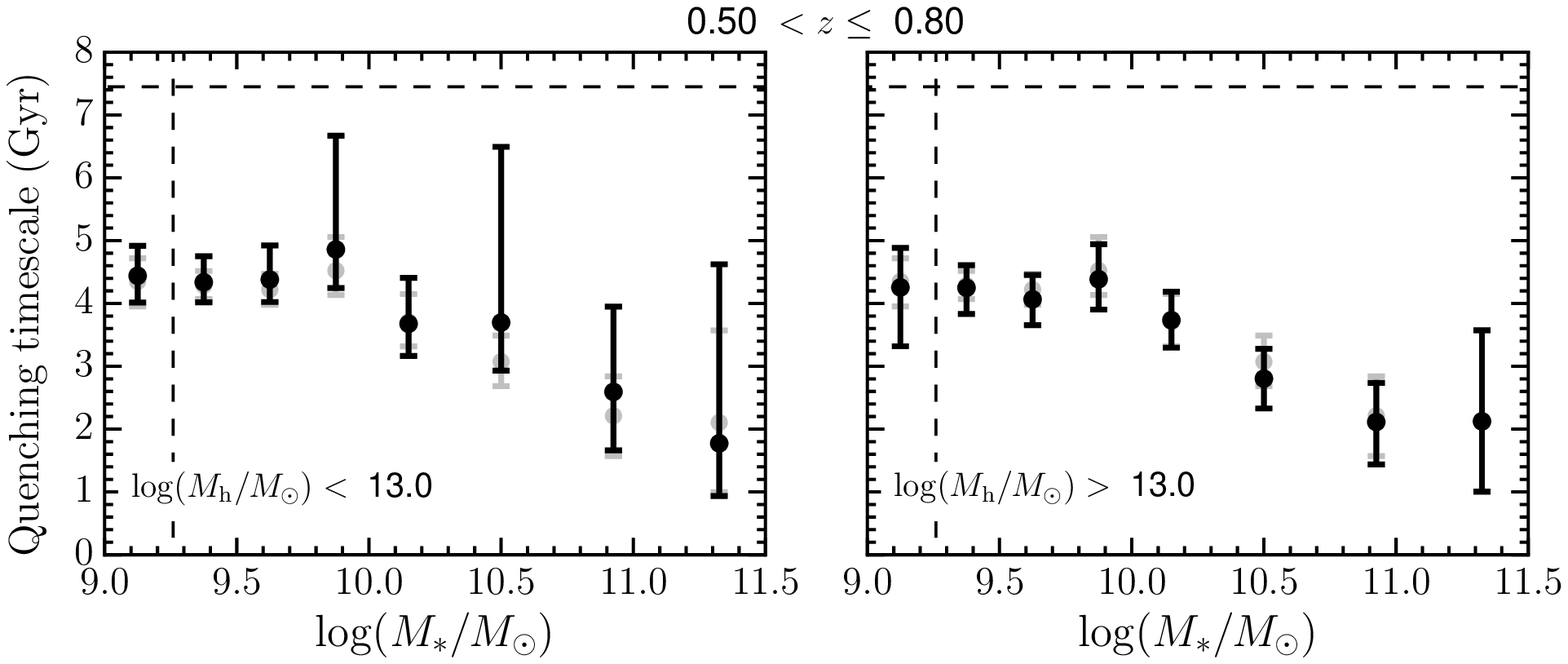}
\includegraphics[width = 15.5cm]{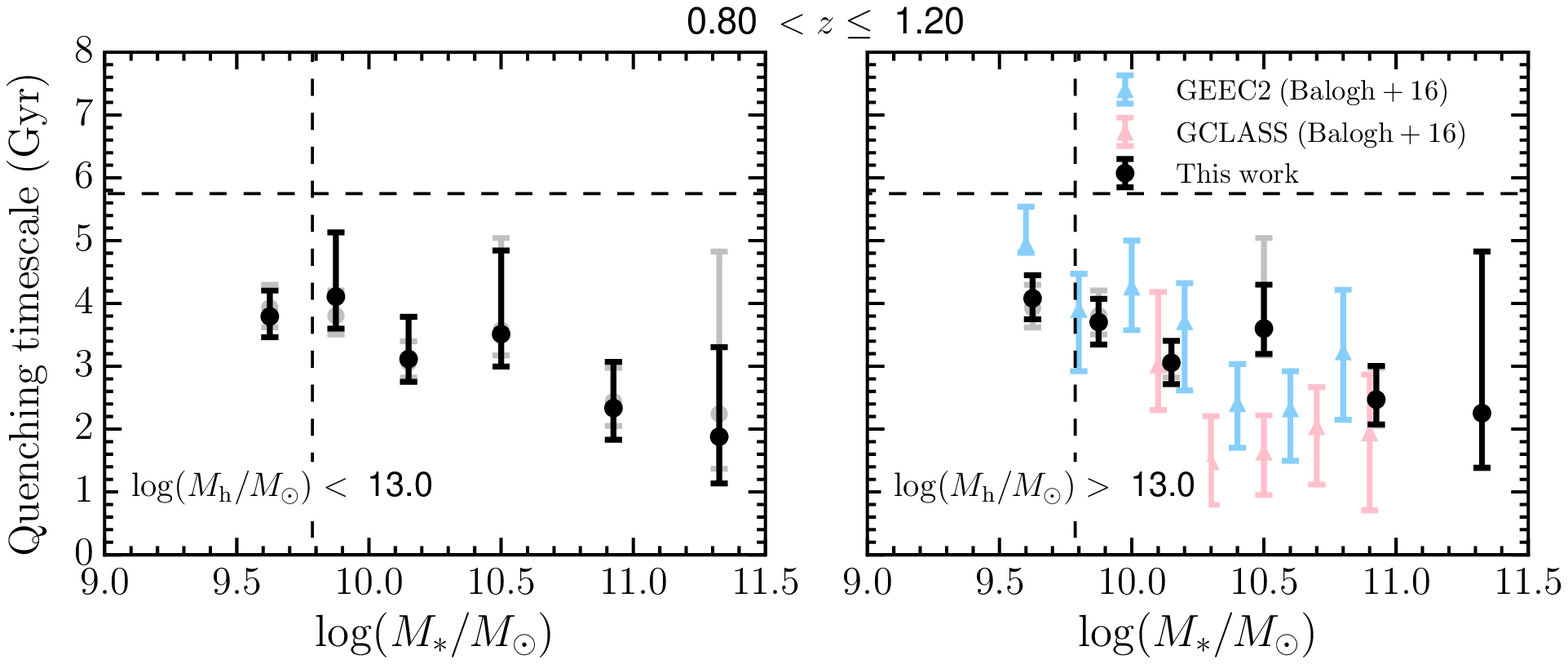}
\includegraphics[width = 15.5cm]{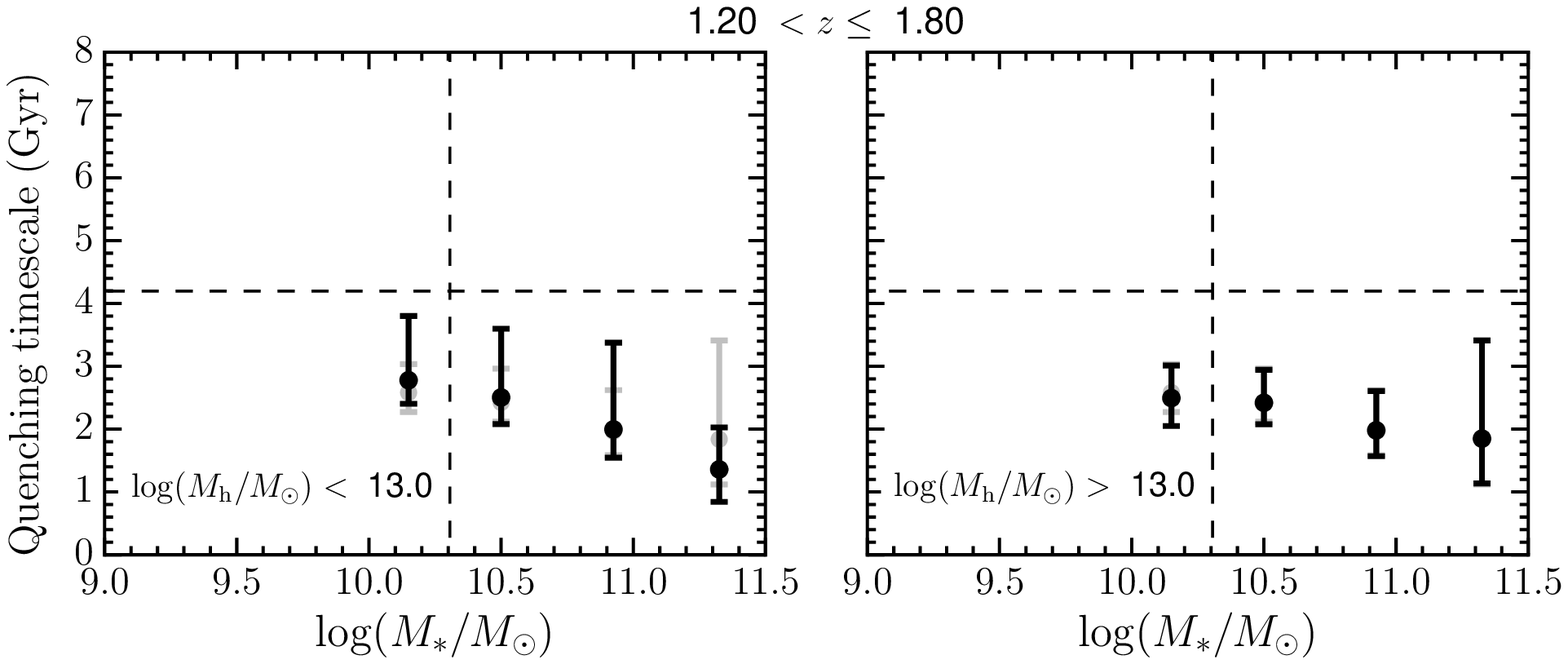}
\end{center}
\caption{Quenching times for satellite galaxies in bins of  $M_{\rm{*}}$ and $M_{\rm{h}}$ obtained from the distributions of $T_{\rm sat}$
from the mock sample. Quenching times are obtained under the assumption that the galaxies which have been satellites the longest 
are those which have been environmentally quenched.  Black points are from this work, while colored points are from previous studies 
in the same redshift and halo mass range. The gray points are obtained from our sample without separating the dataset in two halo mass bins, 
and are therefore identical in the left and right panels. The vertical dashed line marks the stellar mass limit of the volume limited sample. 
The horizontal dashed line is the age of the Universe at the central redshift of each bin.} 
\label{Tquench_Mhalo}
\end{figure*}

\subsection{Satellite quenching efficiency} \label{sec_convfrac_mhalo}
In order to further understand the increased passive fractions for satellite galaxies we compute the ``conversion fractions'' as first 
introduced by \citet{van-den-Bosch08}. This parameter, sometimes called the satellite quenching efficiency, quantifies the fraction of
galaxies that had their star formation activity quenched by environment specific processes since they accreted as satellites into 
a more massive halo \citep[see also][]{Kovac14,Hirschmann14,Balogh16}. It is defined as:
\begin{equation}
f_{\rm conv} (M_*, M_{\rm h}) = \frac{f_{\rm{pass|sat,pure}}(M_*, M_{\rm h})-f_{\rm{pass|cen,pure}}(M_*)}{1-f_{\rm{pass|cen,pure}}(M_*)}
\label{eqConvfrac}
\end{equation}
where $f_{\rm{pass|sat,pure}}(M_*, M_{\rm h})$ and $f_{\rm{pass|cen,pure}}(M_*)$ are the corrected 
fractions of quenched centrals and satellites in a given bin of $M_*$ and  $M_{\rm h}$ obtained as described above.

In equation \ref{eqConvfrac} we compare the sample of centrals at the same redshift as the satellites. This builds on  the assumption that the 
passive fraction of central galaxies only depends on stellar mass and that the effects of mass and environment are independent and separable.  
The conversion fraction then represents the fraction of satellites which are quenched due to environmental processes above what would happen if those 
galaxies would have evolved as centrals of their haloes. A different approach would be to compare the passive fraction of satellites to that of centrals 
at the time of infall in order to measure the total fraction of satellites quenched since they were satellites \citep[e.g.,][]{Wetzel13,Hirschmann14}. 
However this measurement includes the contribution of mass-quenched satellite galaxies, which we instead remove under the assumption
that the physical processes driving mass quenching do not vary in efficiency when a galaxy becomes a satellite.

We also caution the reader that equation \ref{eqConvfrac} has to be taken as a simplification of reality as it does not take into 
account differential mass growth of centrals and satellites which can be caused by tidal phenomena in dense environments or 
different star formation histories.

Figure \ref{Convfrac_Mhalo} shows the conversion fractions in the same bins of $M_{\rm{*}}$, $M_{\rm{h}}$ and redshift as presented in 
Figure \ref{Passfrac_Mhalo}.
Previous results from galaxy groups and clusters from \citet{Knobel13}, and \citet{Balogh16} are plotted in our higher halo mass bin (colored 
points with errorbars). We also add the conversion fractions from \citet{Kovac14} obtained from zCOSMOS data as a function
of local galaxy overdensity. We plot their overdensity bins above the mean overdensity in our higher halo mass bin and the others
in our lower halo mass bin following the overdensity to halo mass conversion given in \citet{Kovac14}. 
The agreement of our measurements with other works is remarkable considering that different techniques to define the 
environment (density and central/satellite status) and passiveness are used in different works. 

The satellite quenching efficiency tends to increase with increasing stellar mass and to decrease with increasing redshift at fixed stellar 
mass. In the lower halo mass bin, we note the presence of similar trends as at higher halo masses although the uncertainties
are larger due to the smaller number of satellites. In our probabilistic approach this is due to the lower $P_{\rm sat}$ in low density 
environments as shown in Figure \ref{PcenPsat}. Moreover, $f_{\rm conv}$ is poorly constrained at $M_* > 10^{11} M_\odot$ due to 
small number statistics of high mass satellites in the 3D-HST fields. 

\begin{figure*}
\includegraphics[width = 17.5cm]{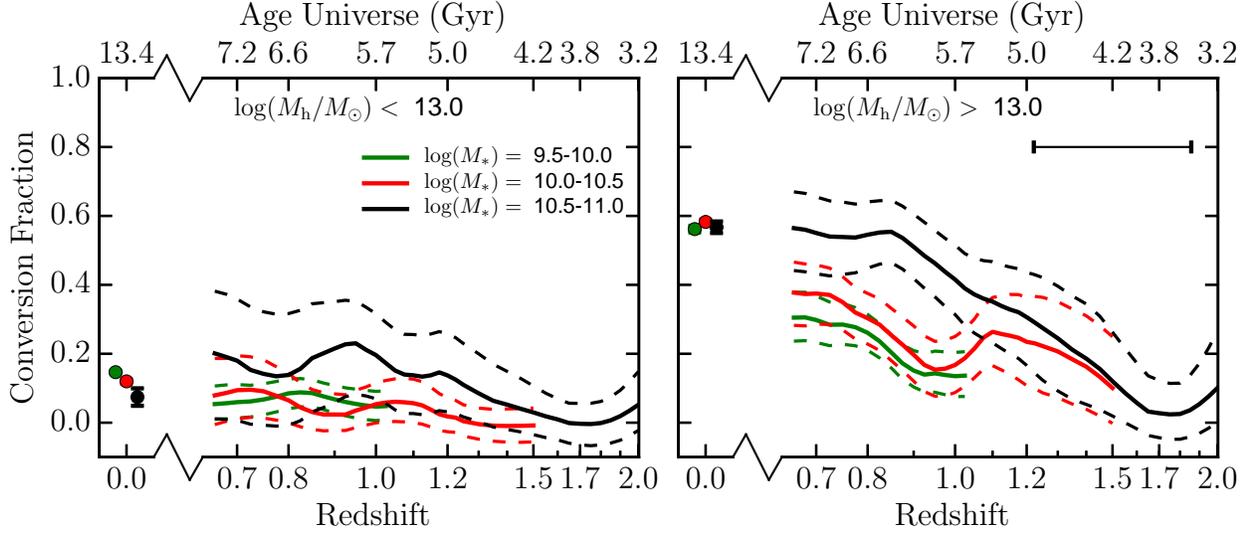}
\caption{Conversion fractions for satellite galaxies as a function of redshift in bins of $M_{\rm{*}}$ and $M_{\rm{h}}$ (solid lines). 
Dashed lines mark the $1\sigma$ confidence levels. The horizontal error bar is the span of the redshift bins (for the 3D-HST sample) 
which is constant in $\Delta z/(1+z) = 0.2$ where $\Delta z$ is the width of the redshift bin and $z$ its center.
Datapoints at $z=0$ are from the SDSS sample and are offset
along the x-axis for clarity if they overlap. } 
\label{Convfracz}
\end{figure*}

\begin{figure*}
\includegraphics[width = 17.5cm]{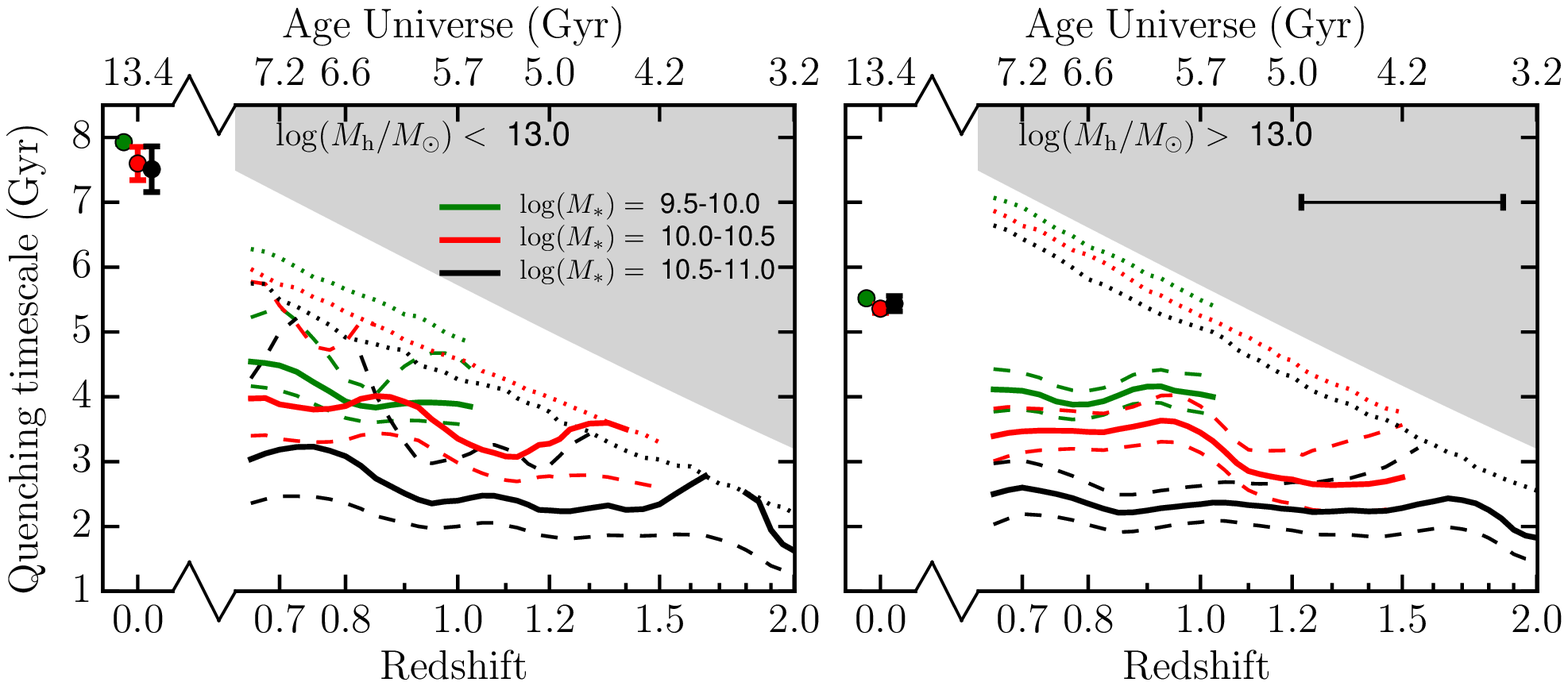}
\caption{Quenching timescales for satellite galaxies as a function of redshift in bins of $M_{\rm{*}}$ and $M_{\rm{h}}$ (solid lines) for
the 3D-HST sample. Dashed lines mark the $1\sigma$ confidence levels. The horizontal error bar is the span of the redshift 
bins (for the 3D-HST sample) which is constant in $\Delta z/(1+z) = 0.2$ where $\Delta z$ is the width of the redshift bin and $z$ its center.
Datapoints at $z=0$ are from the SDSS sample. 
The area where $T_{\rm quench}$ is larger than the Hubble time is shaded in grey. Dotted lines are obtained from the $99^{\rm th}$
percentile of $T_{\rm sat}$ in the mock sample and represent the look-back time at which the first 1\% of the satellite 
population at a given redshift was accreted onto more massive haloes as satellites. We define this limit as the maximum value of 
$T_{\rm quench}$ which would produce a meaningful environmentally quenched satellite population at any given redshift.}
\label{Tquenchz}
\end{figure*}

\subsection{Quenching timescales} \label{sec_tquench_mhalo}
A positive satellite conversion fraction can be interpreted in terms of a prematurely truncated star formation 
activity in satellite galaxies compared to field centrals of similar stellar mass. 

{We define the quenching timescale ($T_{\rm quench}$) 
as the average time elapsed from the first accretion of a galaxy as satellite
to the epoch at which the galaxy becomes passive, and we estimate it by assuming that galaxies 
which have been satellites for longer times are more likely to be quenched \citep{Balogh00, McGee09, Mok14}.
Indeed, the quenching can be interpreted to happen a certain amount  
of time after satellite galaxies cease to accrete material (including gas) from the 
cosmic web (see Section \ref{sec_discussion}).}

In practice, we obtain quenching timescales from the distribution of $T_{\rm sat}$ for satellite galaxies, which we define as the time 
the galaxy has spent as a satellite of haloes of any mass since its first infall \citep[e.g.,][]{Hirschmann14}. 
For each bin of $M_{\rm{*}}$, $M_{\rm{h}}$, 
and redshift we select all satellite galaxies in our mock lightcones which define the distribution of $T_{\rm sat}$. Then we select as the 
quenching timescale the percentile of this distribution which corresponds to $1-f_{\rm conv}(M_*, M_{\rm h})$. This method builds
on the assumption that the infall history of observed satellites is well reproduced by the SAM. Systematic 
uncertainties can arise in the analytic prescriptions used for the dynamical friction timescale of satellites whose parent halo 
has been tidally stripped in the N-body simulation below the minimum mass for its detection (the so-called ``orphan galaxies''). 
When this time is too short, too many satellites merge with the central galaxy and are removed from the sample, 
and vice-versa when the time is too long. 
\citet{Delucia10} explored the dynamical friction timescale in multiple SAMs, finding a wide range of timescales.
However a dramatically wrong dynamical friction recipe impacts the fraction of satellites, the stellar mass functions, 
and the density-mass bivariate distribution, which we found to be well matched between the mocks and the observations.

In principle low stellar mass galaxies ($M_* < 10^{10} M_\odot$) are more affected by the resolution limit of the simulation
and their derived quenching timescales might be subject to a larger uncertainty compared to galaxies of higher stellar mass. We verified 
that this is not the case by comparing the distribution of $T_{\rm sat}$ in two redshift snapshots ($z=1.04$ and $z=2.07$) of \citet{Henriques15} built on the
Millennium-I and the Millennium-II simulations. The latter is an N-body simulation started from the same initial conditions of the original Millennium
run but with a higher mass resolution at the expense of a smaller volume. 
The higher resolution means that the sub-haloes hosting low mass satellites galaxies, which can be tidally stripped, are explicitly tracked 
to lower mass and later times: while these are detected the recipe for dynamical friction is not invoked. We obtain consistent quenching timescales
for Millennium-I and Millennium-II based mock catalogues, and therefore we conclude that the analytical treatment of orphan galaxies 
does not bias our results. 

Figure \ref{Tquench_Mhalo} shows our derived quenching timescales (black points) in the same bins of $M_{\rm{*}}$ and $M_{\rm{h}}$ and 
redshift as presented in Figures \ref{Passfrac_Mhalo} and \ref{Convfrac_Mhalo}. 
The observed trend of $f_{\rm conv}$ with stellar mass that is found in both redshift bins turns into a trend of $T_{\rm quench}$. 
Quenching timescales increase to lower stellar mass in all redshift and halo mass bins, mainly a consequence of the decreasing conversion fraction.
This parameter ranges from $\sim 4-5$ Gyr for low mass galaxies
to $<2$ Gyr for the most massive ones, and is in agreement with that found by \citet{Balogh16}. 

Remarkably, the dependence of quenching 
timescale on halo mass is very weak. We overplot in each panel, as gray symbols, the quenching timescales obtained from our sample with 
the same procedure described above but without separating the data in two halo mass bins. In most of the stellar mass bins we find a good 
agreement, within the uncertainties, between the black and the gray points.

The lack of a strong halo mass dependence is a consequence of the typically shorter time since infall for satellite galaxies in lower mass haloes 
which largely cancels the lower conversion fraction in low mass haloes, and suggests that the physical process responsible for the premature 
suppression of star formation in satellite galaxies (when the Universe was half of its present age) is largely independent of halo mass. 

A mild redshift evolution is also seen when comparing the redshift bins: passive satellites at higher redshift are 
quenched on a shorter timescale.
In the next section we will further explore the redshift evolution of the quenching timescales from $0 < z < 2$ by combining the 3D-HST 
sample with a local galaxy sample from SDSS.

\subsection{Redshift evolution of the quenching timescales} \label{sec_redshiftevo}
Figures \ref{Convfracz} and \ref{Tquenchz} show the evolution of the conversion fraction and the quenching timescale from 
redshift 0 to 2. We now concentrate on three bins of stellar mass, each of 0.5 dex in width, and ranging from $10^{9.5} M_\odot$ to $10^{11} M_\odot$. 

Given that $f_{\rm conv}$ (and consequently $T_{\rm quench}$) are poorly constrained at $M_* > 10^{11} M_\odot$ due to the low 
number statistics of massive satellites, we exclude more massive galaxies from these plots. Similarly, galaxies at $M_* < 10^{9.5} M_\odot$
are only included in the mass limited sample at the lowest end of the redshift range under study, therefore the redshift evolution
of $f_{\rm conv}$, and $T_{\rm quench}$ cannot be derived for those low mass galaxies.
A stellar mass bin appears in Figures \ref{Convfracz} and \ref{Tquenchz} only if the stellar mass range above the mass 
limit is more than half of the entire stellar mass extent of the bin.

The values (solid lines) and their associated uncertainties (dashed lines) are obtained by performing the procedure described in the 
previous sections in overlapping redshift bins defined such that $\Delta z/(1+z) = 0.2$ where $\Delta z$ is the width of the redshift 
bin and $z$ its center. This means we span larger volumes at higher redshift, modulating the decrease in sample density 
(Malmquist bias) and retaining sufficient sample statistics. It is also close to a constant bin in cosmic time. The x-axis of 
both figures is scaled such that the width of the redshift bins is constant and is shown as the horizontal error bar. 
We include only galaxies in a stellar mass complete sample for each redshift bin. 
In addition to the 3D-HST based constraints, we add constraints at $z=0$, obtained using the same method to ensure homogeneity. 
The observational sample is drawn from SDSS and the mock sample 
from the redshift zero snapshot of the \citet{Henriques15} model. We describe the details of how those datasets are processed in Appendix 
\ref{app_SDSS}. For this sample we restrict to stellar masses above $10^{9.5} M_\odot$ to avoid including low mass galaxies with large 
$V_{max}$ corrections.

The evolution of $f_{\rm conv}$ as seen in Figure \ref{Convfrac_Mhalo} is now clearly visible over the 
large redshift range probed by 3D-HST. The fraction of environmentally quenched satellite galaxies is a function of $M_h$, $M_*$ and redshift.
At fixed redshift $f_{\rm conv}$ is higher for higher mass galaxies and at fixed stellar mass it is higher in more massive 
haloes. More notably, the redshift evolution follows a decreasing trend with increasing redshift such that at $z\sim 1.5$
the excess of quenching of satellite galaxies becomes more marginal (at least for massive galaxies) as 
first predicted by \citet{McGee09} using halo accretion models. 
Several observational works reached a similar conclusion. 
\citet{Kodama04}, \citet{DeLucia07a}, and \citet{Rudnick09} found a significant build-up of the faint end of the red sequence (of 
passive galaxies) in cluster environments from $z\sim1$ toward lower redshift. This implies an increase in the fraction of 
quenched satellites with decreasing redshift for low mass galaxies.
Recently, \citet{Darvish16} found that the environmental
quenching efficiency tends to zero at $z>1$, although their analysis is only based on local overdensity and does not separate 
centrals and satellites. With the 3D-HST dataset we cannot rule out that satellite quenching is still efficient for lower mass 
satellites at $z>1.5$; deeper samples are required to robustly assess the satellite quenching efficiency at $z\sim 1.5-2.0$.

Moving to the present day Universe (SDSS data) does not significantly affect the fraction of environmentally quenched 
satellites despite the age of the Universe nearly doubling compared to the lowest redshift probed by the 3D-HST sample. 

The redshift dependence of the quenching timescale originates from the combination of the evolution of $f_{\rm conv}$ 
and the distributions of infall times for satellite galaxies. The redshift evolution of $f_{\rm conv}$ in the high halo 
mass bin is well matched by the halo assembly history (at lower redshift they have been satellites on average 
for more time) and therefore $T_{\rm quench}$ is mostly independent of redshift. 
However, for lower mass galaxies a mild redshift evolution of $T_{\rm quench}$ might be present. 
However the slope is much shallower than the ageing of the Universe. For this reason, going to higher redshift, 
$T_{\rm quench}$ approaches the Hubble time and the satellite quenching efficiency decreases. 

Despite the large uncertainty on the quenching times at low halo mass, their redshift evolution appears to be largely 
independent of halo mass. This means that the halo mass dependence of the conversion fractions may be mostly 
driven by an increase in the time spent as satellites in more massive haloes. At $z=0$ a more significant difference is 
found between the quenching times in the two halo mass bins. In the next section we discuss which 
mechanism can produce these observational signatures.

\section{Discussion} \label{sec_discussion}
There is a growing consensus that the evolution of central galaxies is regulated by the balance between cosmological
accretion, star formation and gas ejection processes in a so-called ``equilibrium growth model'' \citep[e.g.][]{Lilly13}. 
The reservoir of cold gas in each galaxy is replenished by accretion, and will fuel star formation. As the rate of 
cosmological accretion is correlated with the mass of the halo, this regulates mass growth via star formation. As a result
the eventual stellar mass is also tightly correlated with halo mass, driving a tight relation between star formation rate 
and stellar mass for normal star forming galaxies  \citep[the main sequence - MS - of star forming galaxies, e.g.][]{Noeske07}. 

When galaxies fall into a more massive halo the accretion of new gas from the cosmic web is expected to cease: 
such gas will instead be accreted (and shock heated) when it reaches the parent halo \citep{White91}. More recently \citet{Dekel06}
estimate that this process occurs at a minimum halo mass $M_h \sim 10^{12} M_\odot$, which is largely independent
of redshift. This roughly corresponds to the minimum halo mass at which satellites are detected in the 3D-HST survey
(see Figure \ref{Mhalodistr}).

\subsection{Identification of the main mechanism}
There are several additional ways in which a satellite galaxy's gas and stellar content can be modified through 
interaction with its environment, including stripping of the hot or cold gas, and tidal interactions 
among galaxies or with the halo potential itself. An important combined effect is to remove (partially or completely) the 
gas reservoir leading to the quenching of star formation. However, as pointed out by \citet{McGee14}, and \citet{Balogh16},
it might not be necessary to invoke these mechanisms of environmental quenching to be effective. The high SFR 
typical of galaxies at high redshift, combined with outflows, can lead to exhaustion of the gas reservoir in the absence
of cosmological accretion. 

Our approach to link the conversion fractions to the distributions of time spent as satellite is based on the 
assumption that a galaxy starts to experience satellite specific processes at the time of its first infall into a 
larger halo and, in particular, that the cosmological accretion is shut off at that time. 

We now examine whether a pure exhaustion of the gas reservoir can explain the quenching times we observe, 
or whether additional gas-removal mechanisms are required. First we appeal to the similarity of quenching 
times in the two halo mass bins shown in Figure \ref{Tquenchz} to support the pure gas exhaustion scenario. 
Other than at $z=0$, the derived quenching times are indeed consistent within the uncertainties, 
therefore the main quenching mechanism has to be largely independent of halo mass.

Ram pressure stripping is often invoked as the main quenching mechanism
for satellite galaxies in low redshift clusters \citep[e.g.,][]{Poggianti04, Gavazzi13a, Boselli14b}. 
Its efficiency is a function of the intracluster medium (ICM) density and the
velocity of galaxies in the halo. More massive haloes have a denser ICM and satellites move faster through it 
which exerts a stronger dynamical pressure on the gas leading to faster stripping (and shorter quenching times) 
in more massive haloes \citep{Vollmer01, Roediger05}. 
Our 3D-HST dataset does not extend to the extreme high mass end of the halo mass function in which ram pressure effects 
have been clearly observed \citep[e.g.,][]{Sun07, Yagi10, Merluzzi13, Kenney15, Fossati16}, and so the lack of significantly 
shorter quenching times in the higher halo mass bin is consistent with the lack of stripping, and indeed of any strong 
halo-mass dependent gas-stripping process.
However \citet{Balogh16} find a small halo mass dependence of the quenching times comparing their GEEC2 group 
sample ($M_h \sim 10^{13.5} M_\odot$) to the GCLASS cluster sample ($M_h > 10^{14} M_\odot$). 
These evidences might indicate that dynamical stripping can play a minor role in more massive haloes even at $z \sim 1$. 

At $z=0$ instead, thanks to the large area covered by the SDSS dataset, a number of very massive haloes are included in the
higher halo mass bin. This, combined with the presence of hot and dense ICM in massive haloes in the local Universe might be 
sufficient to explain the shorter quenching times in the high halo mass bin. \citet{Haines15}, and \citet{Paccagnella16} 
found quenching timescales which are possibly shorter in massive clusters of galaxies ($\sim 2-5$ Gyr). However, a quantitative
comparison is hampered by the different definitions of the quenching timescale.  



\subsection{Delayed then Rapid or Continuous Slow quenching?} \label{sec_delayedorslow}
Having ruled out gaseous stripping as the main driver of satellite quenching in the range of halo mass commonly probed
by our samples ($M_h \lesssim 10^{14} M_\odot$), we now concentrate on how the gas exhaustion
scenario can explain the observed values of $T_{\rm quench}$. 

\begin{figure}
\includegraphics[width = 9.0cm]{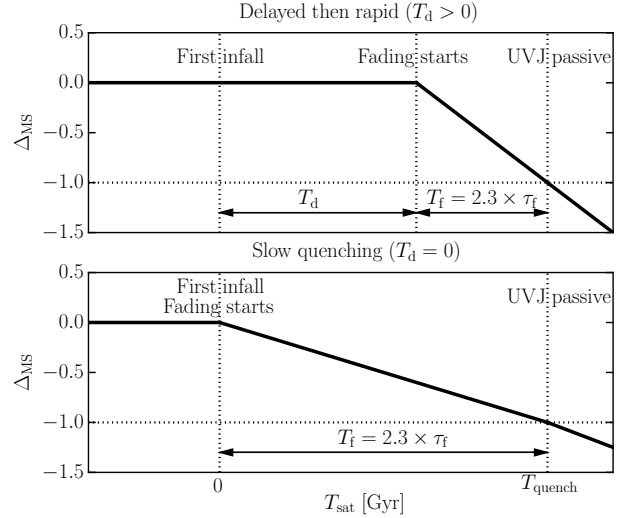}
\caption{Schematic diagram of the evolution of the MS offset for two toy models of satellite quenching as a
function of the time spent as a satellite. In both cases satellite quenching starts at the time of first infall for a 
galaxy at the main sequence mid-line value,  which becomes UVJ passive after $T_{\rm quench}$. 
In the ``delayed then rapid'' model (Top panel) the satellite galaxies evolves on the main sequence for a delay 
time $T_{\rm d}$. Then their SFR drops exponentially with a characteristic timescale $\tau_f$.  
In the ``slow quenching'' model (Bottom panel), $T_{\rm d}=0$ and the galaxy follows a slow(-er) exponential decline 
of the SFR immediately after its first infall into a more massive halo. } 
\label{Quenchmodels}
\end{figure}

To explain the quenching times at $z=0$, \citet{Wetzel13} presented a model dubbed the ``delayed then rapid'' quenching 
scenario, shown in the top panel of Figure \ref{Quenchmodels}.  This model assumes that $T_{\rm quench}$ can be divided 
into two phases. During the first phase, usually called the ``delay time'' ($T_{d}$), the star formation activity of satellites on 
average follows the MS of central galaxies.
After this phase the star formation rate drops rapidly and satellite galaxies become passive on a short timescale called the
``fading time''.  
\citet{Wetzel13} estimated an exponential fading with a characteristic timescale $\tau_f \sim 0.3-0.8$ Gyr which
depends on stellar mass at $z=0$. 
At $z \sim 1 $, \citet{Mok14, Muzzin14}; and \citet{Balogh16} estimated the fading time to be $\tau_f \sim 0.4-0.9$ Gyr, 
by identifying a ``transition'' population of galaxies likely to be transitioning from a star forming to a passive phase. 
These values suggest little redshift evolution of the fading timescale with cosmic time. 

\citet{McGee14} developed a physical 
interpretation of this model. These authors assumed that the long delay times are only possible if the satellite galaxy 
has maintained a multi-phase reservoir which can cool onto the galaxy and replenish the star forming gas 
(typically molecular) at roughly the same rate as the gas is lost to star formation (and potentially outflows).
A constant molecular gas reservoir produces a nearly constant SFR according to the Kennicutt-Schmidt 
relation \citep{Schmidt59, Kennicutt98a}. Then the eventual depletion of this cold gas results in the
rapid fading phase. 

An alternative scenario would be that satellite galaxies retain only their molecular gas reservoirs after infall. In this case, if we 
assume a constant efficiency for star formation we should expect a star formation history which immediately departs from 
the MS, declining exponentially as the molecular gas is exhausted (``slow quenching'' model shown in the bottom panel 
of Figure \ref{Quenchmodels}). By using our data we directly test those two toy models.

We use the star formation rates ($SFR(M_*,z)$) for 3D-HST galaxies presented in the \citet{Momcheva16} catalogue. 
By limiting to galaxies in the redshift range $0.5<z<1.5$,  
stellar mass range $9.5 < \log(M_*) < 11$ and a maximum offset below the main sequence of 0.5 dex, we make sure that the 
SFR estimates are reliable and, for 91\% of the objects, are obtained from {\it Spitzer} 24$\mu$m observations combined with a UV 
monochromatic luminosity to take into account both dust obscured and unobscured star formation. For the remaining 9\% 
SFR estimates are from an SED fitting procedure \citep[see][]{Whitaker14, Momcheva16}.

There is growing evidence of curvature in the MS, which becomes shallower at higher stellar mass. \citet{Whitaker14}, 
\citet{Gavazzi15a}, and \citet{Erfanianfar16} interpreted this as a decline in star formation efficiency caused by 
the growth of bulges or bars in massive galaxies. To study the effects of environment above the internal 
processes driving the star formation efficiency at fixed stellar mass, we convert the
SFR into an offset from this curved MS: $\Delta_{\rm MS,obs} = \log(SFR(M_*,z)/SFR_{\rm MS}(M_*,z)$ 
using the \citet{Wisnioski15} parametrization of the MS from \citet{Whitaker14}. 

\begin{figure}
\includegraphics[width = 8.5cm]{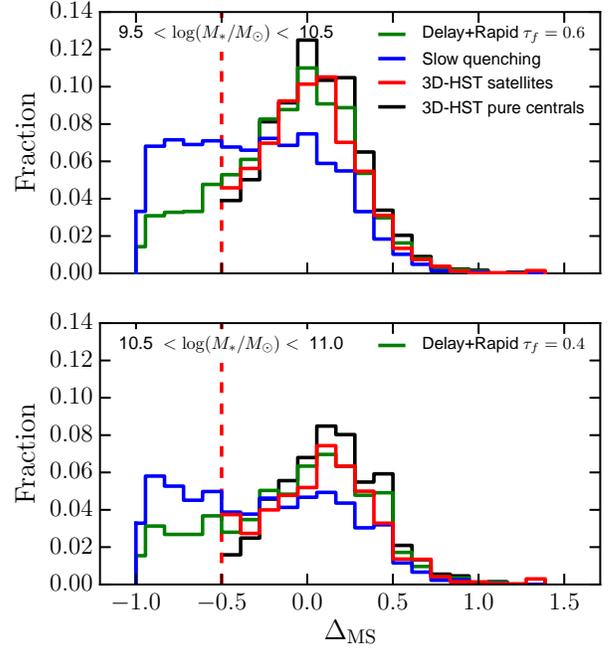}
\caption{Logarithmic offset from the main sequence ($\Delta_{\rm MS}$) in two stellar mass bins for 3D-HST observed 
satellites at $0.5<z<1.5$ (red histogram) and for mock galaxies, in the same redshift range, assuming a 
``slow quenching'' (blue histogram), or a ``delayed then rapid'' (green histogram) scenario (see text 
for the details of those two toy models). The histograms are normalized to the total number of 3D-HST satellites (including UVJ 
passive galaxies). The main sequence offset of a pure sample of observed central galaxies (black histogram) is shown for comparison.
The red vertical dashed line is the limit below the main sequence at which SFR estimates for 
observed galaxies are based predominantly on accurate IR+UV measurements accounting for obscured and unobscured star formation.} 
\label{MSoffset}
\end{figure}

In order to test the two models we again resort to the mock sample. For each central galaxy in the mocks we assign 
a random offset from the main sequence obtained from a pure sample ($P_{\rm cen} > 0.8$) of observed centrals: 
$\Delta_{\rm MS,cen}$. 
For satellite galaxies, instead, their $\Delta_{\rm MS}$ is a function of their time spent as satellites ($T_{\rm sat}$) as follows:
\begin{equation}
\Delta_{\rm MS}= \begin{cases} <-1 & \mbox{if } T_{\rm sat}>T_{\rm quench} \\ \Delta_{\rm MS,cen} & \mbox{if } T_{\rm sat}\le T_{\rm d} \\ \Delta_{\rm MS,cen}+\log(e^{-(T_{\rm sat}-T_{\rm d})/ \tau_f}) & \mbox{if } T_{\rm sat}>T_{\rm d}  \end{cases}
\label{eqMSoffset}
\end{equation}

{where $T_{\rm quench}$ and $T_{\rm d}$ are the total quenching time and the delay time respectively, 
and $\tau_f$ is the characteristic timescale of the exponential fading phase.} The latter is computed for each 
galaxy independently such that the SFR drops 1 dex below the MS\footnote[3]{This is the 
value that defines the typical division between UVJ star forming and passive objects.} in 
($T_{\rm quench}-T_{\rm d}$) Gyr. As we already computed $T_{\rm quench}$, the only free parameter remaining 
in this family of models is $T_{\rm d}$. We define the ``slow quenching'' model 
for $T_{\rm d}=0$, and ``delayed then rapid'' those where $ 0 < T_{\rm d} < T_{\rm quench}$. 

Figure \ref{MSoffset} shows the distributions of $\Delta_{\rm MS}$
for 3D-HST satellites in two stellar mass bins,
obtained as usual by weighting all galaxies by $P_{\rm sat}$, and for the two models obtained from the mock sample 
in the same way. The histograms are normalized to the total number of 3D-HST satellites in the same stellar mass bin (including UVJ 
passive galaxies). We stress that this comparison is meaningful because our models include the cross-talk 
between centrals and satellites. 

In the ``delayed then rapid'' scenario, the value of the delay time that best reproduces the observed data is
 $T_{\rm d}=T_{\rm quench}-1.4 (0.9)$ Gyr for the $10^{9.5}-10^{10.5}$ ($10^{10.5}-10^{11}$) stellar mass bins respectively. 
This means the average satellite fades with an $e$-folding timescale of $\tau_{\rm f} = 0.6 (0.4)$ Gyr. 
Our values are consistent with those from \citet{Wetzel13} at $z=0$ and other independent estimates 
at high-z. {\citet{Tal14} performed a statistical identification of central and satellite galaxies
in the UltraVISTA and 3D-HST fields and found that the onset of satellite quenching occurs 1.5-2 Gyr later 
than that of central galaxies at fixed number density. These values are in good agreement with the delay times 
estimated in our work.}

Conversely the ``slow quenching'' model predicts too many galaxies below the main 
sequence but which are not UVJ passive (``transition'' galaxies). The fraction of 3D-HST satellites for 
which $\Delta_{\rm MS}>-0.5$ is 65\% (46\%), which compares to 67\% (47\%) 
for the ``delayed then rapid'' model; instead it drops to 52\% (39\%) for the ``slow quenching'' model.

We tested that the distributions of $\Delta_{\rm MS}$ and the estimated fading times are not biased
by inaccurate UV+IR SFR for AGN candidates in the sample. Because the CANDELS fields have 
uniform coverage of deep {\it Spitzer}/IRAC observations, we remove the sources selected by the IRAC 
color-color criteria presented in \citet{Donley12}. We find that neither the $\Delta_{\rm MS}$ distributions, 
nor the fading time estimates change appreciably.
 
In conclusion the fading of the star formation activity must be a relatively rapid phenomenon which follows a 
long phase where satellite galaxies have a SFR which is indistinguishable from that of centrals. This is further 
supported by the evidence that the passive and star forming populations are well separated in color and SFR 
and that the ``green valley'' in between them is sparsely populated across different environments 
\citep{Gavazzi10, Boselli14b, Schawinski14, Mok14}.

\subsection{The gas content of satellite galaxies}
Finally, we discuss the implications of the quenching times on the gas content of satellites at the time of infall. 
Because satellite galaxies are not thought to accrete gas after infall, their continued star formation occurs at the 
expense of gas previously bound to the galaxy. 

\begin{figure}
\includegraphics[width = 8.5cm]{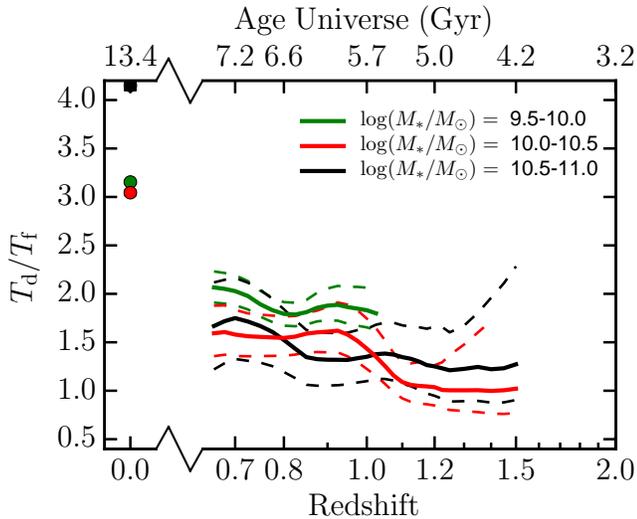}
\caption{Ratio of the delay time ($T_{\rm d}$) to the fading time ($T_{\rm f}$)
as a function of redshift in bins of stellar mass. The values of $T_{\rm f}$ at $z>0.5$ 
are derived in Section \ref{sec_delayedorslow}, while those at $z=0$ are taken 
from \citet{Wetzel13}. The uncertainties on the ratio propagate only the 
errors on the total quenching time.} 
\label{Tquench_Tfade}
\end{figure} 

As previously discussed, \citet{McGee14} explain the fading phase by the depletion of molecular gas.
The depletion time of molecular gas ($T_{\rm depl, H_2}$) has been derived by several authors \citep{Saintonge11,Tacconi13,Boselli14,Genzel15}.
There is general consensus that this timescale (which is an $e$-folding time) is $\sim 1.5$ Gyr at $z=0$ and is
$\sim 0.75$ Gyr at $z=1$. Moreover it is independent of stellar mass. 
In this framework we might expect fading times shorter than (or similar to) $T_{\rm depl, H_2}$, where shorter 
fading times are possible where a fraction of the gas is lost to outflows. 
Our fading times are indeed somewhat shorter than the molecular gas depletion times, consistent 
with this picture, but with a mass dependence which suggests a mass-dependent outflow rate.

In Figure \ref{Tquench_Tfade} we show the ratio of the delay time to the fading time
as a function of redshift in bins of stellar mass. 
Assuming that the fading phase is driven by
depletion of molecular gas (in absence of further replenishment), 
this ratio informs us about the relative time spent refuelling the galaxy to 
keep it on the MS (from a gas reservoir initially in a warmer phase) to 
the time spent depleting the molecular gas. We note that the delay time is 
estimated via the quenching time (which is a function of stellar mass and redshift) 
while the fading time is computed only for 2 stellar mass bins at $0.5<z<1.5$, with $z=0$ 
fading timescales taken from \citet{Wetzel13}. Errors in Figure \ref{Tquench_Tfade} propagate 
only the errors on the total quenching time.

For all stellar mass bins this ratio is above unity,  
which we interpret to mean that gas 
in a non-molecular phase is required to supply fuel for star formation, and this gas is likely 
to exceed the molecular gas in mass. The longer delay times at $z=0$ suggest that 
a smaller fraction of the gas mass is in molecular form. This model also implies
that a significant fraction of the final stellar mass of satellite galaxies is built-up during the
satellite phase. 

A multi-phase gas reservoir is observed in the local Universe in the form of ionized, atomic, and molecular hydrogen. Atomic hydrogen
cools, replenishing the molecular gas reservoir which is depleted by star formation. In the local Universe, using the scaling relations
derived from the {\it Herschel} Reference Sample \citep{Boselli14}, the observed mass of atomic hydrogen is found to be 2-3 times 
larger than the amount of molecular hydrogen for our most massive stellar mass bin. This ratio increases to $\sim8$ for the lower 
mass objects, although with a large uncertainty. 
These numbers are consistent with the picture that much, if not all, of the reservoir required to maintain the satellite on the MS 
during the delay phase at $z\sim0$ can be (initially) in an atomic phase. It is also plausible that much of the gas reservoir 
bound to higher redshift galaxies is contained in a non-molecular form, and that this can be retained and used for star formation 
when the galaxies become satellites.

Assuming that outflows are not only active during the fading phase but rather during the entire quenching time, the 
mass in the multi-phase gas reservoir needs to be even larger, although it is not straightforward to constrain by how much.

In conclusion our work supports a ``delayed then rapid'' quenching scenario for satellite galaxies regulated by star formation, 
depletion and cooling of a multi-phase gas reservoir.

\section{Conclusions} \label{sec_conclusions}
In this work, we have characterized the environment of galaxies in the 3D-HST survey at $z=0.5-3.0$. We used 
the projected density within fixed apertures coupled with a newly developed method for edge corrections to obtain
a definitive measurement of the environment in five well studied deep-fields: GOODS-S, COSMOS, UDS, AEGIS, GOODS-N.
Using a recent semi-analytic model of galaxy formation, we have assigned physical quantities describing the
properties of dark matter haloes to observed galaxies. Our results can be summarized as follows:

\begin{enumerate}
\item The 3D-HST deep fields host galaxies in a wide range of environments, from underdense regions to relatively
massive clusters. This large variety is accurately quantified thanks to a homogeneous coverage of high quality 
redshifts provided by the 3D-HST grism observations. 
\item As described in \citet{Fossati15}, a calibration of density into physically motivated quantities (e.g. halo
mass, central/satellite status) requires a mock catalogue tailored to match the properties of the 3D-HST survey.
We developed such a catalogue and performed a careful match to the observational sample. As a result, 
each 3D-HST galaxy is assigned a probability that it is a central or satellite galaxy with an associated 
probability distribution function of halo mass for each type. This Bayesian approach naturally takes
into account sources of contamination in the matching process. We publicly release our calibrated environment 
catalogue to the community.
\item The 3D-HST sample provides us with a unique dataset to study the processes governing 
environmental quenching from $z\sim 2$ to the present day over a wide range of halo mass. 
As no galaxy has a perfectly defined environment, a Bayesian analysis including forward modelling of 
the mock catalogue allows us to recover ``pure'' passive fractions of central and satellite 
populations. We also estimated robust and realistic uncertainties through a Monte Carlo 
error propagation scheme that takes into account the use of probabilistic quantities.
 \item By computing conversion fractions (i.e. the excess of quenched satellite galaxies compared to 
central galaxies at the same epoch and stellar mass) \citep{van-den-Bosch08}, we find that satellite 
galaxies are efficiently environmentally quenched in haloes of any mass up to $z\sim 1.2-1.5$. Above these 
redshifts the fraction of passive satellites is roughly consistent with that of central galaxies.
\item Under the assumption that the earliest satellites to be accreted become passive first, 
we derive environmental quenching timescales. These are long ($\sim2-5$ Gyr at $z\sim0.7-1.5$; 5-7 Gyr at $z=0$) 
and longer at lower stellar mass. As they become comparable to the Hubble time by $z\sim1.5$, 
effective environmental quenching of satellites is not possible at 
earlier times. More remarkably, their halo mass dependence is negligible. By assuming 
that cosmological accretion stops when a galaxy becomes a satellite, we were able to interpret these 
evidences in a ``gas exhaustion'' scenario \citep[i.e. the ``overconsumption'' model of ][]{McGee14} where 
quenching happens because satellite galaxies eventually run out of their fuel which sustains further star formation. 
\item We tested two toy models of satellite quenching: the ``delayed then rapid'' quenching scenario 
proposed by \citet{Wetzel13} and a continuous ``slow quenching''  from the time of first infall.  By comparing
the observed SFR distribution for 3D-HST satellites to the predictions of these toy models we found that the
scenario that best reproduces the data at $z\sim 0.5-1.5$ is ``delayed then rapid''. Consistently with the
results of \citet{Wetzel13} at $z=0$, we find that the fading of the star formation activity is a relatively rapid 
phenomenon ($\tau_{\rm f} \sim 0.4-0.6$ Gyr, lower at higher mass) which follows a long phase where satellite 
galaxies have a SFR which is indistinguishable from that of centrals.
\item By linking the fading to the
depletion of molecular gas we conclude that the ``delayed then rapid'' scenario is best explained, even at high redshift, 
by the presence of a significant multi-phase reservoir which can cool onto the galaxy and replenish the star forming gas 
at roughly the same rate as the gas is turned into stars. 
\end{enumerate}

This analysis of satellite quenching is only one of  many possible analyses that can be performed with the
environmental catalogue built in this work. In the future, the advent of the {\it James Webb Space 
Telescope, WFIRST} and {\it Euclid} space missions, as well as highly multiplexed spectroscopic instruments 
from the ground (e.g., {\it MOONS} at VLT; {\it PFS} at Subaru), 
will provide excellent redshift estimates for fainter objects
over a much larger area, to which similar techniques to calibrate environment can be applied. 
This, in combination with deeper scaling relations for the atomic and molecular 
gas components from the {\it Square Kilometer Array} and {\it ALMA} will revolutionize
measurements to constrain how galaxies evolve and quench as a function of their environment.

\acknowledgements
It is a pleasure to thank Nikhil Arora for his help in a preliminary examination of the local SDSS data, 
to Alessandro Boselli, Gabriella de Lucia, Sean McGee, Greg Rudnick, and Andrew Wetzel for useful discussion, to
Michael Balogh and Alexis Finoguenov for providing catalogs in useful format, and to
Stefano Zibetti and Tam{\'a}s Budav{\'a}ri for the work in helping deriving densities for the SDSS sample 
as presented in \citet{Wilman10}, and \citet{Phleps14}. We thank the anonymous referee
for his/her comments that improved the quality of the manuscript.

MFossati and DJW acknowledge the support of the Deutsche Forschungsgemeinschaft via 
Projects \mbox{WI 3871/1-1}, and \mbox{WI 3871/1-2}.
JCCC acknowledges the support of the Deutsche Zentrum f{\"u}r Luft- und Raumfahrt
(DLR) via Project ID 50OR1513. KEW gratefully acknowledge support by NASA through Hubble Fellowship 
grant \#HF2-51368 awarded by the Space Telescope Science Institute, which is operated by the 
Association of Universities for Research in Astronomy, Inc., for NASA.

This work is based on observations taken by the 3D-HST Treasury Program (GO 12177 and 12328) with the 
NASA/ESA HST, which is operated by the Association of Universities for Research in Astronomy, Inc., under 
NASA contract NAS5-26555. 

Based on observations made with ESO Telescopes at the La Silla or Paranal Observatories under 
programme ID(s) 175.A-0839(B), 175.A-0839(D), 175.A-0839(I), 175.A-0839(J), 175.A-0839(H), 175.A-0839(F), 
092.A-0091, 093.A-0079, 093.A-0187,  094.A-0217, 094.A-0287, 095.A-0047, 095.A-0109, 096.A-0093, 
and 096.A-0025.

The MOSDEF data were obtained at the W.M. Keck Observatory, which is operated as a scientific partnership 
among the California Institute of Technology, the University of California and the National Aeronautics and 
Space Administration. The Observatory was made possible by the generous financial support of the W.M. 
Keck Foundation. We recognize and acknowledge the very significant cultural role and reverence that the 
summit of Mauna Kea has always had within the indigenous Hawaiian community. We are most fortunate 
to have the opportunity to conduct observations from this mountain.

We acknowledge all the teams and observatories that provided datasets included in the 
photometric and spectroscopic catalogues used in this work.

Funding for SDSS-III has been provided by the Alfred P. Sloan Foundation, the Participating Institutions, the 
National Science Foundation, and the U.S. Department of Energy Office of Science. The SDSS-III web site 
is http://www.sdss3.org/.

SDSS-III is managed by the Astrophysical Research Consortium for the Participating Institutions of the 
SDSS-III Collaboration including the University of Arizona, the Brazilian Participation Group, Brookhaven 
National Laboratory, Carnegie Mellon University, University of Florida, the French Participation Group, the 
German Participation Group, Harvard University, the Instituto de Astrofisica de Canarias, the Michigan 
State/Notre Dame/JINA Participation Group, Johns Hopkins University, Lawrence Berkeley National 
Laboratory, Max Planck Institute for Astrophysics, Max Planck Institute for Extraterrestrial Physics, New 
Mexico State University, New York University, Ohio State University, Pennsylvania State University, 
University of Portsmouth, Princeton University, the Spanish Participation Group, University of Tokyo, 
University of Utah, Vanderbilt University, University of Virginia, University of Washington, and Yale University.


\bibliographystyle{apj}


\appendix

\section{Extended catalogs for edge corrections in the GOODS-S, COSMOS, and UDS fields} \label{edgecor_cats}

\subsection{GOODS-S}
The GOODS-S field is part of a larger field known as the Extended Chandra Deep Field South \citep[ECDFS,][]{Lehmer05}. This field
has been covered by the Multiwavelength Survey by Yale-Chile \citep[MUSYC,][]{Gawiser06} in 32 broad and medium bands from 
the optical to the medium infrared wavelengths. The broadband data originates from various sources \citep{Arnouts01, Moy03, Taylor09} 
and a consistent reduction and analysis is performed by the MUSYC team \citep{Cardamone10}.
The source extraction is performed on a deep combined image of three optical filters ($BVR$) and reaches a depth of $\sim 25.5$ mag.
Stars are removed from the catalogue by using the \texttt{star\_flag} parameter.

In order to select galaxies in a consistent way as for 3D-HST  we need deep observations in a filter with a central 
wavelength as close as possible to that of WFC3/F140W ($1.4 \mu m$). However, the near infrared observations from MUSYC are shallow 
and only reach a depth of $J = 22.4$ mag. We therefore match the MUSYC catalogue with the Taiwan ECDFS Near-Infrared Survey 
\citep[TENIS,][]{Hsieh12}. This survey provides deep $J$ and $Ks$ images of the ECDFS area with limiting magnitudes of 24.5 and 23.9 
respectively. Hereafter, where sky coordinates matching between different catalogues is required we select the closest match within 
a 1 arcsec radius. The comparison of $J$ band magnitudes 
from the two surveys for sources above the sensitivity limit of the MUSYC data shows a remarkable agreement.
We then match the MUSYC and 3D-HST/GOODS-S catalogue, again by sky coordinates. Using the galaxies that are present in both surveys
we fit a linear function between \jhhst\ and $J_{\rm TENIS}$ magnitudes. Given the significant overlap between the filters we neglect
color terms in the fit. The best bisector fit \citep{Isobe90} is $JH_{\rm 140} =  0.99 \times J_{\rm TENIS} + 0.22$. Then we use this function to 
generate \jhhst\ magnitudes for all objects in the MUSYC+TENIS catalogue.

We evaluate the depth of the resulting catalogue by inspecting the histogram of the number counts in the \jhhst \ band. Since this is obtained 
from deep  $J_{\rm TENIS}$ data (24.5 mag), the limiting factor will be the depth of the $BVR$ selection band of MUSYC.
The number counts increase linearly in log 
space up to $JH_{\rm 140} \sim 23.5$ and we therefore adopt this value for the selection. Since this limit is brighter than the one we set for the primary 
sample in 3D-HST, a fraction of the neighbours are missed. We correct for this by assigning to each MUSYC galaxy in eq. \ref{eqSigma} 
a weight $w = 1.42$  that corresponds to the ratio of the cumulative luminosity function at $JH_{\rm 140} = 23.5$ and $JH_{\rm 140} = 24$ mag 
from the deeper 3D-HST catalogue.

The most recent calculation of photometric redshifts in ECDFS is presented by \citet{Hsu14}. These authors combined the MUSYC 
photometry with TENIS and HST/CANDELS \citep{Guo13} in the GOODS-S area. We match our catalog with their catalog 
based on MUSYC ID and we find a match for each source. \citet{Hsu14} also present a compilation of spectroscopic redshifts 
available in the literature which we use whenever available. Photo-zs are computed using \textit{LePhare} \citep{Arnouts99, Ilbert06} 
and their accuracy depends primarily on the availability and depth of multiwavelength photometry. 
The GOODS-S area has deep HST coverage from CANDELS, but those galaxies are already present in our primary 
3D-HST catalog. Therefore we are primarily interested in sources outside the CANDELS/3D-HST area. In the ECDFS footprint 
which is not covered by HST more than 30 photometric bands are available and photo-zs are quite accurate\footnote[4]{We measure the photo-z accuracy using the normalized median absolute deviation (NMAD): $\sigma_{\rm NMAD} = 1.48 \times {\rm median}(|\Delta z|/(1 + z))$, where $\Delta z$ is the difference between the spectroscopic and the photometric redshift.}:
$\sigma_{\rm NMAD} \sim 3000~(4000)~\rm{km~s^{-1}}$ for galaxies with $H<23 (H>23)$ respectively. Those values degrade where continuum 
spectral features (e.g. Balmer break) are redshifted outside the range observed with medium band filters ($z>1.5$), although 
low number statistics hampers a robust determination of the photometric redshift quality.

Stellar masses are computed using the photometric data and the redshift information by choosing the same grid of templates 
used by \citet{Skelton14} for the 3D-HST fields. We assess the quality of the stellar masses by comparing to those from 
\citet{Skelton14}, where MUSYC and 3D-HST overlap and we find a median offset of 0.01 dex and a scatter of 0.15 dex.
The scatter is driven by the scatter in photometric redshifts in the two catalogs.

As a last step we remove from this catalog galaxies in the 3D-HST footprint that satisfy the selection criteria for the primary 
environment sample, to obtain a pure catalogue that we use only for the edge corrections.

\subsection{COSMOS}
The entire COSMOS 2deg$^2$ field has been observed in 30 photometric bands from UV to medium infrared (including
several medium bands) \citep{Sanders07, Taniguchi07, Erben09, Bielby12}. Photometric redshifts are computed using \textit{LePhare}
and are presented by \citet{Ilbert09} for sources with $i^+ < 25$ mag. We include spectroscopic redshifts from 
zCOSMOS-bright \citep{Lilly07} where available. 

The photometric redshift uncertainty is evaluated by \citet{Ilbert09} comparing photo-z to spec-z and is 
$\sigma_{\rm NMAD} \sim 2100~(9000)~\rm{km~s^{-1}}$ for galaxies with $i^+<22.5 (i^+>23)$ respectively. The latter value must be 
taken with caution as it is calibrated using a small number of objects. We remake this comparison by using
3D-HST spec-z and grism-z as a reference (restricting our analysis to the 3D-HST/COSMOS field). We divide the sample
into bright ($i^+<22.5$ mag) and faint ($i^+ \ge 22.5$ mag) for $0.5 < z \le 1.5$ and irrespective of magnitude for $1.5 < z \le 3.0$.
We note that for faint sources the effective magnitude limit is that of the grism redshift extraction $JH_{140} < 24$ mag
and the comparison is limited by the degraded accuracy of grism redshifts for faint sources with no emission line 
detection (see Figure \ref{fig1}).
The redshift accuracy is: $\sigma_{\rm NMAD} \sim 3000~(7500)~\rm{km~s^{-1}}$ for the bright (faint) sample at low redshift
and $\sigma_{\rm NMAD} \sim 8500~\rm{km~s^{-1}}$ for the high redshift sample.
Those values are consistent with the determination by \citet{Ilbert09} and the reduced accuracy at high redshift is due to 
the lack of narrow bands in the NIR. 

To overcome this limitation the Newfirm Medium Band Survey \citep[NMBS,][]{Whitaker11} observed the COSMOS field
with 5 medium band filters in the $J$ and $H$ bands and a broadband filter in $K$. As a result the accuracy of photometric
redshifts is significantly improved \citep[see Section 5 in ][]{Whitaker11} and we use those photo-z where they are available. 

Deep $J$ band magnitudes are provided by the UltraVISTA survey \citep{Mccracken12}. After matching their catalog via 
sky coordinates we generate synthetic $JH_{\rm 140}$ magnitudes as described in the previous section and using the best fit:
$JH_{\rm 140} =  0.98 \times J_{\rm UltraVISTA} + 0.31$. The depth of our catalog is limited by the depth of the $i^+$ selection band
from \citet{Ilbert09}. The number counts increase linearly in log space until $JH_{140} \sim 23.0$ and we therefore adopt this 
value for the selection limit. As for the MUSYC catalog this limit is brighter than the one we set for the primary 
sample in 3D-HST therefore we assign to each galaxy in eq. \ref{eqSigma} a weight ($w = 2.06$).
Stars are removed from the catalog by using the \texttt{type} flag from \citet{Ilbert09}. 

We compute stellar masses as described in the previous section. The agreement with stellar masses from \citet{Skelton14}
is remarkable, with a median offset of 0.02 dex and a scatter of 0.20 dex. Lastly we remove the primary 3D-HST sources 
from this edge correction sample.

\begin{figure*}
\begin{center}
\includegraphics[width = 16cm]{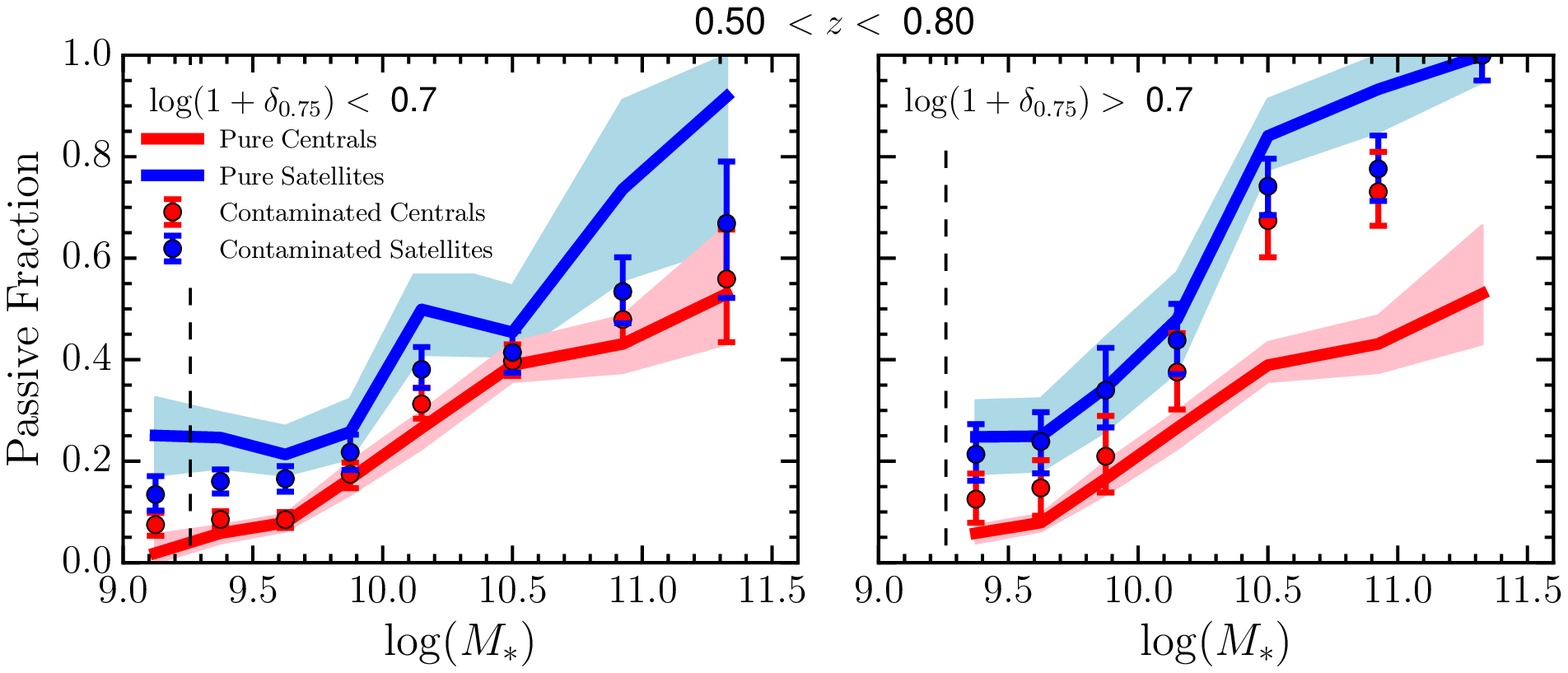}
\includegraphics[width = 16cm]{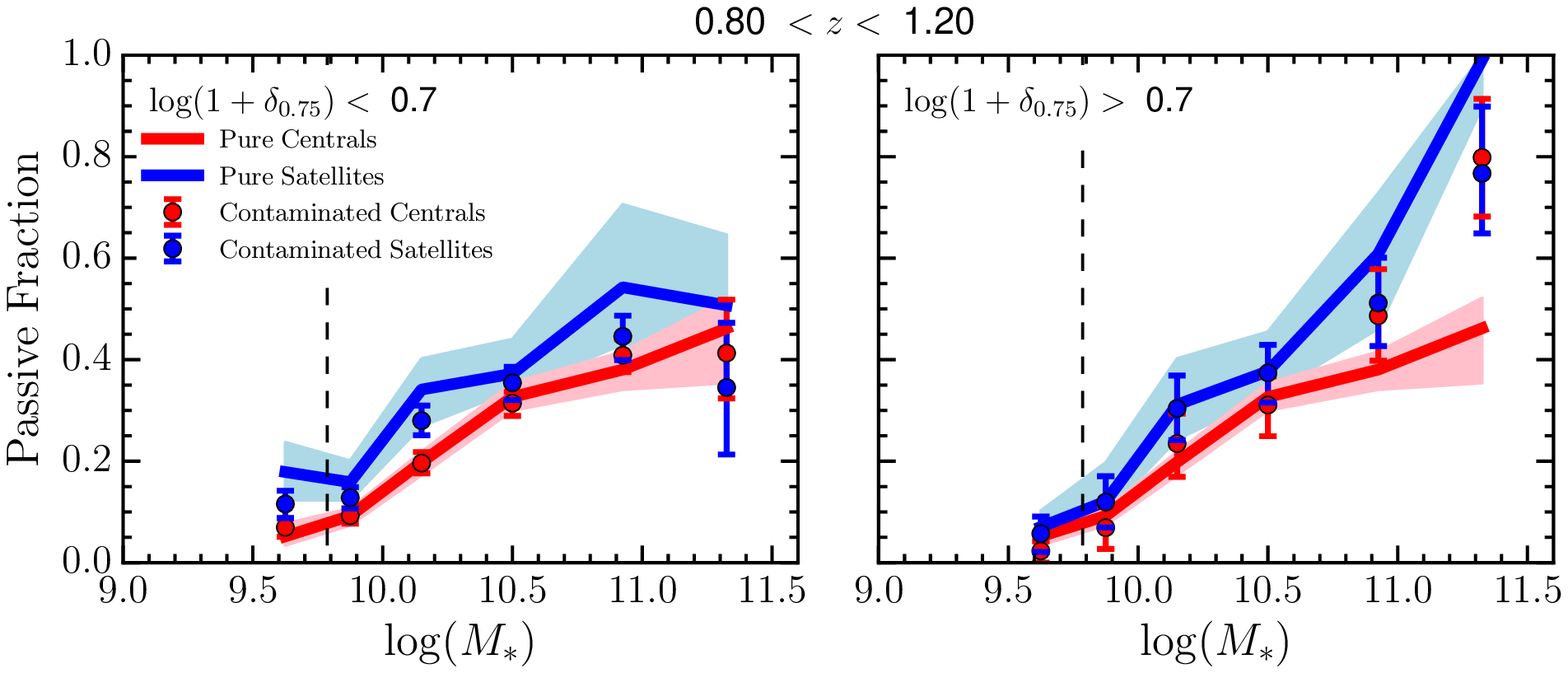}
\includegraphics[width = 16cm]{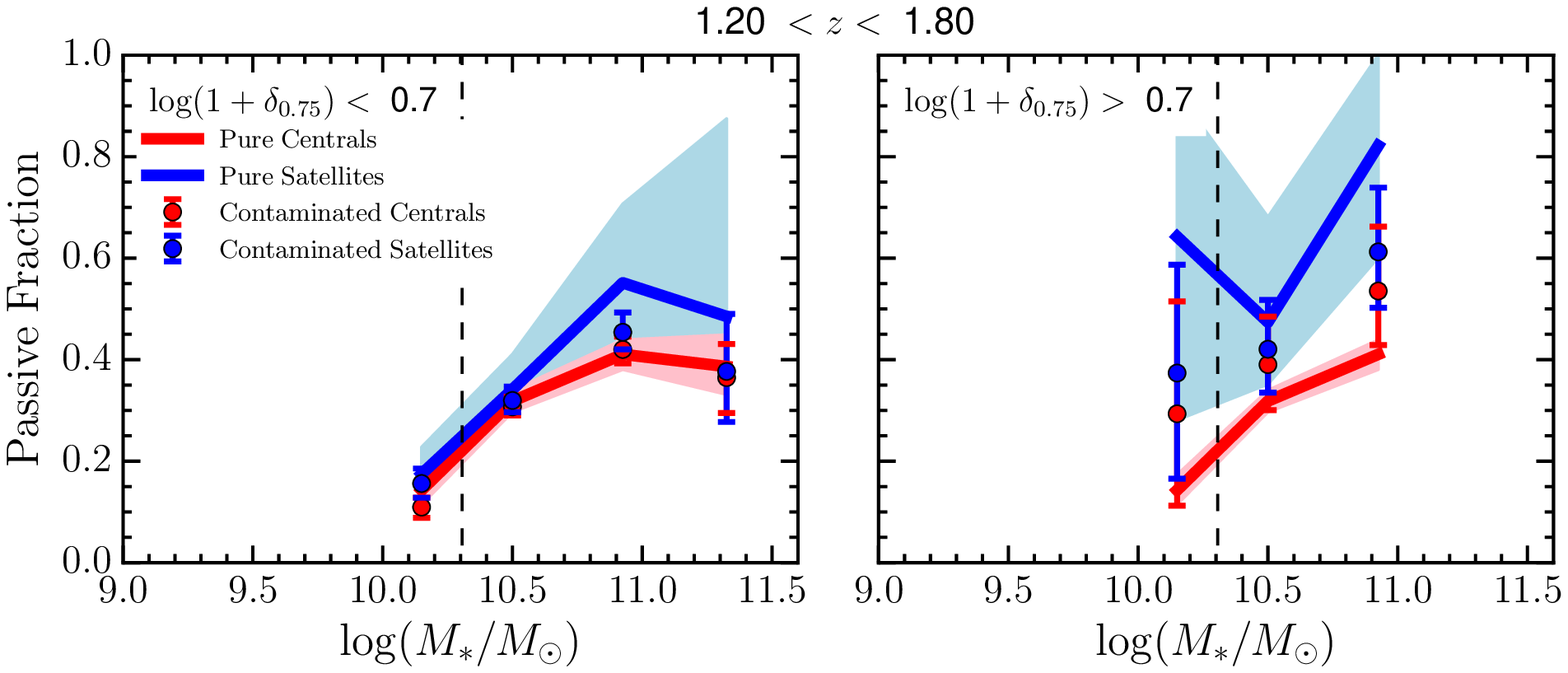}
\end{center}
\caption{Passive fraction for central and satellite galaxies in bins of  $M_{\rm{star}}$, density contrast  $\log(1+\delta_{0.75})$, 
and redshift. Datapoints are the observed passive fractions with uncertainties derived from Monte Carlo resampling of the mock sample. Thick blue 
and red lines are the ``pure'' passive fractions with $1\sigma$ confidence intervals as shaded regions. The vertical 
dashed line marks the stellar mass of the volume limited sample.} 
\label{Passfrac_Dens}
\end{figure*}

\subsection{UDS}
The 3D-HST UDS field is part of a larger field known as UKIDSS UDS. This field features deep near 
infrared $J$, $H$ , and $K$ observations with the UKIDSS telescope (Almaini et al. in prep) complemented 
by optical and medium infrared data \citep{Furusawa08, Ashby13}. 

The UDS/DR8 catalog selection is performed in $K$ band and the completeness limit is $K\sim 24.6$ mag. 
As for the previous fields we exclude stars and we compute synthetic \jhhst\ magnitudes using the best fit relation:
$JH_{\rm 140} =  0.98 \times J_{\rm UKIDSS} + 0.19$. The depth of our catalog matches the limiting magnitude for 
the primary 3D-HST sample, thus we do not apply any statistical weight for the UKIDSS UDS galaxies when 
computing the density. 

Photometric redshifts (W. Hartley private comm.)
have a typical accuracy of $\sigma \sim 9000~\rm{km~s^{-1}}$ due to the lack of narrow- or medium band 
photometry in this field. As for the other fields we compute stellar masses
using the FAST code and we find a good agreement with the values from \citet{Skelton14} for the 3D-HST/UDS field
with an offset of -0.03 dex and a scatter of 0.22 dex. Again the last step is to remove the 3D-HST primary sources via 
positional matching with the \citet{Skelton14} catalogue. 


\section{Passive fraction as a function of density} \label{app_passfrac_dens}

Halo mass is the parameter which most easily allows the interpretation of environmental 
effects across cosmic time. It also allows for easier and less biased comparisons across different works. 
Moreover it can be directly linked to models (either semi-analytic or hydrodynamical) 
allowing a better understanding of which physical processes are most relevant at different halo masses. 
Density, on the other hand, depends on the depth (and to some extent the observing strategy) of each survey.
Detailed and quantitative comparisons are also made difficult by different approaches to density 
\citep[e.g.,][]{Muldrew12, Haas12, Etherington15}. However it is a parameter directly obtained from 
the observed redshift space coordinates of the population of galaxies 
under investigation. In this respect it is less sensitive to the quality and uncertainties in the calibration of 
halo mass.

In this Appendix we derive the passive fraction of galaxies in two bins of density and compare them to those obtained
in Figure \ref{Passfrac_Mhalo} using halo mass.
The observed fractions of passive centrals and satellites in bins of $M_{\rm{*}}$ and density contrast $\log(1+\delta_{0.75})$ 
are given by
\begin{equation}
f_{\rm{pass|ty}} = \frac{\sum_i \left(\delta_{\rm pass,i} \times \delta_{M_* \rm{,i}}  \times \delta_{\log(1+\delta_{0.75}),i} \times P_{\rm{ty,i}}\right)}{\sum_i \left(\delta_{M_* \rm{,i}} \times \delta_{\log(1+\delta_{0.75}),i} \times P_{\rm{ty,i}} \right)}
\label{eqpassive_dens}
\end{equation}
where ty refers to a given type (centrals or satellites), $\delta_{\rm pass,i}$ is 1 if a galaxy is UVJ passive and 0 otherwise, 
$\delta_{M_* \rm{,i}}$ is 1 of a galaxy is in the stellar mass bin and 0 otherwise,  $\delta_{\log(1+\delta_{0.75}),i}$ is 1 if a galaxy is 
in the density bin and 0 otherwise, and $P_{\rm{ty,i}}$ is the probability that a galaxy is of a given type. 
In this equation the only uncertain property for each object is its central/satellite status, while the cross talk between multiple density
bins is not present (as it was for halo mass). 

We therefore perform a simpler decontamination procedure. For each density, stellar mass and redshift bin, we assign to real centrals 
in the mocks a probability of being passive equal to the passive fraction of the pure sample of observed central galaxies 
$f_{\rm{pass|cen,pure}}(M_*,\delta_{0.75})$, while the passive fraction of satellites $f_{\rm{pass|sat,pure}}(M_*,\delta_{0.75})$ is a free parameter. 
Then we use equation \ref{eqpassive_dens} to compute the observed passive fractions for mock galaxies 
(therefore contaminating the ``pure'' values). 
We solve for  $f_{\rm{pass|sat,pure}}(M_*,\delta_{0.75})$ by maximising the likelihood that the contaminated passive fractions for mock 
galaxies match the observed passive fractions (jointly for centrals and satellites).
This procedure is repeated 500 times in a Monte Carlo fashion in order to propagate the uncertainties on the datapoints to the 
``pure'' (decontaminated) passive fractions. 

The decontaminated values of the passive fraction for centrals and satellites shown in Figure \ref{Passfrac_Dens} are qualitatively 
similar to those obtained in bins of halo mass in the same redshift slices (see Figure \ref{Passfrac_Mhalo}). 

We conclude that the dependence of environmental quenching when binned on local density is similar to that in bins of halo 
mass, where density is a more directly observed quantity.

\section{An example of the fitting procedure to recover the passive fraction of pure satellites} \label{app_decontamination}
In this Appendix we illustrate the results of the fitting process described in Section \ref{sec_decontamination} for a single 
redshift bin.

\begin{figure}
\includegraphics[width = 1.00\columnwidth]{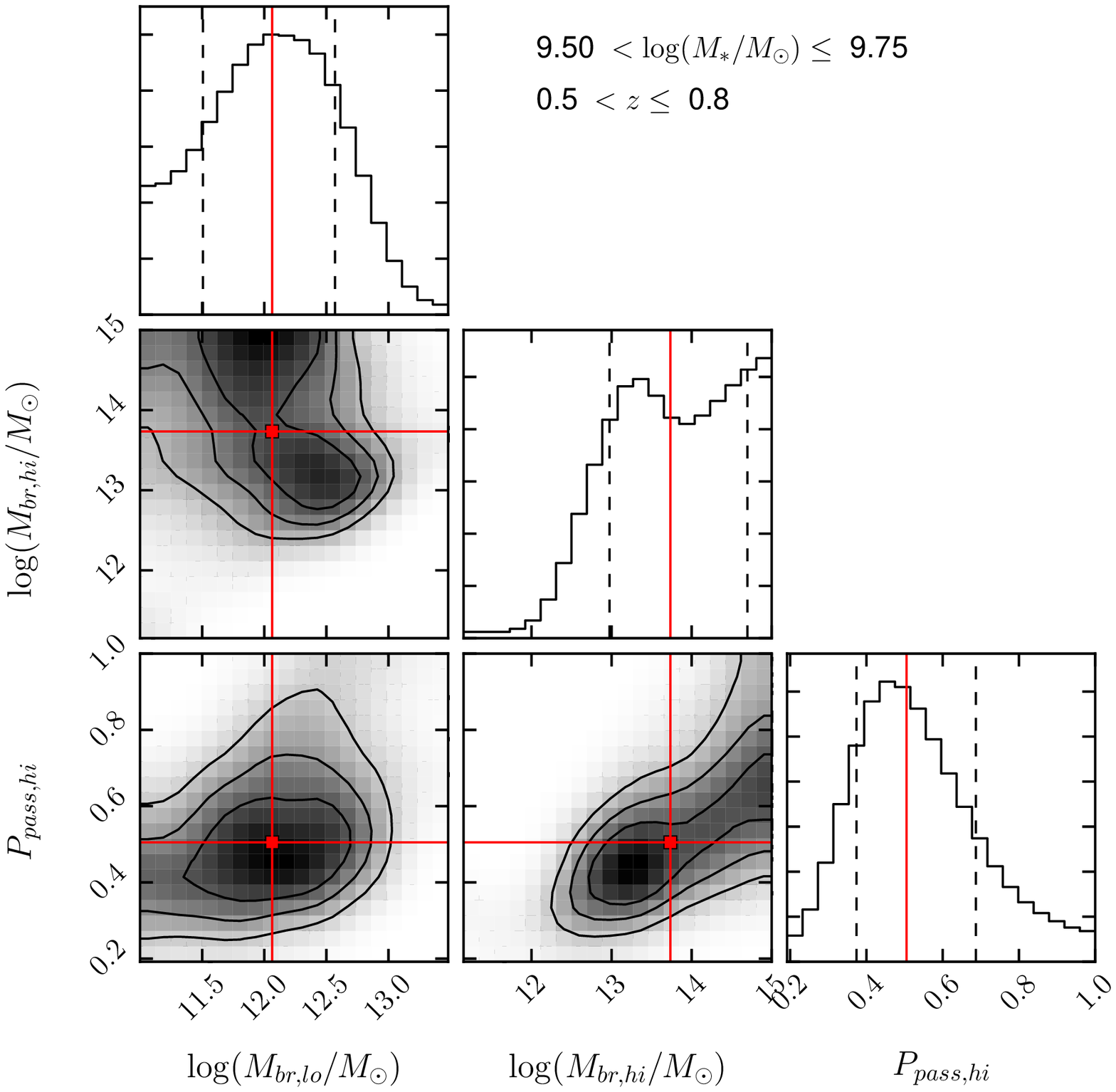}
\caption{Marginalized likelihood distributions for individual model parameters (panels along the diagonal) and 
marginalized maps for pairs of parameters, for the stellar mass bin $9.50 < \log(M_*/M_\odot) \leq 9.75$ and 
redshift bin $0.5 < z \leq 0.8$. The red lines show the median value for each parameter (which may be distinct 
from the global maximum likelihood value). The vertical dashed lines in the histograms show the $1\sigma$ confidence 
intervals. The black contours in the two dimensional maps show the $1\sigma$, $1.5\sigma$, and $2\sigma$ 
confidence intervals. } 
\label{Corner_plot_example}
\end{figure}

Figure \ref{Corner_plot_example} presents the constraints on the model parameters 
(marginalised over the nuisance parameter $P_{\rm{pass|cen}}$) for a single stellar mass and redshift bin.
The panels along the diagonal show the marginalised posterior distributions for each for the 
three parameters $(M_{\rm br,lo},M_{\rm br,hi},P_{\rm pass,hi})$. The red solid lines show the median value of each 
parameter, and the black dashed lines show the $1\sigma$ confidence intervals. 
The off-diagonal panels show the marginalised posterior distributions for a pair of model
parameters. The black contours show the $1\sigma$, $1.5\sigma$, and $2\sigma$ confidence
intervals. The fits for the other stellar mass bins give qualitatively similar results. 

\begin{figure*}
\centering
\includegraphics[width = 17cm]{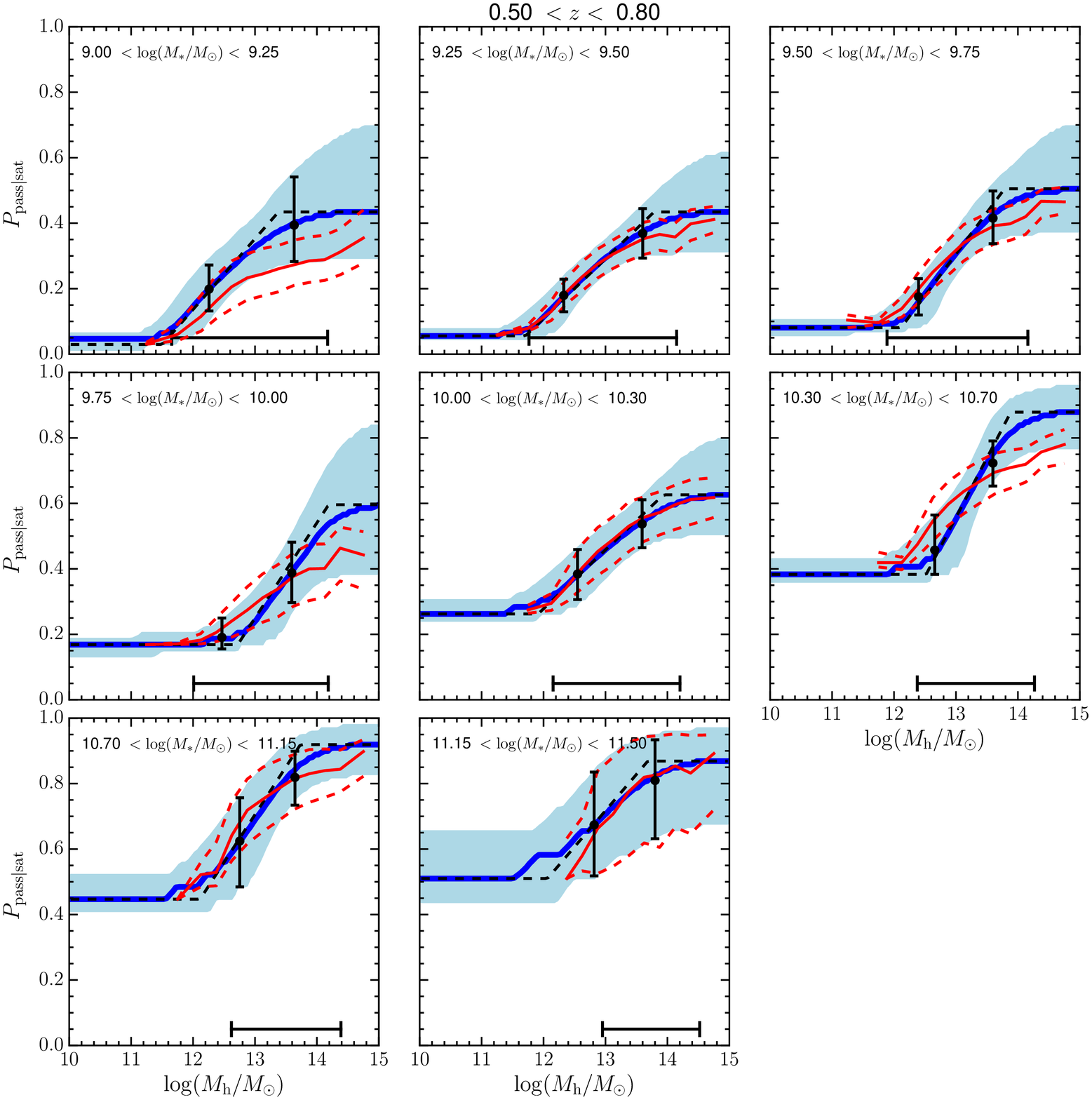}
\caption{Median (thick blue lines) and $1\sigma$ confidence intervals (shaded areas) of the probability that a 
satellite galaxies is passive ($P_{\rm{pass|sat}}$) as a function of halo mass from our fitting method in 
different stellar mass bins at $0.5 < z \leq 0.8$.  The black dashed line is the best fit model with
values obtained from the marginalised distributions in each parameter.
The horizontal black lines show the halo mass 
range that includes $90\%$ of the satellites in each stellar mass bin.  The black points with errorbars show the
average value of $P_{\rm{pass|sat}}$ (and its 1$\sigma$ uncertainty) for galaxies in haloes above and below
$10^{13}M_\odot$. The red lines are the
median prediction (solid) and $1\sigma$ confidence intervals (dashed lines) 
for $P_{\rm{pass|sat}}$ under the assumption of a quenching timescale independent of halo mass.
In most of the stellar mass bins this assumption well reproduces the best fit of $P_{\rm{pass|sat}}$.} 
\label{Psat_model_example}
\end{figure*}

Figure \ref{Psat_model_example} shows the 
median value (thick blue lines) of $P_{\rm{pass|sat}}$ as a function of log halo mass and $1\sigma$ confidence 
intervals in each stellar mass bin. Despite the significant covariance of the model parameters, 
the shape of the passive fraction models for satellites is
well determined. The horizontal black lines show the halo mass 
range that includes $90\%$ of the satellites in each stellar mass bin.

The average passive fractions in the two halo mass bins above and 
below $10^{13}M_\odot$, presented as the thick blue lines in Figure \ref{Passfrac_Mhalo}, 
are shown for each stellar mass bin in Figure \ref{Psat_model_example} by the black points.

We add in Figure \ref{Psat_model_example} an additional test of the result presented in Section 
\ref{sec_tquench_mhalo} that the quenching time is largely independent of halo mass. We compute
a single quenching time per stellar mass bin without binning the data in halo mass. Then we 
compute which fraction of mock galaxies have $T_{\rm sat} > T_{\rm quench}$ as a function of 
halo mass. This is converted in a probability of being passive as a function of halo mass which we show as
solid red lines (with $1\sigma$ confidence intervals as dashed lines) in Figure \ref{Psat_model_example}.
The agreement with the best fit values of $P_{\rm{pass|sat}}$ is remarkable in most of the stellar mass bins,
further supporting the result of a quenching timescale that is independent of halo mass.

\section{A z=0 sample from SDSS}  \label{app_SDSS}
\subsection{Observational data}
The $z=0$ points in Figures \ref{Convfracz} and \ref{Tquenchz} are obtained from a sample of galaxies in the local Universe 
selected from the SDSS \citep{York00} survey. We use the data from the SDSS DR8 database \citep{Aihara11} cross correlated with an updated 
version of the multi-scale density catalog from \citet{Wilman10} \citep[with densities computed according to equation \ref{eqSigma}; updated DR8 catalog 
as used by][]{Phleps14, Hirschmann14}. 
SDSS DR8 includes 5 color $ugriz$ imaging of 14555 square degrees.  The spectroscopic part of the survey provides redshifts for 
77\% of objects brighter than a limit of $r=17.77$  across 8032 square degrees.  Our sample is derived from the spectroscopic database. Luminosities are 
computed by k-correcting and adding the distance modulus to the Petrosian r-band magnitude. k-corrections are performed using the 
{\sc k-correct idl} tool \citep{Blanton07}. We select as primary galaxies those with $M_r < -18$ mag and $0.015 < z < 0.08$. In contrast to the method we use at high redshift the 
sample of neighbours (galaxies used to calculate the density in equation \ref{eqSigma}) is restricted to $M_r < -20$ mag.
This ensures a volume limited sample for the neighbours in this redshift range, while for the primary galaxies we correct for volume incompleteness 
using $V_{\rm max}$ corrections. The primary sample numbers $\sim3\times10^5$ galaxies.  
Stellar masses and star formation rates are obtained from the JHU-MPA\footnote[5]{http://www.mpa-garching.mpg.de/SDSS/DR7/} catalogues updated to DR7 \citep{Brinchmann04, Kauffmann03}. 

For this work we use the density computed on a fixed scale of 1 Mpc, with a velocity cut of $dv = \pm1000 \rm{km~s^{-1}}$. This scale is larger than what we use in the 
3D-HST sample in order to take into account the growth of structure with cosmic time. We stress that our results do not significantly depend on the scale 
chosen because the halo mass calibration is performed self consistently and we only compare calibrated quantities across the two samples. 
We have further computed stellar mass ranks for each primary galaxy in the adaptive aperture as described in Section \ref{sec_censat}.

\begin{figure}
\includegraphics[width = 0.95\columnwidth]{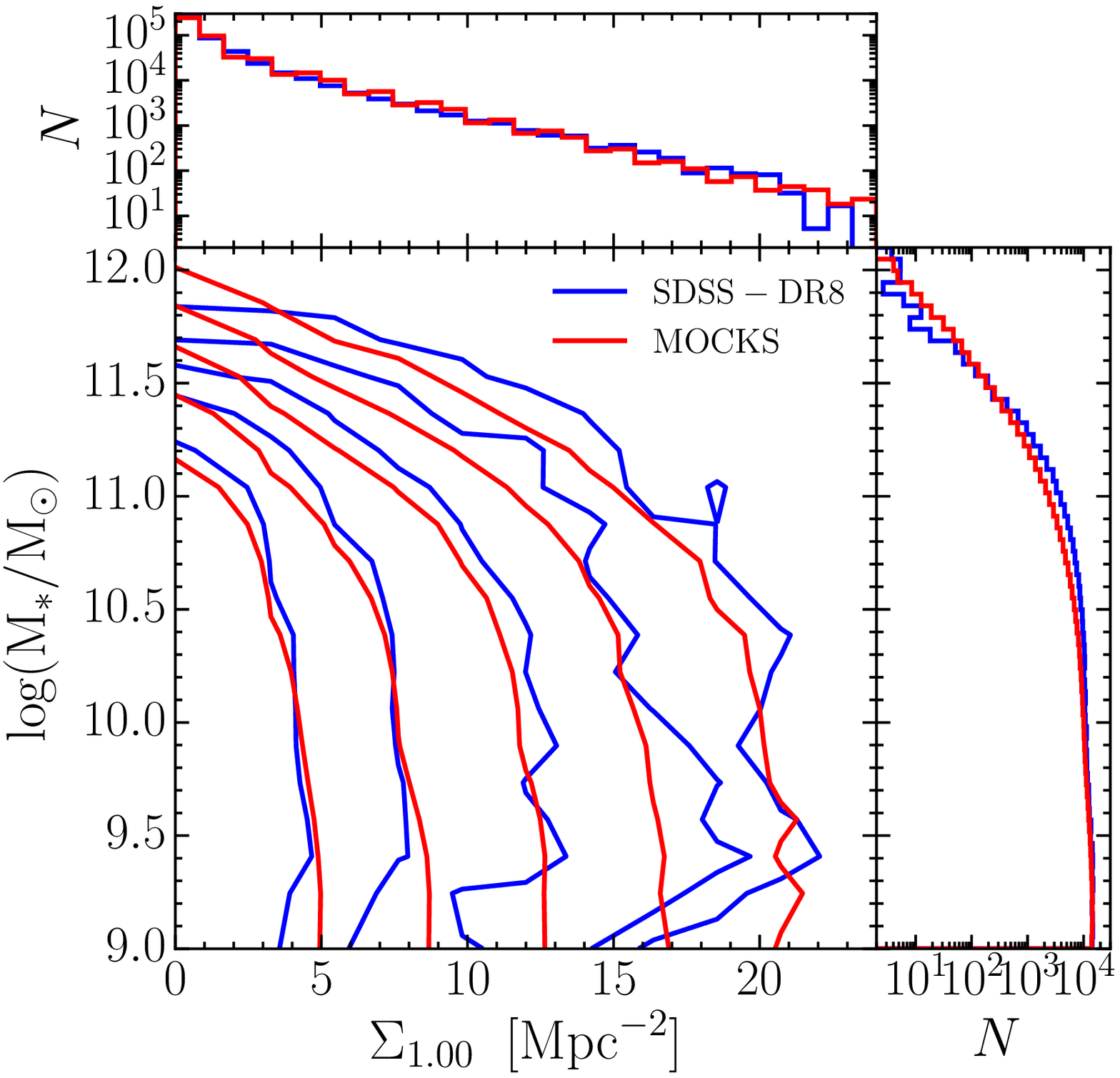}
\caption{Main panel: bivariate distribution of density on the 1.00 Mpc scale and stellar mass for the SDSS sample (blue) and
the mock sample (red). The mock contours have been scaled to account for the ratio of volumes between the simulation box 
and the data. The contours are logarithmically spaced with the outermost contour at 4 objects per bin and the innermost 
at 300 objects per bin. Upper panel: marginalized distributions of density on the 1.00 Mpc scale for the SDSS and the mock samples. 
The counts refer to the SDSS sample while the mock histogram has been normalized by the ratio of the volumes. Right-hand panel:
same as above but marginalized over the stellar mass.} 
\label{Mstar_dens_compare_sdss}
\end{figure}

One limitation of the SDSS spectroscopic strategy is that not all the spectroscopic targets can be actually observed because two fibers cannot be
placed closer than $55"$ on the sky and each patch of the sky is only observed once (although with small overlaps between adjacent spectroscopic plates). 
As a result the spectroscopic catalogue does not contain all the
sources detected in the imaging. Spectroscopic incompleteness is taken into account in the computation of the densities as described by \citet{Wilman10},
and we further consider it when we match to the mock galaxy sample. 
Passive galaxies are selected using the specific star formation rate ($sSFR$) as a tracer. For consistency with previous studies \citep[e.g.][]{Hirschmann14}
we define passive galaxies those with $sSFR < 10^{-11} {\rm yr}^{-1}$. We note that this corresponds to a $\sim1$ dex offset from the main sequence 
of star forming galaxies at $z=0$, which is consistent with the division of UVJ star forming from UVJ passive galaxies adopted in Section \ref{sec_passfrac_mhalo}.

\subsection{The model sample}
We generate a model galaxy sample that matches the stellar mass and density distributions of the SDSS observational catalogue. To do so we take
the SAM from \citet{Henriques15} at the $z=0$ snapshot of the Millennium simulation. In this case we do not use lightcones but a three dimensional box because
of the large area covered by SDSS and the single redshift bin. Densities are computed by projecting one of the
axes of the box into a redshift axis as described in \citet{Fossati15}.  We set an aperture size of 1 Mpc, a velocity cut $dv = \pm1000 \rm{km~s^{-1}}$, 
and we compute densities according to equation \ref{eqSigma}. 

\begin{figure*}
\begin{center}
\includegraphics[width = 15.5cm]{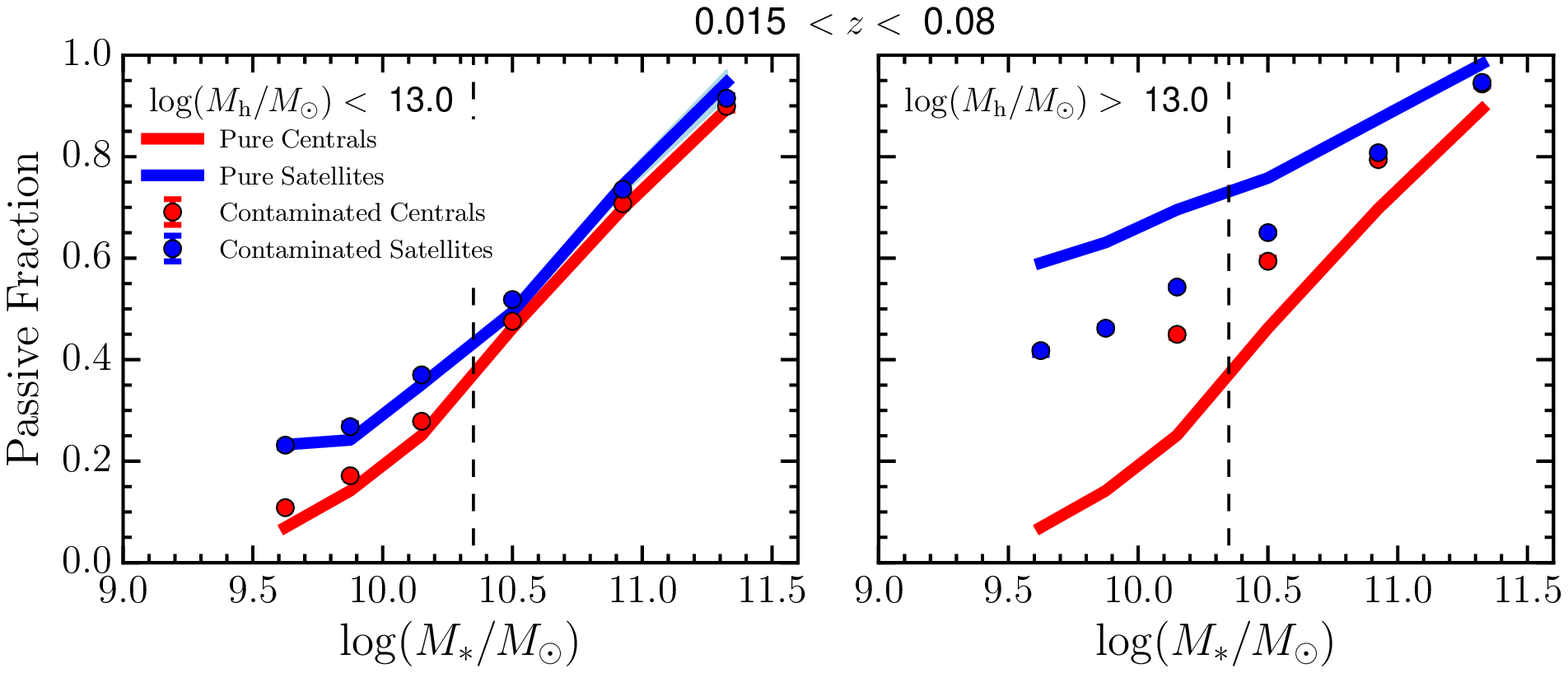} 
\includegraphics[width = 15.5cm]{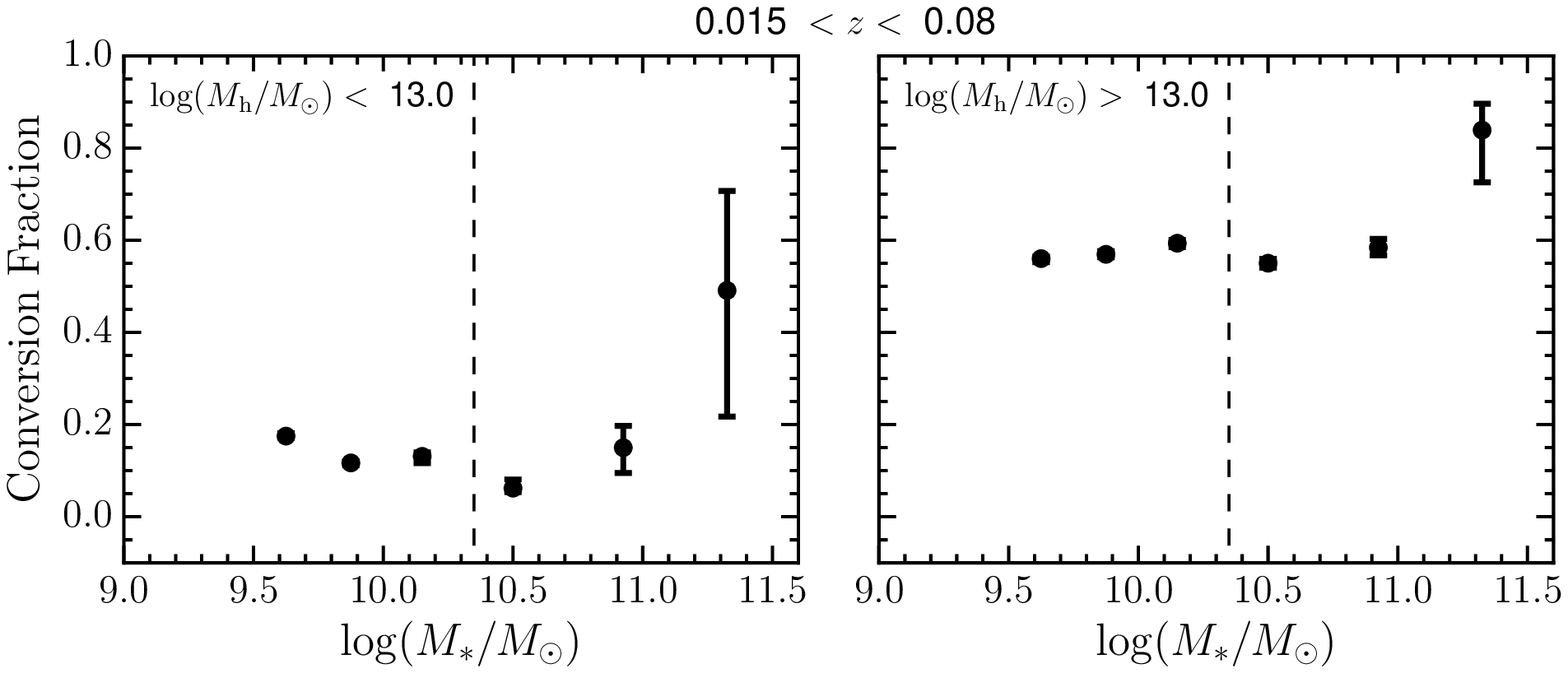} 
\includegraphics[width = 15.5cm]{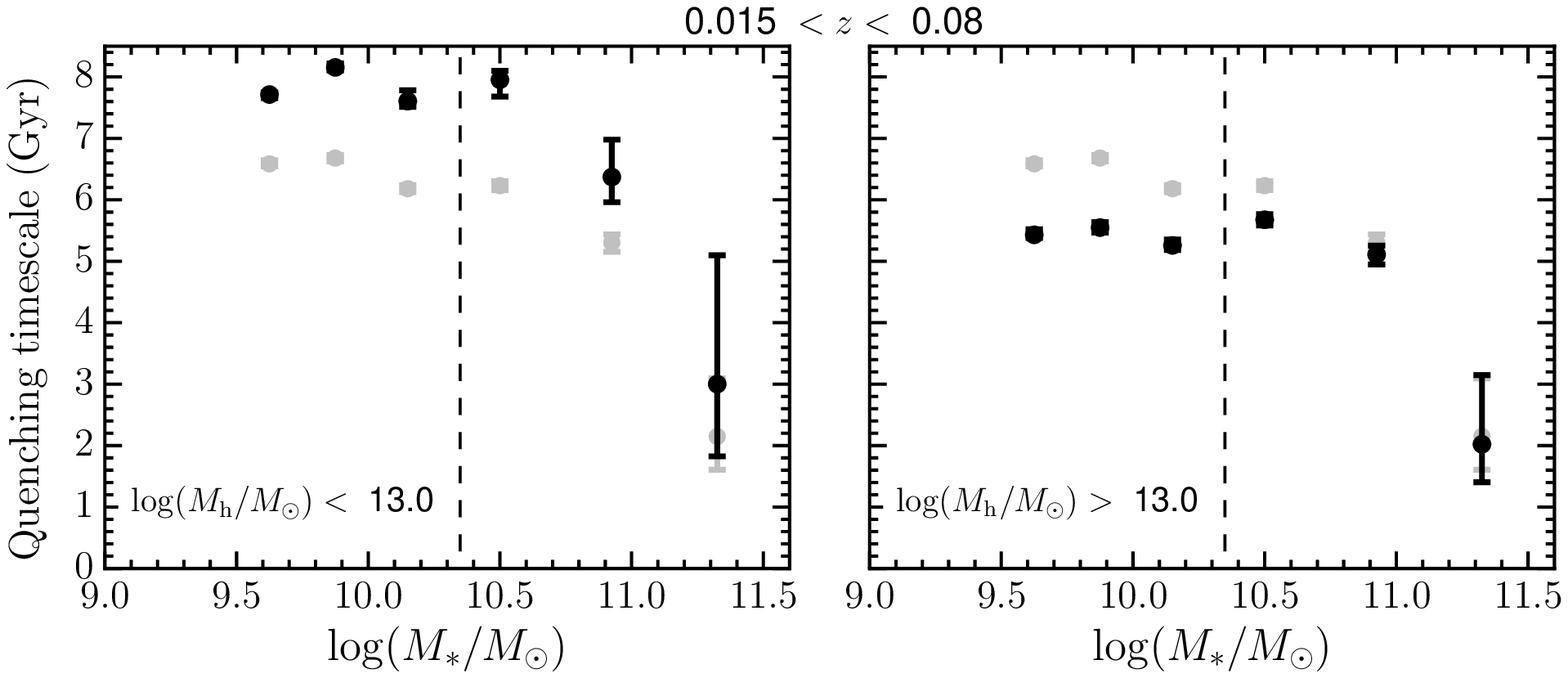} 
\end{center}
\caption{Top Panels: passive fraction for central and satellite galaxies in bins of  $M_{\rm{*}}$ and $M_{\rm{halo}}$ for the SDSS sample. 
The median (log) halo masses for satellites are 12.39, 13.77 for the lower and higher halo mass bin respectively. Points and lines are color coded as in 
Figure \ref{Passfrac_Mhalo}. Middle panels: conversion fractions for satellite galaxies in bins of  $M_{\rm{*}}$ and $M_{\rm{h}}$ obtained from equation 
\ref{eqConvfrac} for the SDSS sample. Bottom Panels: quenching times for satellite galaxies in bins of  $M_{\rm{*}}$ and $M_{\rm{h}}$ for the SDSS
sample.} 
\label{threeplots_Mhalo_SDSS}
\end{figure*}

The model sample does not suffer from spectroscopic incompleteness; on the other hand the distribution of $r$-band magnitudes does not 
match perfectly the one obtained from the observations. To overcome both those issues at once we employ a method that iterates on the 
magnitude limits for the primary and the neighbours samples until the number density and the density distribution of the selected sample 
match the observational data. 
Before doing that we need to derive the total number of photometric galaxies in the SDSS DR8 footprint (more precisely in the area followed 
up by spectroscopy) that would have been observed if fiber collisions were not a limitation. We query the SDSS database for the number of 
galaxies in the spectroscopic database and the number of galaxies in the photometric database that would satisfy the criteria for 
spectroscopic follow-up. The ratio of those values is 0.769. Therefore the number density of mock galaxies needs to be 
$\rho_{\rm mod} = 1.3 \times \rho_{\rm SDSS,sp}$ where $ \rho_{\rm SDSS,sp}$ is the number density of primary galaxies in our observational 
catalogue once we account for $V_{\rm max}$ corrections. The absolute magnitude cuts we set in the models using this iterative method are 
$M_r < -17.6$ mag and $M_r < -19.0$ mag for the primary and the neighbour samples respectively. We note that these cuts are up to 1 mag 
deeper than those used in the SDSS sample. This difference arises in a non perfect match of the $r$-band luminosity function, while stellar mass
functions are better matched between the SAM and the SDSS data. Figure \ref{Mstar_dens_compare_sdss} 
shows that, with this choice of magnitude limits, both the density and the stellar mass distributions are well matched. 
This is a critical step to trust our Bayesian approach to halo mass and central/satellite status. 

As a last step we assign to each SDSS galaxy (and to model galaxies) a probability that it is central ($P_{\rm{cen}}$) or satellite ($P_{\rm{sat}}$) and
the halo mass PDFs $P_{M_h | {\rm{cen}}}$ and $P_{M_h | {\rm{sat}}}$ as described in Section \ref{sec_halomass}.

Figure \ref{threeplots_Mhalo_SDSS} shows the passive fraction for centrals and satellites, conversion fractions and satellite quenching timescales derived for the SDSS 
sample as described in Sections \ref{sec_passfrac_mhalo}, \ref{sec_convfrac_mhalo}, and \ref{sec_tquench_mhalo}. 
Section \ref{sec_redshiftevo} contains the scientific discussion of these results in the context of the evolution
of satellite quenching efficiency and timescales from $z=0$ to $z\sim 2$.

\section{Description of the environment catalogue for the 3D-HST sample} \label{envcatalogue}
The environmental properties of 3D-HST galaxies are made available at \url{http://dx.doi.org/10.5281/zenodo.168056}. Conditional halo mass PDFs given that 
each galaxy is a central or a satellite and covering the range $10 < \log(M_h/M_\odot) < 15$ with 100 
uniform bins are also available as separate tables in the same repository. Table \ref{envtableexample} gives an example of the quantities provided in the 
catalog and the description of the columns follows:
\begin{itemize}
\itemsep0em
\item (1) 3D-HST field
\item (2) 3D-HST photometric ID from \citet{Skelton14}
\item (3) 3D-HST spectroscopic (grism) ID from \citet{Momcheva16}
\item (4) fraction of the 0.75 Mpc aperture in the photometric catalogue
\item (5) density of galaxies in an aperture of 0.75 Mpc radius (see eq. \ref{eqSigma})
\item (6) overdensity of galaxies in an aperture of 0.75 Mpc radius (see eq. \ref{eqoverdens})
\item (7) stellar mass rank in the adaptive aperture
\item (8) and (9) probability that the galaxy is a central or a satellite
\item (10), (11), and (12) $16^{\rm th}$, $50^{\rm th}$, and $84^{\rm th}$ percentile of the log halo mass cumulative PDF given that the galaxy is a central
\item (13), (14), and (15) $16^{\rm th}$, $50^{\rm th}$, and $84^{\rm th}$ percentile of the log halo mass cumulative PDF given that the galaxy is a satellite
\end{itemize}

\begin{table*}
\begin{center}
\resizebox{18.cm}{!} {
  \begin{tabular}{c c c c c c c c c c c c c c c}
    \hline
        Field & PhotID & SpecID & $f_{\rm area,0.75}$ &  $\Sigma_{0.75}$ & $\delta_{0.75}$ & $M_{\rm rank}$ & $P_{\rm CEN}$ & $P_{\rm SAT}$ & $M_{\rm h,16|CEN}$ & $M_{\rm h,50|CEN}$ & $M_{\rm h,84|CEN}$ &$M_{\rm h,16|SAT}$ &$M_{\rm h,50|SAT}$ & $M_{\rm h,84|SAT}$ \\
 (1) & (2) & (3) & (4) & (5) & (6) & (7) & (8) & (9) & (10) & (11) & (12) & (13) & (14) & (15) \\
    \hline
        COSMOS & 22162  & cosmos-16-G141\_22162 & 1.00 & 3.96 & 0.984 & 1 & 0.911 & 0.089 & 11.968 & 12.195 & 12.524 & 12.429 & 12.754 & 13.134 \\
        UDS         & 19166  & uds-07-G141\_19166 & 1.00 & 18.67 & 7.937 & 5 & 0.199 & 0.801 & 11.774 & 11.967 & 12.399 & 13.168 & 13.697 & 14.049 \\
        AEGIS     & 19285   & aegis-09-G141\_19285 & 1.00 & 2.83 & 1.078 & 4 & 0.544 & 0.456 & 11.867 & 12.087 & 12.510 & 12.537 & 12.917 & 13.352 \\
       
    ..... \\
    \hline
  \end{tabular}}
  \caption{Example of the environmental catalogue table made available with this work.}
  \label{envtableexample}
\end{center}
\end{table*}


\end{document}